%% file: main.tex
\documentclass[fleqn,usenatbib]{mnras}

\usepackage{amssymb}
\usepackage{newtxtext,newtxmath}
\usepackage{subcaption}
\usepackage{graphicx}
\usepackage{xspace}
\usepackage[T1]{fontenc}
\usepackage{fix-cm}
\usepackage{longtable}
\usepackage{amsmath}

\DeclareRobustCommand{\VAN}[3]{#2}
\let\VANthebibliography\thebibliography
\def\thebibliography{%
    \DeclareRobustCommand{\VAN}[3]{##3}%
    \VANthebibliography
}

\newcommand{\mass}{\mbox{M$_{\odot}$}}
\newcommand{\mzams}{$\mathrm{M}_{\mathrm{ZAMS}}$\xspace}

\title[]{Core-Collapse Supernovae from Single and Binary-Stripped Massive Stars -- I. Explodability and Nucleosynthesis}

\pubyear{2026}

\author[J. M. Pollin et al.]{J. M. Pollin,$^{1}$\thanks{E-mail: pollinj@oregonstate.edu}
S. Curtis,$^{1}$
C. Fr\"ohlich,$^{2}$
E. J. Griffith,$^{3}$\thanks{NASA Hubble Fellow}
A. P. Ji,$^{4,5,6}$
D. H. Weinberg,$^{7,8}$
J. A. Johnson,$^{7,8}$
\newauthor
R. A. Patton,$^{9}$
and C.~E.~Fields$^{10}$\\
\parbox[t]{\textwidth}{%
\raggedright
$^{1}$Department of Physics, Oregon State University, 301 Weniger Hall, Corvallis, OR 97331-6507, USA\newline
$^{2}$Department of Physics, North Carolina State University, Raleigh, NC 27606, USA\newline
$^{3}$Center for Astrophysics and Space Astronomy, Department of Astrophysical and Planetary Sciences, University of Colorado, 389~UCB, Boulder, CO 80309-0389, USA\newline
$^{4}$Kavli Institute for Cosmological Physics, University of Chicago, 5640 S Ellis Avenue, Chicago, IL 60637, USA\newline
$^{5}$Department of Astronomy \& Astrophysics, University of Chicago, 5640 S Ellis Avenue, Chicago, IL 60637, USA\newline
$^{6}$NSF-Simons AI Institute for the Sky (SkAI), 172 E.\ Chestnut St., Chicago, IL 60611, USA\newline
$^{7}$Department of Astronomy, The Ohio State University, 140 West 18th Avenue, Columbus, OH 43210, USA\newline
$^{8}$Center for Cosmology and AstroParticle Physics, The Ohio State University, 191 West Woodruff Avenue, Columbus, OH 43210, USA\newline
$^{9}$Department of Physics and Astronomy and Pittsburgh Particle Physics, Astrophysics and Cosmology Center (PITT PACC), University of Pittsburgh, 3941 O'Hara Street, Pittsburgh, PA 15260, USA\newline
$^{10}$Department of Astronomy and Steward Observatory, University of Arizona, 933 North Cherry Avenue, Tucson, AZ 85721-0065, USA
}%
}

\begin{document}
\label{firstpage}
\pagerange{\pageref{firstpage}--\pageref{lastpage}}
\sloppy
\maketitle

\begin{abstract}
Binary interaction can substantially alter massive-star evolution and pre-supernova structure, changing core-collapse outcomes and nucleosynthetic yields.
We compare explodability and nucleosynthesis in 66 non-rotating, solar-metallicity single-star and binary-stripped progenitors spanning zero-age main-sequence (ZAMS) masses of $11$--$45\,{\rm M}_\odot$, using the spherically symmetric PUSH framework and nucleosynthesis post-processing.
Both populations explode up to $22\,{\rm M}_\odot$, with binary stripping altering explodability at higher ZAMS masses.
Successful explosions span similar energies ($\sim$0.4--1.3$\,$B) and remnant masses ($\sim$1.4--2.0$\,{\rm M}_\odot$).
The largest nucleosynthetic differences occur at low ZAMS masses, particularly for material undergoing explosive O and Si burning.
The $^{56}$Ni yields and iron-group abundances broadly agree with observations of core-collapse supernovae and the metal-poor star HD~84937.
When compared with calculations using time-dependent mixing-length theory and a two-parameter explodability criterion for the same progenitors, our results reveal substantial differences in explodability, with agreement on successful explosions for only 5 cases across all three approaches.
Nevertheless, elemental yield patterns can remain broadly similar for models that explode successfully across treatments.
Salpeter initial-mass-function-weighted yields of most elements differ by $\sim$20\% between populations, with K enhanced by a factor of $\sim$3.4 in the binary-stripped population, driven by increased production in the lowest-mass progenitors.
Binary-stripped models show closer agreement with empirical core-collapse S, K, and Ca abundance ratios relative to Mg inferred from APOGEE, while Na remains overproduced in both populations.
Together, our results highlight the need to account for binarity when interpreting supernova abundance patterns. 

\end{abstract}

\section{Introduction}
\label{sec:Introduction}

Core-collapse supernovae (CCSNe) mark the energetic deaths of massive stars. At lower zero-age main-sequence masses (\mzams $\simeq 8$--$10\,\mass$), they may arise from the collapse of an ONeMg core, while more massive stars undergo collapse after forming an Fe core. During these explosions, newly synthesised material is ejected into the surrounding environment, enriching the interstellar medium with intermediate-mass and iron-group elements. CCSNe therefore provide an important source of chemical enrichment and feedback, playing a central role in the chemical evolution of galaxies.

A large fraction of massive stars are found in stellar systems with a companion \citep{Abt1983,Mason2009}. Mass transfer is expected to be common during their evolution \citep{Sana2012,Moe2017}. Such interactions can fundamentally alter the evolutionary pathways of these systems \citep{Podsiadlowski1992,Brown1996,Wellstein1999,Eldridge2008,Langer2012,demink2013}.
Roche-lobe overflow can change the primary's subsequent evolution by removing part or all of its envelope. Accretion can also modify the companion's evolution, so binary interaction affects both the pre-collapse structure of the stars and their contributions to chemical enrichment \citep{Braun1995a,Laplace2021,Farmer2021,Farmer2023}.

Studies of binary-stripped progenitors have begun to establish how binary interaction affects explosive nucleosynthesis \citep{Braun1995b,Vartanyan2021,Farmer2023,Boccioli2026,Jalin2026}, although calculations for single-star progenitors remain more extensive \citep[e.g.,][]{Woosley1993,Heger2001,Heger2002a,Heger2002b,Woosley2007a,Pignatari2016,Limongi2018,push3,push4,Andrews2020,Sieverding2023}.
A major challenge has been the production of suitably detailed binary progenitor models that retain an accurate description of the stellar interior through the final burning stages to the onset of collapse.
Producing these progenitor models has been limited by the computational cost of the detailed nuclear reaction networks needed to follow late-stage burning and electron captures \citep{Farmer2016,Laplace2021}. An accurate treatment is important as even small differences in the pre-collapse structure can affect the subsequent explosion outcome \citep{Vartanyan2018,Laplace2021}.

The difficulty of modelling CCSNe does not end with producing suitably detailed pre-collapse progenitor grids, as the explosion mechanism itself also remains challenging to model. It requires the coupled treatment of hydrodynamics, gravity, nuclear equations of state, neutrino transport, nuclear reaction rates, and possibly neutrino flavour conversion. Although major progress has been made in multidimensional simulations over the last decade, fully self-consistent calculations remain computationally expensive and are not yet practical for population-level studies. A growing number of multidimensional simulations do now follow the evolution from collapse to shock revival \citep[e.g.][]{janka12,Burrows2013,Nakamura2015,Janka2016,Bruenn2016,Burrows2018,Burrows2020,Vartanyan2023,Sandoval2021,Burrows2024,Calvert2025,Burrows2026,Neopane2026,Calvert2026,Jalin2026}, but uncertainties remain in which progenitors explode, what explosion energies they produce, and how these results vary across large progenitor grids \citep[e.g.][]{Takiwaki2014,Lentz2015,Melson2015a,Melson2015b,Muller2015a,Janka2016,Hix2016}. For a recent review of advances in multidimensional CCSN modelling, see \citet{Janka2025}.

In one-dimensional (1D), self-consistent, spherically symmetric simulations, CCSNe generally do not explode through the delayed neutrino-driven mechanism alone, except at the low-mass end of the progenitor population \citep[e.g.][]{Liebendoerfer2004}. Successful shock revival for more massive progenitors appears to require multidimensional hydrodynamic effects, particularly neutrino-driven convection and accretion shock instability, which increase the efficiency of neutrino heating behind the stalled shock \citep[e.g.][]{Herant1994,Janka1996,Blondin2003,Foglizzo2007}. Alternatively, rapid rotation and magnetic-field amplification may produce magnetorotational explosions in a subset of strongly magnetised progenitors \citep{Winteler2012,Mosta2014,Mosta2015,Nishimura2015,Nishimura2017,Halevi2018,Kuroda2020,Obergaulinger2020,Aloy2021}.

As multidimensional simulations remain computationally expensive, several 1D approaches have been developed to parameterise the effects that facilitate successful explosions and apply them systematically across large progenitor grids. Simple prescriptions such as piston and thermal-bomb explosions, in which the shock is launched by moving an inner mass shell or by depositing thermal energy into the innermost stellar layers, have been widely used for nucleosynthesis calculations \citep[e.g.][]{Aufderheide1991,Woosley1995,Sawada2019}. These approaches remain in use in recent studies \citep[e.g.,][]{Farmer2023} and provide an efficient way to calculate nucleosynthetic yields across large progenitor grids. However, key explosion properties (i.e., the explosion energy and mass cut) are imposed rather than emerging from the explosion dynamics, limiting their ability to predict the nucleosynthesis of the innermost ejecta \citep[e.g.][]{thielemann96,Woosley1995,rauscher02,limongi06,Woosley2007a,umeda08,limongi12}. More physically motivated 1D approaches parameterise the processes that facilitate shock revival. The P-HOTB approach employs parameterised proto-neutron-star (PNS) core contraction and neutrino emission through an inner boundary \citep{Ugliano2012,Ertl2016,Sukhbold2016}, while the Supernova Turbulence In Reduced-dimensionality (STIR) model incorporates neutrino-driven convection and turbulence through a time-dependent mixing-length prescription \citep{Couch2020,Boccioli2021}.
The PUSH method, which is used in this study, instead introduces an additional source of neutrino heating to facilitate shock revival while retaining detailed neutrino transport \citep[e.g.,][]{push1,push2}. These approaches provide a framework for systematically studying explosion outcomes, remnant properties, and yields across large progenitor grids.

Assessing the impact of different stellar-evolution pathways requires considering not only how they alter the explodability of the resulting progenitors, but also how they affect the composition of the ejecta and the resulting observables \citep[e.g.][]{nomoto13,Bruenn2016,Caffau2013}. Nucleosynthetic yields therefore provide an important link between the pre-supernova evolution, the explosion dynamics, and the final composition of the ejecta. These yields depend both on the pre-supernova composition and stratification of the progenitor and on which regions ultimately become unbound during the explosion. In the outer ejecta, the final abundances are inherited from hydrostatic burning during stellar evolution, with the explosion determining whether these layers are successfully ejected. This material contains the products of H, He, C, Ne, and O burning, including He, C, O, Ne, Mg, and other intermediate-mass elements (IMEs). In contrast, the composition of the innermost ejecta is considerably more sensitive to the explosion itself, with the production of iron-group elements depending on the shock strength, the thermodynamic histories of the ejecta, the location of the mass cut, and the electron fraction, $Y_e$ \citep[][]{push3}. Binary stripping can therefore influence nucleosynthetic yields both by changing the progenitor’s pre-collapse composition and stratification and by altering the subsequent explosion conditions through these structural changes.

In this work, we use the PUSH framework to simulate the single-star and binary-stripped progenitors of \citet{Farmer2023}, allowing us to compare their explosion properties and nucleosynthetic yields consistently. We investigate how binary stripping alters the explodability of the progenitors, the resulting explosion energies and remnant masses, and the composition of the ejecta. The paper is structured as follows. In Section~\ref{sec:Method}, we describe the progenitor models, the PUSH framework, and the nucleosynthesis post-processing. In Section~\ref{sec:explosion_properties_across_the_mass_range}, we show the explosion outcomes and properties across the progenitor grid and examine their dependence on ZAMS mass, compactness, and the adopted PUSH calibration. 
In Section~\ref{sec:Nucleosynthesis}, we explore the nucleosynthetic yields and the differences between the single-star and binary-stripped models. 
In Section~\ref{sec:Comparison_With_Observations}, we compare our predictions with observational constraints from CCSNe and metal-poor stars, and with empirical core-collapse abundance ratios inferred from APOGEE stellar abundances.
Finally, we summarise our main findings and conclusions in Section~\ref{sec:Discussion_and_Conclusions}.

\section{Method and Inputs}
\label{sec:Method}
\subsection{Input Models}
\label{sec:Input_Models}

In this study, we consider the two progenitor sets generated by \citet{Farmer2023}: isolated single stars (the ``Ms'' series) and binary-stripped stars that have undergone Roche-lobe overflow (the ``Mb'' series). Both sets were calculated with the MESA stellar evolution code \citep{Paxton2011,Paxton2013,Paxton2015,Paxton2018,Paxton2019,Jermyn2023}. Importantly, the Mb progenitors differ from their single-star counterparts not only in their envelope masses, but also in their core masses, composition gradients, and subsequent shell-burning evolution.

The adopted grids contain non-rotating progenitors with ZAMS masses of 11--45~$\mass$, sampled at 1~$\mass$ intervals. Both populations share the same initial abundance pattern, based on \citet{Grevesse1998}, with a solar metallicity of $Z = 0.0142$. For both progenitor sets, the initial helium mass fraction follows $Y = 2Z + 0.24$ \citep{Pols1995,Tout1996}, corresponding to $Y = 0.2684$. The Mb models with ZAMS masses of 24, 25, 28, and 29$\,\mass$ are excluded from our analysis due to numerical difficulties preventing their evolution to core collapse in \textsc{MESA} \citep[see][]{Farmer2023}.
Throughout this work, we place our results in the context of previous solar-metallicity PUSH calculations using the s-series \citep{Woosley2002} and w-series \citep{Woosley2007a} progenitors presented by \citet{push2,push3}. We refer to these progenitor sets as WHW02 and WH07, respectively.
Note that these single-star models were calculated with the \textsc{kepler} code, presented in \citet{Woosley2002} and \citet{Woosley2007a}.

\begin{figure}
    \centering
    \includegraphics[width=0.89\columnwidth]{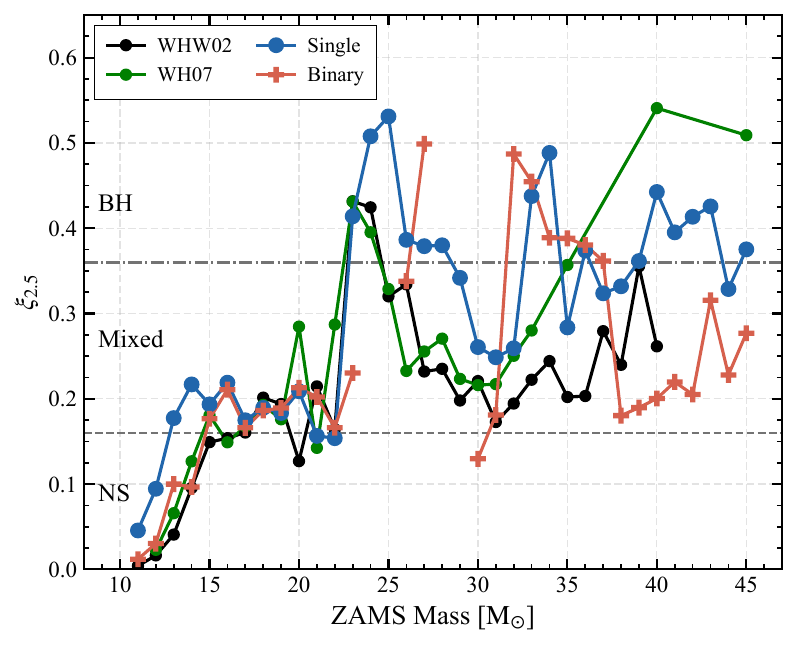}
\caption{
Compactness at $2.5\,\mathrm{M}_\odot$ ($\xi_{2.5}$) as a function of ZAMS mass for the single and binary-stripped progenitors. Horizontal lines at $\xi_{2.5}=0.16$ and $0.36$ mark the approximate boundaries between neutron-star, mixed, and black-hole outcome regimes \citep{Farmer2023}. Solar-metallicity progenitors from \citet{Woosley2002} (WHW02) and \citet{Woosley2007a} (WH07) are shown for comparison.
}
\label{fig:Method_compactness}
\end{figure}

Following previous studies, we characterise the pre-collapse core structure using the compactness parameter introduced by \citet{oconnor11}. For consistency with the explodability analysis of \citet{Farmer2023}, we evaluate the compactness at an enclosed mass of $2.5\,\mass$:
\begin{equation}
\xi_{2.5} = \frac{2.5}{R(M = 2.5\,\mass)/1000\,\mathrm{km}},
\end{equation}
where $R(M = 2.5\,\mass)$ is the radius enclosing a baryonic mass of $2.5\,\mass$. Figure~\ref{fig:Method_compactness} shows $\xi_{2.5}$ as a function of ZAMS mass for the Ms, Mb, WHW02 and WH07 progenitor series. The compactness varies non-linearly with ZAMS mass, reflecting the sensitivity of the final core structure to shell-burning evolution, semiconvection, convective overshooting, and mass loss. Note that the PUSH calibration described below is based on the compactness at core bounce, $\xi_{2.0}$, following \citet{push2}. The corresponding $\xi_{2.0}$ trends with ZAMS mass are shown in Appendix~\ref{apen:compactness}.

\begin{figure*}
    \centering

    \begin{minipage}[t]{0.49\textwidth}
        \centering
        \includegraphics[width=0.89\linewidth]{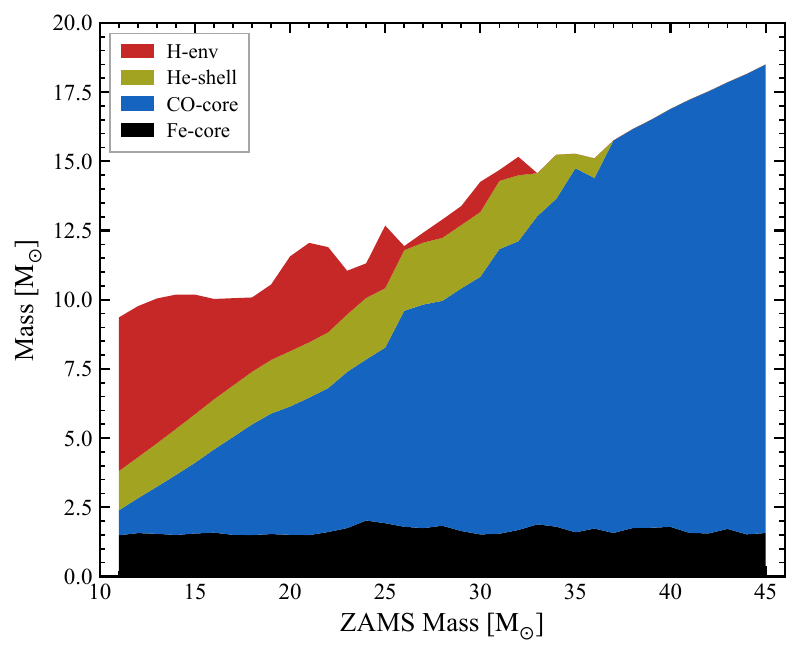}
        \caption{  
        Pre-collapse structure of the single-star progenitors from \citet{Farmer2023} as a function of ZAMS mass. The coloured regions distinguish the Fe core (black), the surrounding CO-rich layers (blue), the He shell (yellow), and the H-rich envelope (red).
        }
        \label{fig:mass_components_single}
    \end{minipage}
    \hfill
    \begin{minipage}[t]{0.49\textwidth}
        \centering
        \includegraphics[width=0.89\linewidth]{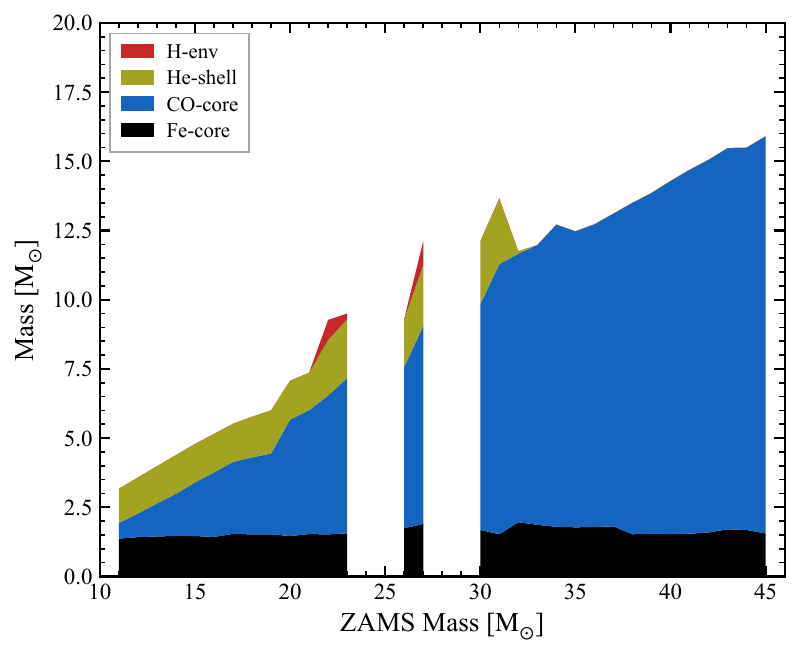}
        \captionof{figure}{
        Same as Figure~\ref{fig:mass_components_single}, but for the binary-stripped progenitors \protect\citep{Farmer2023}.
        }
        \label{fig:mass_components_binary}
    \end{minipage}

\end{figure*}

\subsection{The PUSH Framework}
\label{sec:push_framework}

We adopt the standard PUSH setup used in previous studies to follow the collapse, core bounce, shock revival, and the subsequent post-bounce evolution \citep[see][]{push2,push3,push4}. The resulting simulations provide the explosion properties, remnant masses, and thermodynamic trajectories required for nucleosynthesis post-processing. 
PUSH is based on the delayed neutrino-driven explosion mechanism. To mimic the enhanced neutrino heating associated with convection and accretion in multidimensional simulations, an additional heating term is employed within the gain region of spherically symmetric models \citep{push1,push2}. This term uses a fraction of the energy carried by heavy-flavour neutrinos ($\nu_x = \nu_\mu, \bar{\nu}_\mu, \nu_\tau, \bar{\nu}_\tau$) to support shock revival.

The method \citep{push1} employs the general-relativistic hydrodynamics code \texttt{Agile}, coupled to neutrino transport. Electron-flavour neutrinos are treated using the isotropic diffusion source approximation \citep[\texttt{IDSA};][]{Liebendoerfer.Agile}, while heavy-flavour neutrinos are handled with the advanced spectral leakage \citep[\texttt{ASL};][]{push1} scheme. For matter in nuclear statistical equilibrium (NSE), we adopt the finite-temperature HS(DD2) equation of state \citep{Hempel.SchaffnerBielich:2010}. Outside NSE, the matter is described by an ideal-gas equation of state coupled to an $\alpha$-network.

The PUSH heating term, $Q_{\mathrm{push}}^+(t,r)$, transfers a fraction of the heavy-flavour neutrino energy  to matter behind the shock. It depends on the neutrino spectral energy flux and includes both spatial and temporal components, with two free parameters: $k_{\mathrm{push}}$ and $t_{\mathrm{rise}}$. 
We adopt the standard calibration of \citet{push2}, based on SN~1987A, with $t_{\mathrm{rise}} = 400\,\mathrm{ms}$ and a compactness-dependent scaling
\begin{equation}
k_{\mathrm{push}} = a\xi^2 + b\xi + c ,
\label{eqn:push}
\end{equation}
where $a = -23.99$, $b = 13.22$, and $c = 2.5$. Here, $\xi_{2.0}$ denotes the compactness at core bounce.

The simulations include material from the centre to the helium layer and are evolved for 15\,s unless black-hole formation occurs or a failed explosion is identified earlier\footnote{For successful explosions, all quantities are extracted from the timestep immediately before the shock leaves the computational domain.}. The outcome of each model is classified as (i) a successful explosion if the explosion energy saturates at a positive value, (ii) black-hole formation if the central density exceeds $\sim10^{15}\,\mathrm{g\,cm^{-3}}$, or (iii) a failed explosion if the final explosion energy remains negative after PUSH becomes inactive. The explosion energy is computed following \citet{push1} as the sum of thermal, kinetic, and gravitational energy integrated from the mass cut to the surface, with the mass cut emerging self-consistently from the hydrodynamic evolution rather than being imposed.

\subsection{Nucleosynthesis Post-Processing}
\label{sec:Method_nucleosynthesis}

We calculate the nucleosynthetic yields of successful explosions by post-processing their thermodynamic trajectories with \texttt{CFNET} \citep{frohlich2006b}, following the setup of \citet{push2,push3}. The network tracks 2,902 isotopes, from free nucleons to $^{211}$At, and includes nuclei on both sides of the valley of $\beta$-stability. Electron and positron captures are treated using the rates of \citet{lmp}, while $\beta^{\pm}$-decay rates are obtained from NuDat2\footnote{http://www.nndc.bnl.gov/nudat2/} and \citet{Moller1995}. Neutrino and antineutrino captures on free nucleons are also included. For the nuclear reaction rates, we use \texttt{REACLIB} \citep{reaclib}, with experimentally determined rates where available and theoretical neutron-, proton-, and $\alpha$-capture rates from \citet{Rauscher.FKT:2000} otherwise.

We divide the ejecta into tracers of $10^{-3}\,\mass$ and perform nucleosynthesis calculations for those reaching peak temperatures above 1.75\,GK. Each tracer records the thermodynamic evolution from the onset of collapse to the final hydrodynamic timestep, up to 15\,s or until the shock leaves the computational grid. The starting conditions for post-processing depend on the peak temperature \citep{push3}. For tracers heated above 10\,GK, we initialise the composition in NSE as they cool through this threshold. For cooler tracers, we use the abundances from the nuclear network coupled to the hydrodynamics and begin post-processing at the start of the hydrodynamic evolution. In both cases, the initial electron fraction is taken from the hydrodynamic simulation, with its subsequent evolution calculated self-consistently within the reaction network.
Beyond the final hydrodynamic timestep, we extrapolate the tracer trajectories following \citet{push2,push3}:
\begin{equation}
r(t) = r_{\mathrm{final}} + t\,v_{\mathrm{final}}
\end{equation}
\begin{equation}
\rho(t) = \rho_{\mathrm{final}} \left(\frac{t}{t_{\mathrm{final}}}\right)^{-3}
\end{equation}
\begin{equation}
T(t) = T[s_{\mathrm{final}}, \rho(t), Y_e(t)]
\end{equation}
where $r$, $v$, $\rho$, $T$, $s$, and $Y_e$ denote radius, velocity, density, temperature, entropy per baryon, and electron fraction, respectively. The subscript ``final'' refers to values at the end of the hydrodynamic simulation. The temperature evolution is obtained using the equation of state of \citet{Timmes.Swesty:2000}. Nucleosynthesis calculations are terminated once the temperature drops below 0.05\,GK.

We note that we exclude the three innermost tracer particles immediately adjacent to the mass cut as the late-time neutrino-driven wind is poorly resolved, and therefore do not draw conclusions about its detailed composition. Our hydrodynamic calculations also omit neutrino-electron scattering, which contributes to neutrino thermalisation and spectral down-scattering and may consequently affect the electron fraction of the wind \citep{Fischer2012}. The post-processing network also does not account for neutrino interactions that can synthesise some rare isotopes through the $\nu$-process in the outer ejecta \citep{Woosley1990,Heger2005}. See \citet{push3} for an in-depth discussion of the nucleosynthetic production channels and modelling approach.

\section{Explosion Properties across the Mass Range}
\label{sec:explosion_properties_across_the_mass_range}

In this section, we discuss the explosion outcomes and properties of both progenitor sets, spanning ZAMS masses of $11$--$45\,\mass$. Four Mb models ($24$, $25$, $28$, and $29\,\mass$) did not reach core collapse and are therefore excluded. We first examine successful and failed explosions, followed by trends in explosion energies, remnant masses, and explosion times with ZAMS mass and bounce compactness $\xi_{2.0}$.

\subsection{Explosion Outcomes}
\label{sec:explosion_outcomes}

\begin{figure}
    \centering
    \includegraphics[width=0.49\textwidth,height=0.36\textheight,keepaspectratio]{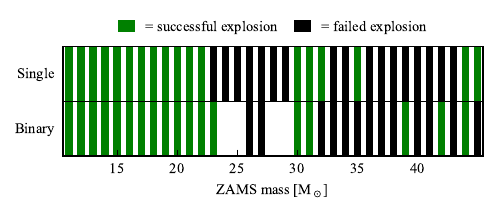}
\caption{
Explosion outcomes as a function of ZAMS mass for the single-star (top) and binary-stripped (bottom) progenitor series, showing successful and failed explosions. Gaps indicate masses for which no binary-stripped progenitor reached core collapse.
}
    \label{fig:explosion_map_ms_mb}
\end{figure}

\begin{figure}
    \centering
    \noindent\includegraphics[width=\columnwidth]{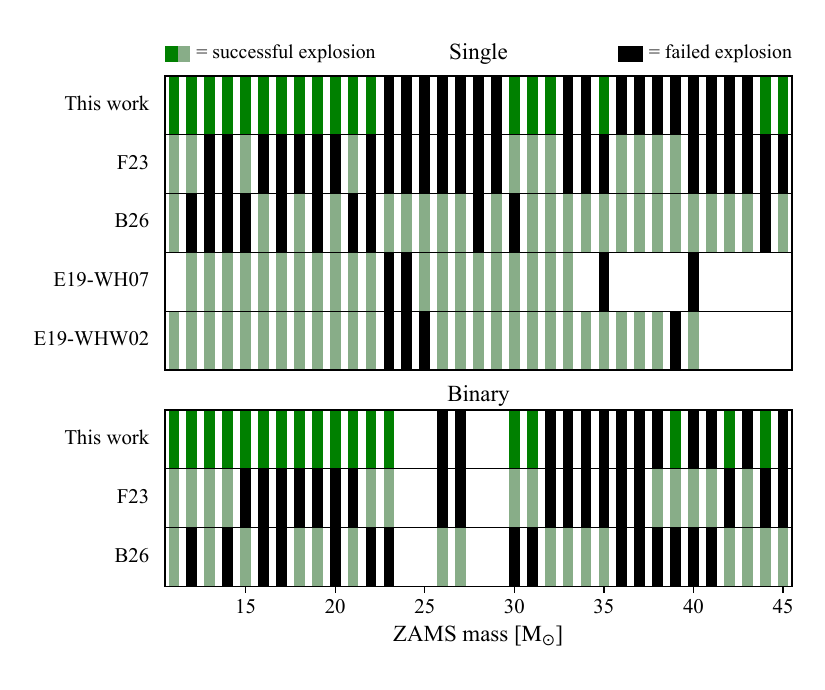}
    \caption{
        Similar to Figure~\ref{fig:explosion_map_ms_mb}, but comparing explosion outcomes for single-star (top) and binary-stripped (bottom) progenitors across different studies. Rows show this work; \citet{Farmer2023} (F23), using the two-parameter explodability criterion of \citet{Ertl2016}; and \citet{Boccioli2026} (B26), using neutrino-driven explosion simulations with time-dependent mixing-length theory. These three studies use the F23 progenitors. The single-star panel also includes PUSH outcomes from \citet{push2} for progenitors from \citet{Woosley2007a} (E19-WH07) and \citet{Woosley2002} (E19-WHW02) at integer ZAMS masses. Green indicates successful explosions, with muted green for literature results; black indicates failed explosions. Gaps indicate unavailable models.
    }
    \label{fig:explosion_map_comparison}
\end{figure}

Figure~\ref{fig:explosion_map_ms_mb} summarises the explosion outcomes for the two progenitor series across the full ZAMS-mass range. We find that 18 of the 35 Ms models investigated here explode, corresponding to a success rate of 51.4 per cent, while 18 of the 31 Mb models explode, giving a slightly higher success rate of 58.1 per cent. The two series behave similarly at low ZAMS masses, with all models up to and including $22\,\mass$ exploding. Above this mass, however, their outcomes begin to diverge: the first Ms failure occurs at $23\,\mass$, whereas the first Mb failure occurs at $26\,\mass$. The two series produce opposite outcomes at $23\,\mass$, $32\,\mass$, $35\,\mass$, $39\,\mass$, $42\,\mass$, and $45\,\mass$, while both explode at $30\,\mass$, $31\,\mass$, and $44\,\mass$.

Figure~\ref{fig:explosion_map_comparison} compares our results with those of \citet{Farmer2023}, using the two-parameter explodability criterion of \citet{Ertl2016}, and \citet{Boccioli2026}, using neutrino-driven explosion simulations with time-dependent mixing-length theory. For the Ms series at $\mathrm{M}_{\rm ZAMS}\leq22\,\mass$, all of our models explode, while \citet{Farmer2023} predict explosions only at $11$, $12$, $15$, and $21\,\mass$, and \citet{Boccioli2026} at $11$, $16$, $18$, and $20\,\mass$. Between $23$ and $29\,\mass$, none of our Ms models explode, in agreement with \citet{Farmer2023}, whereas \citet{Boccioli2026} predict successful explosions for all but one model. At higher ZAMS masses, \citet{Boccioli2026} again predict explosions for most models, while ours show more frequent failures and somewhat better agreement with \citet{Farmer2023}. A similarly mixed picture is found for the Mb series. All of our models up to and including $23\,\mass$ explode, whereas both \citet{Farmer2023} and \citet{Boccioli2026} predict a mixture of successful and failed explosions and frequently disagree over individual models. Between $26$ and $37\,\mass$, our results are broadly consistent with \citet{Farmer2023}, producing a mixture of outcomes. At the highest ZAMS masses, the three approaches again diverge. Overall, the limited agreement across both progenitor series highlights the strong sensitivity of the explodability landscape to the adopted explosion treatment.

Figure~\ref{fig:explosion_map_comparison} also shows the corresponding explosion outcomes for the \citet{Woosley2002} and \citet{Woosley2007a} progenitor sets, WHW02 and WH07. These models provide a useful comparison as they were used in the original PUSH calibration and nucleosynthesis studies \citep{push2,push3}, but were evolved using different stellar-evolution assumptions from the \citet{Farmer2023} models explored here. Since the same explosion prescription is applied, differences between the model sets reflect variations in the pre-collapse progenitor structures produced by the underlying stellar-evolution calculations. At low ZAMS masses, the WHW02 and WH07 progenitors show broadly similar behaviour to the Ms and Mb series, with all models up to and including $22\,\mass$ exploding. The first failed explosion in both the WHW02 and WH07 sets occurs at $23\,\mass$, matching the Ms series. The clearest difference between the progenitor grids appears between $25\,\mass$ and $30\,\mass$, where the WHW02 and WH07 sets produce more successful explosions than the Ms and Mb series. This behaviour is likely related to differences in the compactness distributions of the progenitor grids. In the standard PUSH calibration, $k_{\rm push}$ depends on the bounce compactness $\xi_{2.0}$ and reaches its largest values at intermediate compactness, around $\xi_{2.0}\approx0.3$ \citep[see figures~2 and 8 of][]{push2}. Progenitor grids containing compactness troughs, or a larger number of models with intermediate compactness in this mass range, therefore receive stronger PUSH heating, allowing a greater fraction of models to explode. The role of this calibration is discussed further in Section~\ref{sec:kpush_sensitivity}.
Compactness nevertheless provides only a one-parameter characterisation of the core structure: models with similar $\xi_{2.0}$ can have different density profiles, shell structures, and accretion histories, affecting their explosion outcomes and yields. We examine such structural differences in Section~\ref{sec:binary_stripping_yields}.

\subsection{Explosion Properties as a Function of ZAMS Mass}
\label{sec:explosion_zams_trends}

\begin{figure}
    \centering
    \includegraphics[width=0.45\textwidth,height=0.36\textheight,keepaspectratio]{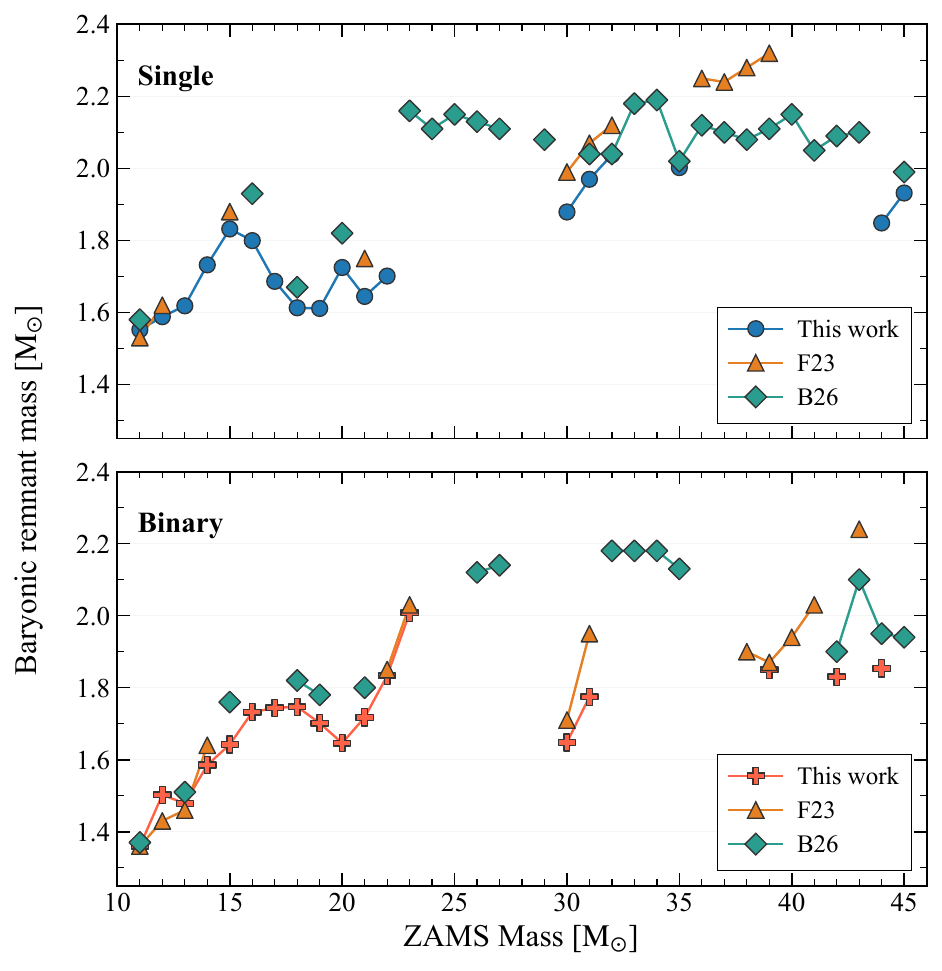}
\caption{
Baryonic remnant mass as a function of ZAMS mass for successfully exploding single-star (top) and binary-stripped (bottom) models from this work, \citet{Farmer2023} (F23), and \citet{Boccioli2026} (B26). Lines connect neighbouring successful models, with gaps indicating failed explosions or unavailable progenitors.
}

\label{fig:ZAMS_vs_remnant_mass_comparison}
\end{figure}

In Figure~\ref{fig:ZAMS_vs_remnant_mass_comparison}, we compare the remnant masses obtained for the Ms and Mb progenitor series in this work with those of \citet{Farmer2023} and \citet{Boccioli2026}. Although all three report a remnant mass, this quantity is determined differently in each approach. 
In both PUSH and the neutrino-driven simulations of \citet{Boccioli2026}, the mass cut emerges self-consistently from the hydrodynamic evolution. PUSH retains the PNS within the computational domain. In contrast, \citet{Farmer2023} excise the material interior to the mass coordinate where the entropy per baryon reaches $s=4$ before initiating the explosion and define the remnant mass as the compact-object mass at shock breakout, neglecting any fallback.
The three methods also adopt different equations of state, with PUSH using HS(DD2) \citep{Hempel.SchaffnerBielich:2010}, \citet{Boccioli2026} using SFHo \citep{Steiner2013}, and \citet{Farmer2023} using the standard MESA EOS. 
These differences in numerical treatments may therefore contribute to small systematic offsets between the calculations.

Despite the differences between the three methods, the predicted remnant masses are remarkably similar at low ZAMS masses when the same progenitor successfully explodes. Below $\sim21$--$22\,\mass$, the remnant masses generally agree to within $\sim0.15\,\mass$, with \citet{Boccioli2026} typically predicting the largest values, \citet{Farmer2023} intermediate values, and PUSH the smallest. At higher ZAMS masses, the predictions diverge more strongly. Where direct comparisons between successful Ms models are possible, PUSH consistently produces less massive remnants than \citet{Farmer2023} and \citet{Boccioli2026}. The same trend is found for the Mb series, although the differences become particularly large above $\sim40\,\mass$.

\begin{figure*}
    \centering

    \begin{minipage}[t]{0.49\textwidth}
        \vspace{0pt}
        \centering

        \hfill
        \includegraphics[width=0.93\linewidth]{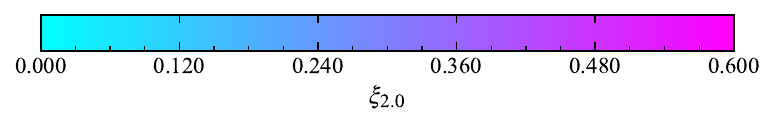}
        \par

        \includegraphics[width=0.93\linewidth]{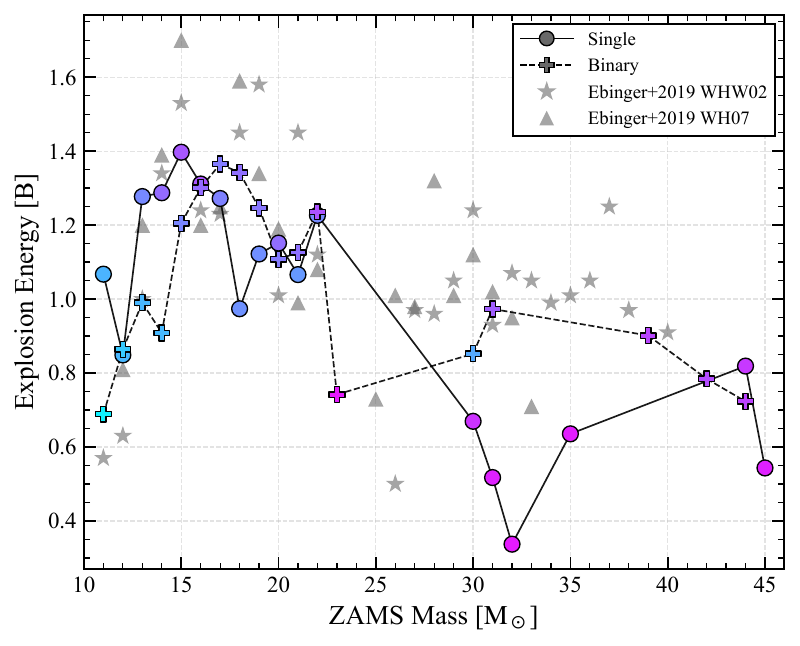}
        \par

        \includegraphics[width=0.93\linewidth]{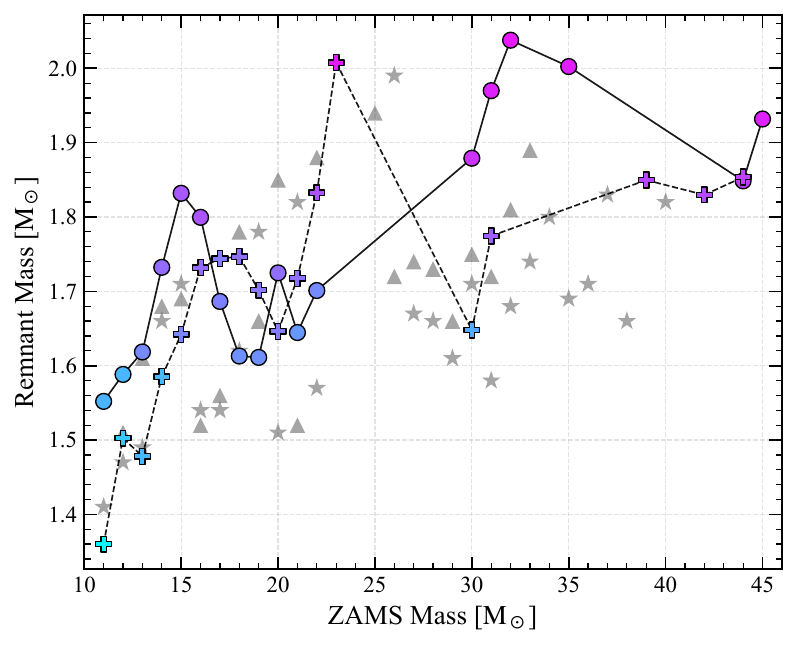}
        \par

        \includegraphics[width=0.93\linewidth]{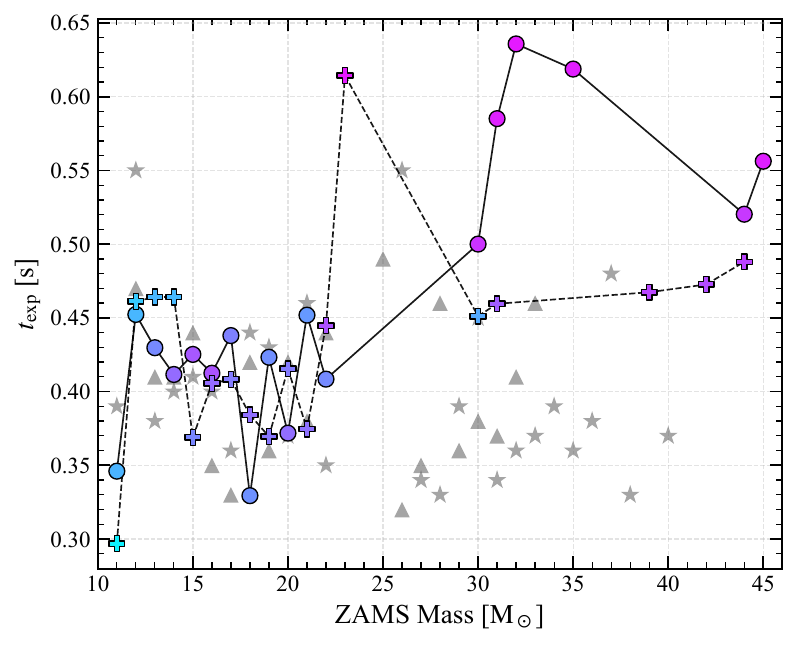}
        \par

        \caption{
        Explosion energy (top), remnant mass (middle), and explosion time (bottom) as functions of ZAMS mass for successfully exploding single-star and binary-stripped models. The WHW02 and WH07 models from \citet{push2} are included for comparison. Lines connect models within each progenitor series and symbol colour denotes the compactness at bounce, $\xi_{2.0}$, for the models from this work.
        }
        \label{fig:explosion_zams_trends}
    \end{minipage}
    \hfill
        \begin{minipage}[t]{0.49\textwidth}
        \vspace{0pt}
        \centering

        \makebox[\linewidth][c]{%
            \hspace*{0.03\linewidth}%
            \includegraphics[width=0.89\linewidth]{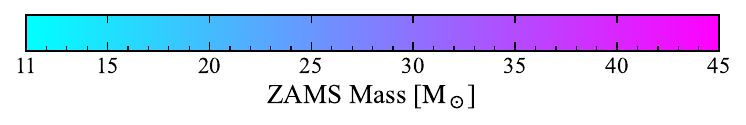}%
            \hspace*{-0.03\linewidth}%
        }
        \par

        \includegraphics[width=0.93\linewidth]{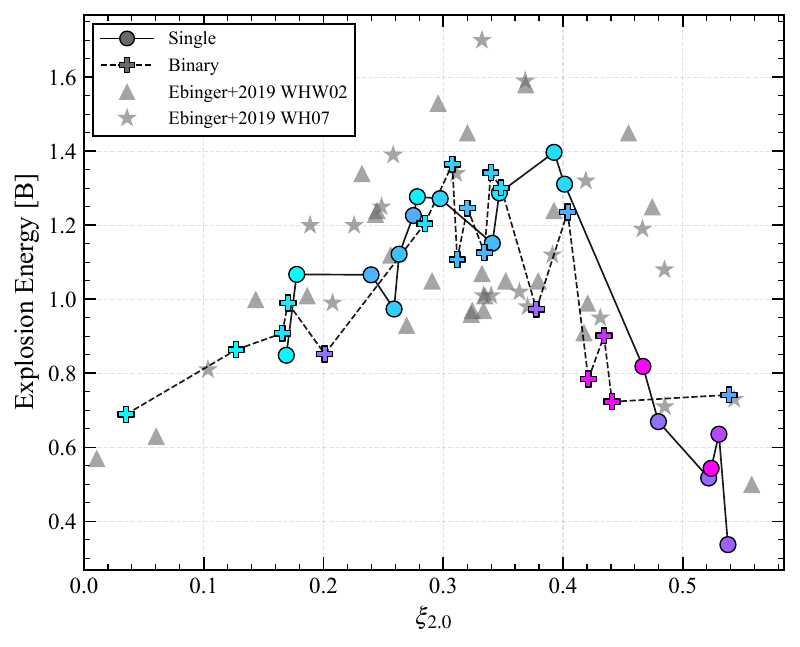}
        \par

        \includegraphics[width=0.93\linewidth]{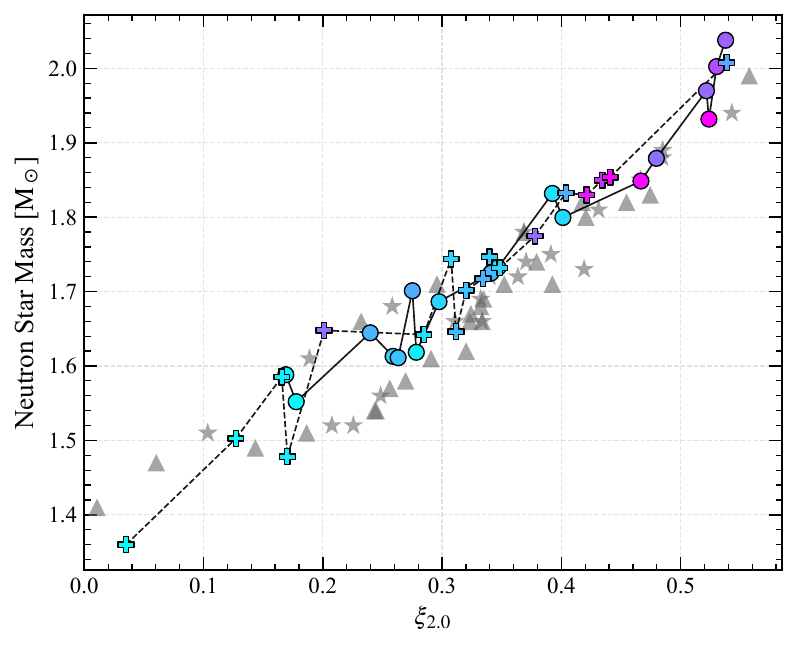}
        \par

        \includegraphics[width=0.93\linewidth]{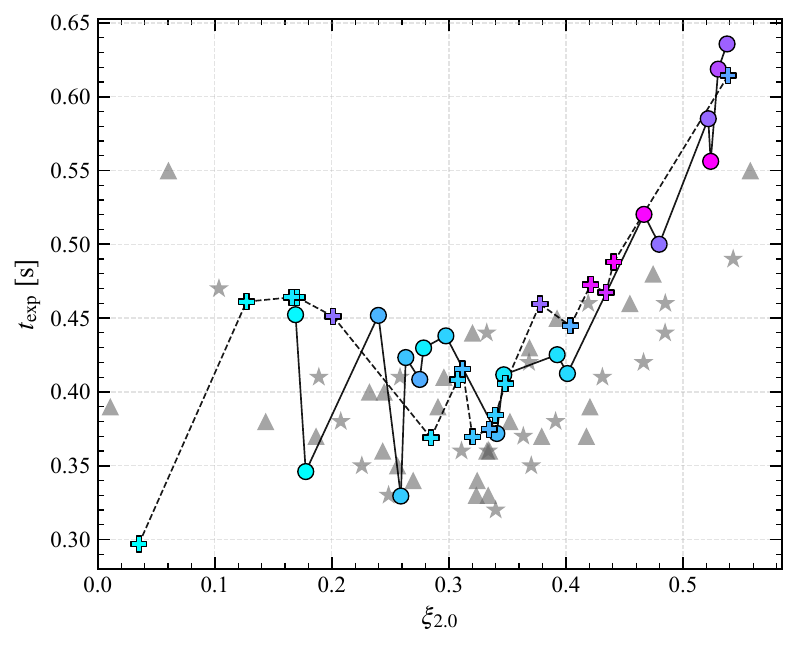}
        \par

        \caption{
        Similar to Figure~\ref{fig:explosion_zams_trends}, but with the explosion properties shown as functions of the compactness at bounce, $\xi_{2.0}$, with symbol colour denoting the ZAMS mass.
        }
        \label{fig:explosion_compactness_trends}
    \end{minipage}

\end{figure*}

Figure~\ref{fig:explosion_zams_trends} shows the explosion energy, remnant mass, and explosion time of the successfully exploding Ms and Mb models as a function of ZAMS mass, with the compactness of each model indicated by the colour bar. The successfully exploding WHW02 and WH07 models from \citet{push2} are included for comparison. Overall, the Ms and Mb models occupy similar regions of parameter space, with explosion energies of $\sim0.4$--$1.3$\,B, remnant masses of  $\sim1.4$--$2.0\,\mass$, and explosion times of $\sim0.30$--$0.65$~s. At low ZAMS masses, the Ms and Mb series produce similar remnant masses and explosion times. The main difference occurs between $11$ and $14\,\mass$, where the Mb models produce slightly lower explosion energies. These models also have lower compactness values, corresponding to lower values of $k_{\rm push}$ and therefore weaker additional heating. 
Between $15$ and $22\,\mass$, the two series give very similar values for all three properties.
Larger differences emerge among the more massive exploding models, particularly above $\sim30\,\mass$, where the successful Ms models generally produce lower explosion energies, larger remnant masses, and later explosion times than the corresponding Mb models. The Ms models in this range also typically have higher compactness values. 
It is important to note that the differences we find at high ZAMS masses should be interpreted with caution, as uncertainties in stellar evolution affect the progenitor structures and, consequently, the resulting explosion properties \citep[e.g.][]{Hirschi2015,Sukhbold2016}.

The explosion properties of the Ms and Mb models are broadly consistent with those obtained previously for the WHW02 and WH07 progenitor sets, although the latter generally explode earlier. This likely reflects differences in both the underlying pre-collapse structures and in the exact evolutionary stage at which the progenitors are mapped into the hydrodynamic simulations. Finally, at a fixed ZAMS mass, the differences between the single-star and binary-stripped progenitors can be comparable to the scatter among the different progenitor sets. We therefore examine the same quantities as a function of compactness in the following section.

\subsection{Explosion Properties as a Function of Compactness}
\label{sec:explosion_compactness_trends}

Figure~\ref{fig:explosion_compactness_trends} shows the explosion energy, remnant mass, and explosion time as a function of the compactness at bounce, $\xi_{2.0}$. Compared with ZAMS mass, compactness reveals clearer trends across all three explosion properties. We find that remnant mass increases with $\xi_{2.0}$ in both progenitor series, giving the clearest relation among the three properties examined. This agrees with the previous PUSH results of \citet{push2} and reflects the greater mass accreted by the PNS in more compact progenitors before a successful explosion is launched \citep{push1}.

The explosion energy also shows a broad dependence on compactness. In both series, the most energetic explosions occur at intermediate compactness, $\xi_{2.0}\sim0.25$--$0.4$, while models at both lower and higher compactness generally produce less energetic explosions. This behaviour is closely related to the standard PUSH calibration, in which the compactness-dependent $k_{\rm push}$ reaches its largest values at intermediate $\xi_{2.0}$ and decreases towards both lower and higher compactness. The resulting energy trend therefore partly reflects the imposed compactness-dependent heating prescription, although substantial scatter remains at fixed compactness.

The explosion time shows a weaker dependence on compactness. Models with larger $\xi_{2.0}$ generally explode later than the lowest-compactness models, although there is considerable scatter within this relation. The scatter in remnant mass also varies with compactness: it is largest at low and intermediate compactness, while above $\xi_{2.0}\sim0.4$ the remnant masses become more tightly clustered across the model sets. This differs from the behaviour reported by \citet{Boccioli2026}, who find substantially larger scatter in the mass cut of the high-compactness Ms progenitors from \citet{Farmer2023}.

We note that in general the Ms and Mb models also show substantial overlap in compactness trends with the WHW02 and WH07 models, indicating that binary-stripped progenitors do not occupy a separate region of PUSH explosion parameter space. Compactness therefore provides a useful organising parameter for the explosion properties, but it does not capture all aspects of the pre-collapse structure. Models with similar $\xi_{2.0}$ can still differ in their density profiles, shell locations, and accretion histories, producing differences in both the timing and energetics of shock revival. In particular, \citet{Boccioli2026} find that the explodability of lower-compactness progenitors depends strongly on the density drop associated with the Si/O interface, whereas high-compactness explosions are less often triggered by accretion of this interface and instead arise from increasingly efficient neutrino heating. We also find that differences in the Si- and O-burning regions have an important impact on the resulting nucleosynthetic yields, which we discuss further in Section~\ref{sec:Nucleosynthesis}.

\subsection{Sensitivity to the Compactness-Dependent \texorpdfstring{$k_{\rm push}$}{kpush} Calibration}
\label{sec:kpush_sensitivity}

The comparisons above show that Ms and Mb progenitors with the same ZAMS mass can have different explosion outcomes.
These differences cannot be interpreted from ZAMS mass alone as in the standard PUSH calibration, the strength of the additional heating is itself progenitor dependent. The parameter $k_{\rm push}$ is set by the compactness at bounce, so differences in compactness between an Ms progenitor and an Mb progenitor affect the amount of additional PUSH heating they receive and therefore their hydrodynamic evolution.

To test how strongly the explosion outcome depends on this calibration, we considered the first pair with different outcomes, Ms23 and Mb23. Their bounce compactnesses are $\xi_{2.0}=0.64$ and $0.54$, respectively. With the standard calibration, Ms23 fails to explode while Mb23 explodes successfully. We then repeated both calculations while keeping the progenitor structures unchanged but swapping the compactness values used to determine $k_{\rm push}$. In other words, Ms23 was assigned the PUSH factor corresponding to Mb23, while Mb23 was assigned that of Ms23. The explosion outcomes reversed: Ms23 exploded with the Mb23 PUSH factor, whereas Mb23 failed with the Ms23 value. We performed the same test for the Ms32 and Mb32 pair, for which the standard calculations give the opposite outcome: Ms32 explodes while Mb32 fails. Swapping their PUSH factors again reversed the explosion outcomes.

These tests highlight an important caveat of the adopted parameterisation of $k_{\rm push}$: for both model pairs considered, changing only the PUSH heating factor is sufficient to reverse the explosion outcome.
The predicted explodability therefore depends strongly on the assumption that compactness is the appropriate and dominant quantity with which to calibrate the additional heating. The differences between the Ms and Mb series must therefore be interpreted within the context of this calibration. Comparisons between progenitors with similar compactnesses are consequently particularly valuable, since they reduce differences arising from the parameterisation and instead provide a clearer test of how the differences in progenitor structure affect the explosion properties and nucleosynthetic yields.

\section{Nucleosynthesis Calculations}
\label{sec:Nucleosynthesis}

We compute detailed nucleosynthetic yields for all successfully exploding models in the Ms and Mb series. Our analysis follows several diagnostic comparisons presented by \citet{push3}, which we apply here to investigate the effects of binary stripping. We refer the reader to that work for a detailed discussion of the nucleosynthetic production channels.
We begin by examining the Ms18 and Mb18 models in detail in Section~\ref{sec:binary_stripping_yields}. These models are selected as their ZAMS mass is close to that inferred for the progenitor of SN~1987A. Their abridged isotopic yields are presented in Table~\ref{tab:ms18_mb18_isotopic_yields} of Appendix~\ref{MsandMbyields}.
We then examine the impact of binary stripping on elemental yields in Section~\ref{Production Factors}, comparing our results with previous studies and calculating yields weighted by a Salpeter initial mass function (IMF).
Finally, we investigate how selected isotopic and iron-group yields vary with progenitor compactness and pre-explosion structure in Section~\ref{sec:yield_trends}.

\subsection{Two Representative Models: Ms18 and Mb18}
\label{sec:binary_stripping_yields}

\begin{figure*}
\centering

\begin{minipage}{0.49\textwidth}
    \centering
    \includegraphics[width=\linewidth]{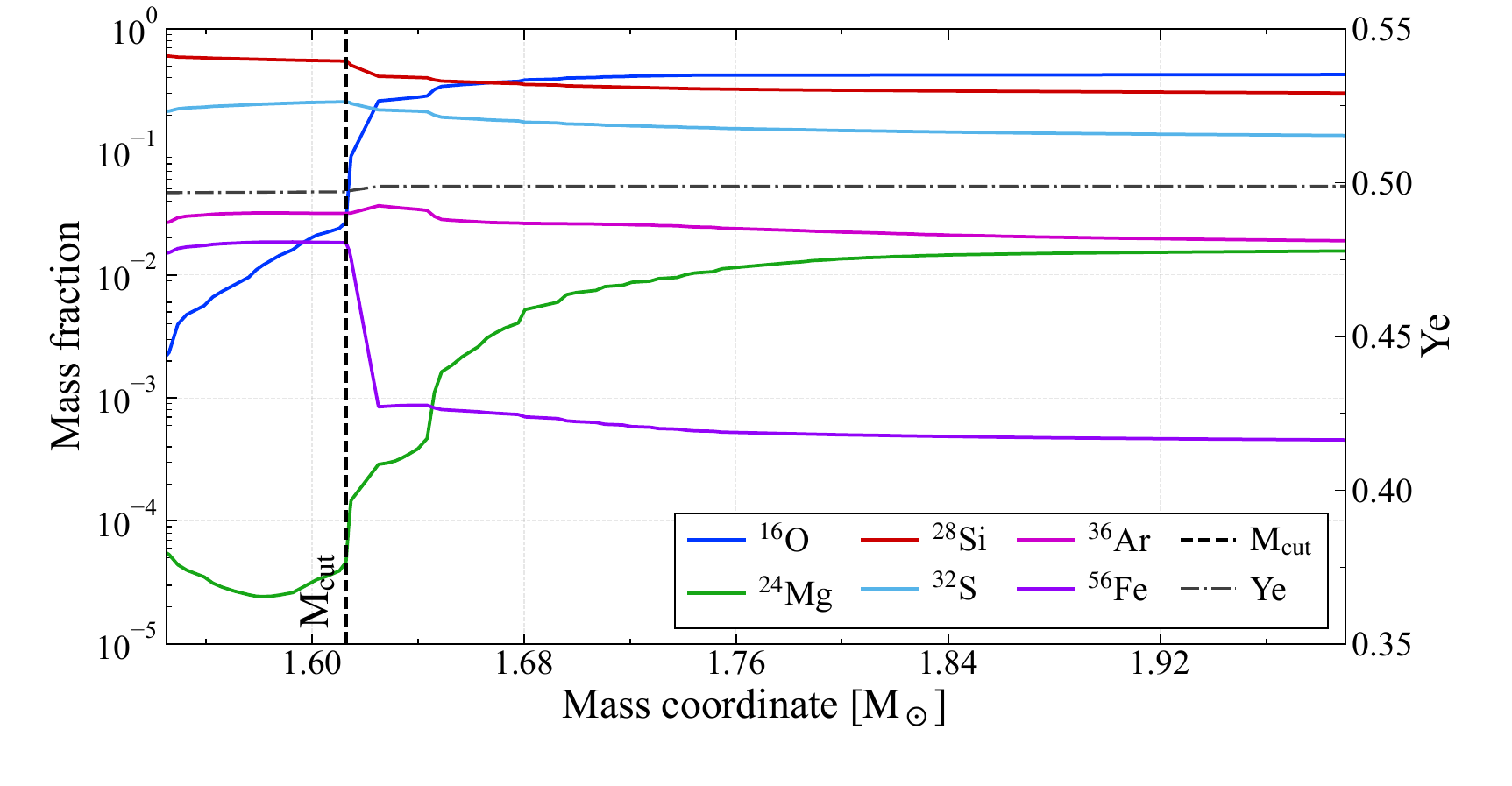}
\end{minipage}
\hfill
\begin{minipage}{0.49\textwidth}
    \centering
    \includegraphics[width=\linewidth]{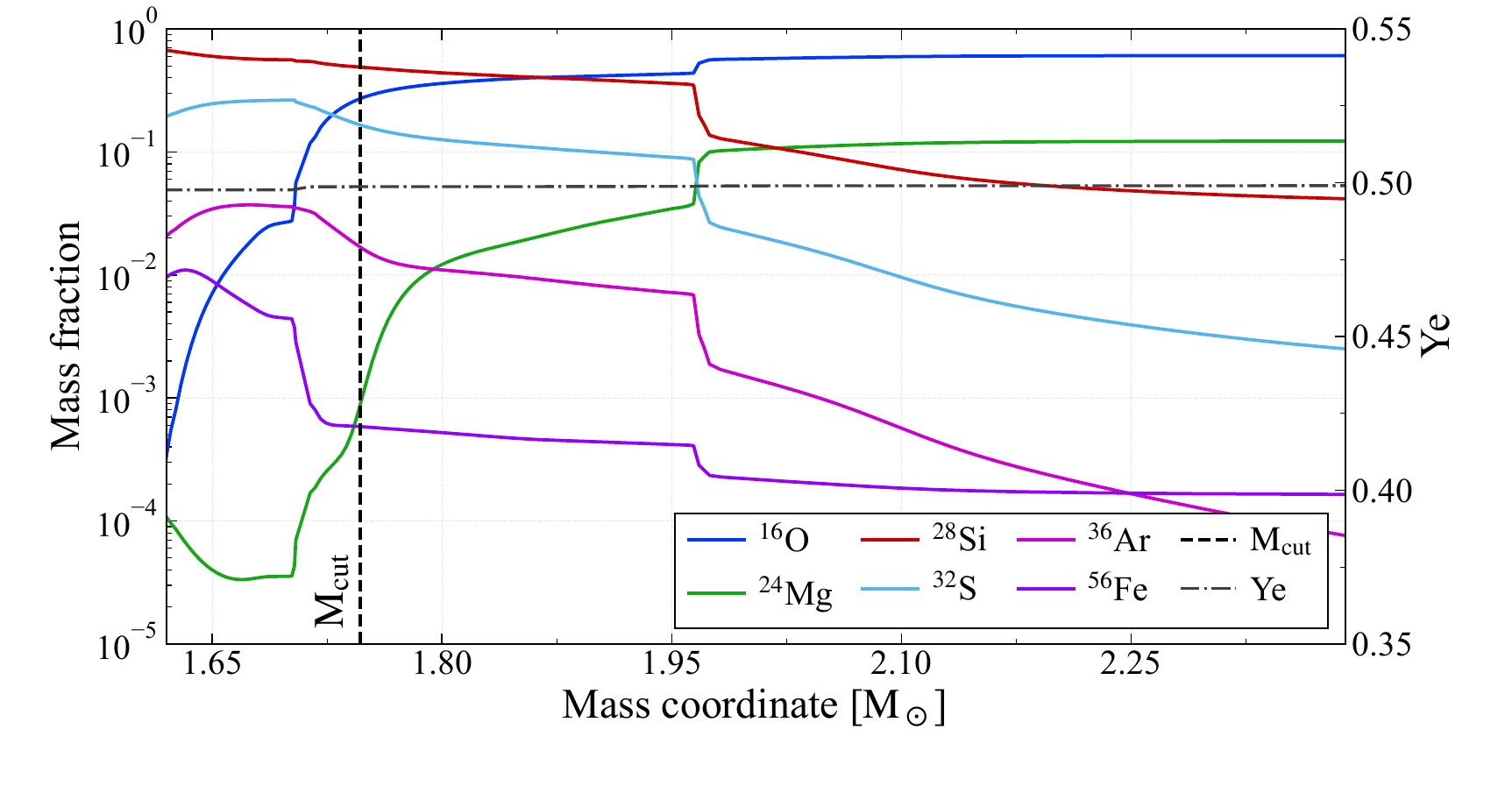}
\end{minipage}

\vspace{0.25cm}

\begin{minipage}{0.49\textwidth}
    \centering
    \includegraphics[width=\linewidth]{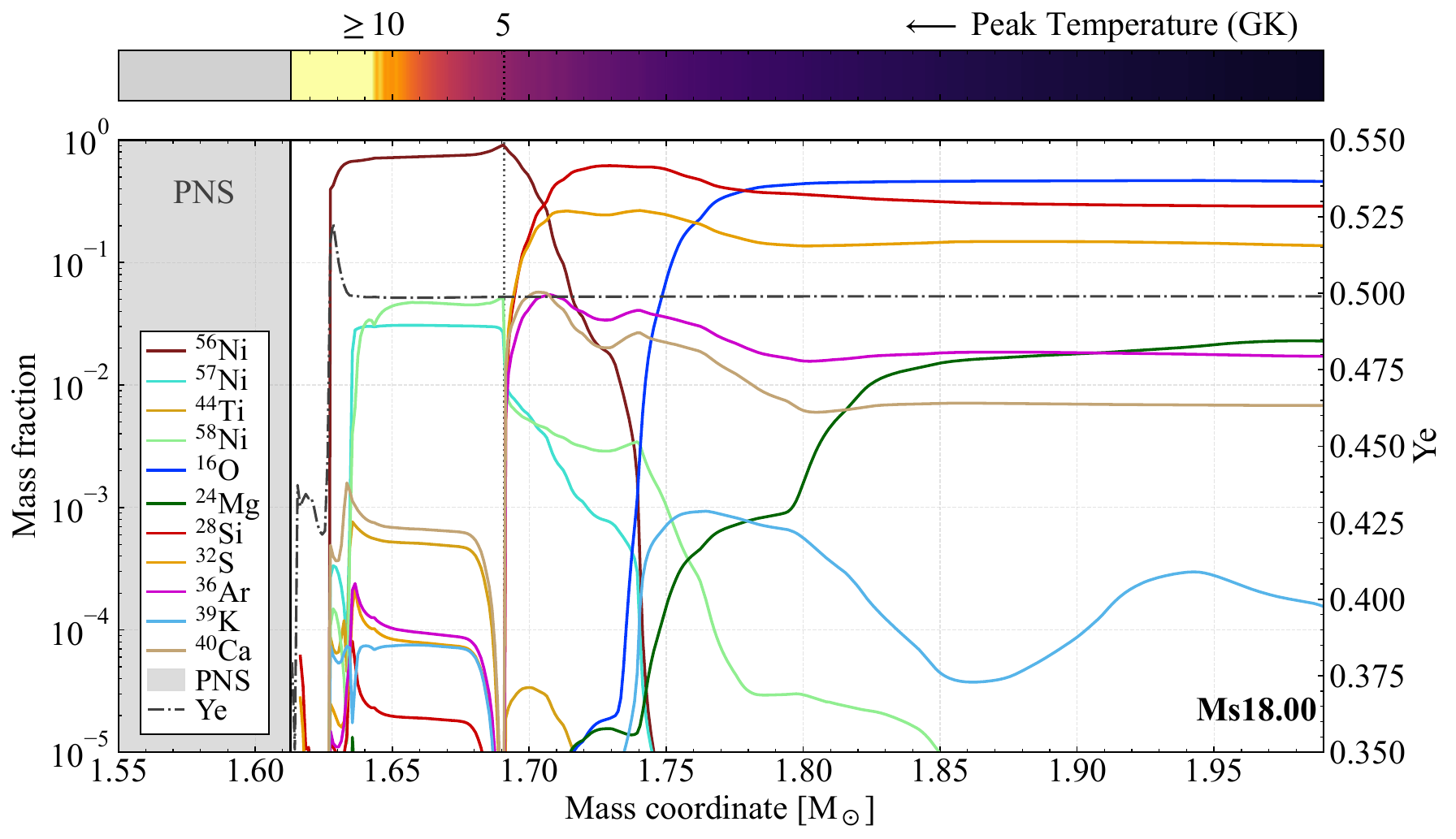}
\end{minipage}
\hfill
\begin{minipage}{0.49\textwidth}
    \centering
    \includegraphics[width=\linewidth]{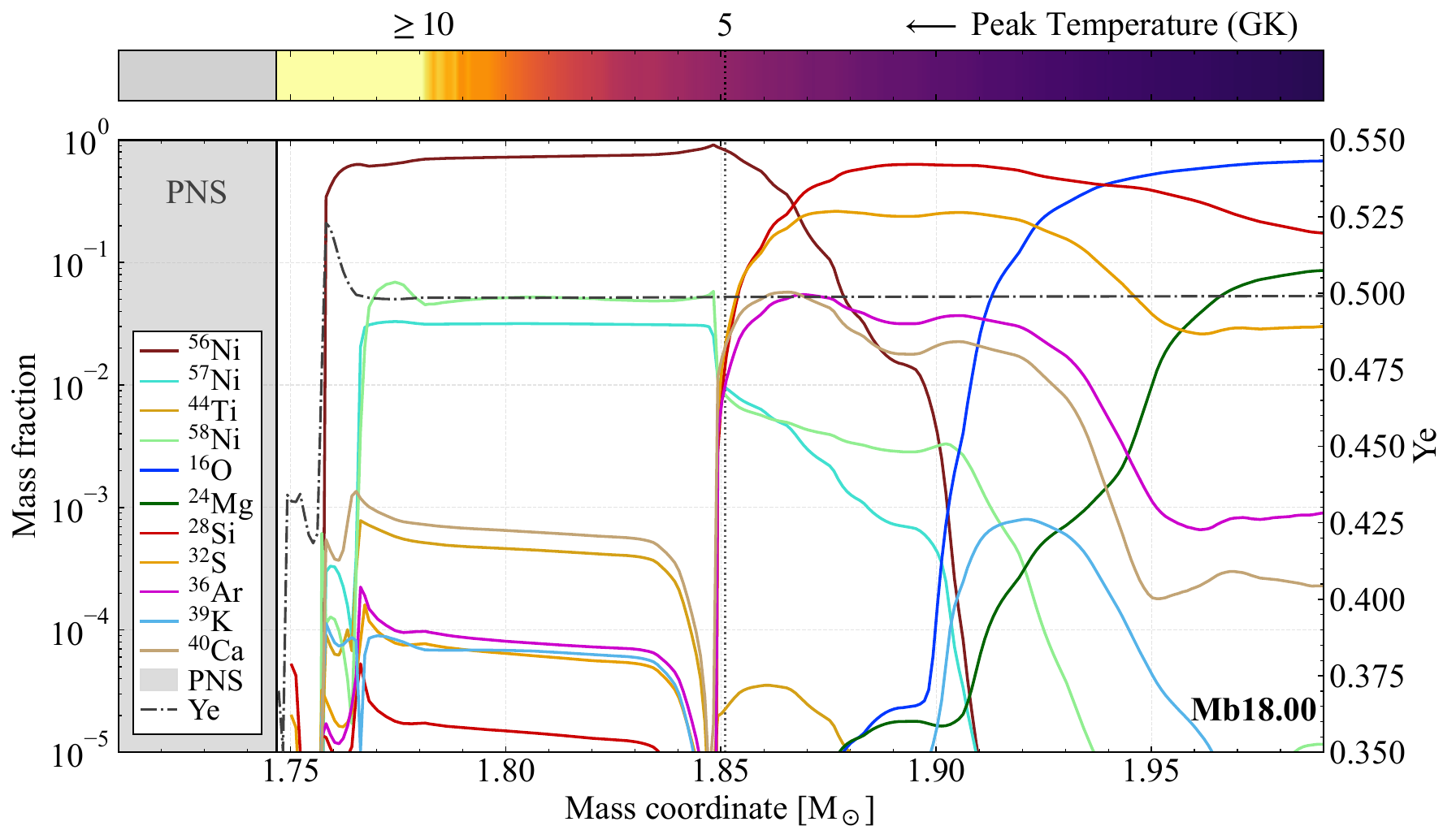}
\end{minipage}

\caption{
Top panels: pre-explosion composition of the inner layers of the $18\,\mathrm{M}_\odot$ single-star (left) and binary-stripped (right) models.
Bottom panels: final composition following explosive nucleosynthesis, with the colour scale indicating the peak temperature attained by each layer. Vertical dotted lines mark $5$\,GK, above which complete silicon burning occurs. Vertical dashed lines mark the mass cuts, while dot-dashed curves show the corresponding electron fraction.
}
\label{fig:inner_composition}

\end{figure*}

\begin{figure}
    \centering
    \includegraphics[width=\columnwidth]{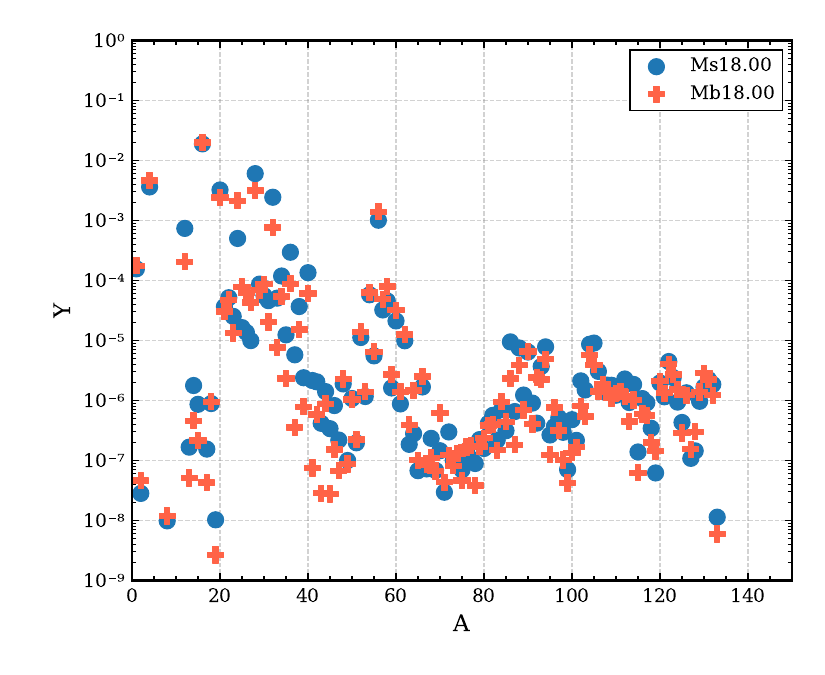}
\caption{Mass-integrated final abundances, Y, as a function of mass number, A, for the $18\,\mathrm{M}_\odot$ single-star (blue circles) and binary-stripped (red pluses) models, for material reaching peak temperatures above $1.75$\,GK.}

\label{fig:integrated_final_abundances}
\end{figure}

Before explosion, Ms18 and Mb18 have broadly similar chemical stratification, as illustrated by the $\alpha$-chain mass-fraction profiles in the top panels of Figure~\ref{fig:inner_composition}. In both progenitors, Si- and S-rich material surrounds the iron core, with O- and Mg-rich layers farther out.
However, the locations and extent of these regions differ, with Mb18 showing sharper transitions between layers. The iron-rich region extends to $\sim1.6\,\mass$ in Ms18 and $\sim1.7\,\mass$ in Mb18, beyond which the iron-group mass fraction falls below $10^{-3}$. The Si-, S-, and Ar-rich layers extend farther outwards in Ms18, becoming subdominant to O- and Mg-rich material at $\sim2.2\,\mass$, compared with $\sim1.9\,\mass$ in Mb18. The differences in pre-explosion structure become more pronounced farther out. In Mb18, the O-rich region declines sharply at $\sim4\,\mass$ and is followed by a relatively small He-rich layer, with no H-rich envelope. Ms18 instead retains a more extended O-rich region followed by substantial He- and H-rich outer layers. These outer layers are not shown in full in Figure~\ref{fig:inner_composition}, which focuses on the inner regions. Binary stripping therefore modifies not only the total mass and compactness of the progenitors, but also the composition and extent of the layers available for explosive processing and ejection.

We note that the PUSH simulations follow the collapse, core bounce, shock revival, and early shock propagation through the inner progenitor. The extended envelopes are therefore excluded from the hydrodynamic calculations and reattached afterwards \citep[see][]{Curtis2021}. The resulting ejecta can then be evolved with radiation-hydrodynamics and Monte Carlo radiative transfer codes, such as \textsc{snec} \citep{Morozova2015} and \textsc{supernu} \citep{Wollaeger2013}, to generate synthetic observables.

The bottom panels of Figure~\ref{fig:inner_composition} show the post-processed compositions of the innermost ejecta exterior to the mass cut, together with the peak temperature and electron fraction reached by each layer. 
In both models, the innermost ejecta reach peak temperatures above $\sim5$\,GK, undergoing complete silicon burning and producing predominantly iron-group isotopes, particularly $^{56}$Ni, $^{57}$Ni, and $^{58}$Ni.
The stable $^{58}$Ni-rich material is concentrated interior to the $^{56}$Ni-rich region, although its distribution extends to somewhat larger mass coordinates and higher mass fractions in Ms18 than in Mb18. Such differences could influence the synthetic observables, particularly the line profiles associated with stable and radioactive Ni, although detailed non-local thermodynamic equilibrium (NLTE) radiative transfer calculations would be required to quantify this effect \citep[e.g.,][]{Dessart2025}. 
The mass cuts in Ms18 and Mb18 both fall within the Si/O-rich shell, with Mb18 having the larger mass cut and consequently ejecting less of its innermost Si-rich material. At bounce, Mb18 has a compactness of $\xi_{2.0}=0.34$, compared with $0.26$ for Ms18. These results are consistent with the increase in remnant mass with compactness found across our model grid in Section~\ref{sec:explosion_compactness_trends} and in previous PUSH studies \citep{push2,push3}.
The progenitor properties, explosion properties, and selected isotopic yields of Ms18 and Mb18 are summarised in Table~\ref{tab:Ms18_pre_and_explosive_yields}.

Following \citet{push3}, we compare the integrated final abundances of material reaching peak temperatures above $1.75\,\mathrm{GK}$ in Figure~\ref{fig:integrated_final_abundances}, corresponding to processed masses of $0.72\,\mass$ and $0.68\,\mass$ for Ms18 and Mb18, respectively.
Both models show broadly similar abundance patterns, with most isotopic abundances agreeing to within an order of magnitude. The largest differences occur for light and intermediate-mass nuclei, reflecting variations in the initial composition and stratification, and in the material undergoing explosive O and incomplete Si burning ($24 \lesssim A \lesssim 44$). Initially, Ms18 and Mb18 contain $0.294\,\mass$ and $0.324\,\mass$ of $^{28}$Si, respectively, with Si-rich material more centrally concentrated in Mb18 (see Figure~\ref{fig:inner_composition}). The larger mass cut in Mb18 therefore removes a greater fraction of this Si-rich material from the ejecta, while the material that is ejected also undergoes more extensive explosive processing. This results in the net destruction of $0.104\,\mass$ and $0.178\,\mass$ of $^{28}$Si, and final yields of $0.191\,\mass$ and $0.146\,\mass$ for Ms18 and Mb18, respectively. Thus, despite its larger initial $^{28}$Si mass, Mb18 ejects less $^{28}$Si. The higher abundance at $A=56$ in Mb18 reflects its larger $^{56}$Ni yield, consistent with more extensive complete Si burning.

\begin{table*}
\centering
\caption{
Progenitor, explosion, and nucleosynthetic properties of the Ms18 and Mb18 models. The table columns give the ZAMS mass, $\xi_{2.0}$ at bounce, total progenitor mass at collapse, iron-core mass, carbon--oxygen core mass, helium-core mass, hydrogen-rich envelope mass, explosion energy, remnant mass, and the nucleosynthetic yields of selected isotopes.
The iron-core mass is defined by the outer boundary of the central region with $Y_{\mathrm{e}}<0.495$. The CO-core mass is the enclosed mass at the outer boundary where the helium mass fraction satisfies $X_{\mathrm{He}}\leq0.2$. The helium-core mass is the enclosed mass at the outer boundary where the hydrogen mass fraction satisfies $X_{\mathrm{H}}\leq0.2$. The hydrogen-rich envelope mass is the mass beyond the helium core.
}

\label{tab:Ms18_pre_and_explosive_yields}
\begin{tabular}{lccccccccccccc}
\hline
Model &
$\mathrm{M}_{\mathrm{ZAMS}}$ &
$\xi_{2.0}$ &
$\mathrm{M}_{\mathrm{prog}}$ &
$\mathrm{M}_{\mathrm{Fe}}$ &
$\mathrm{M}_{\mathrm{CO}}$ &
$\mathrm{M}_{\mathrm{He}}$ &
$\mathrm{M}_{\mathrm{env}}$ &
$\mathrm{E}_{\mathrm{expl}}$ &
$\mathrm{M}_{\mathrm{remn}}$ &
$^{56}\mathrm{Ni}$ &
$^{57}\mathrm{Ni}$ &
$^{58}\mathrm{Ni}$ &
$^{44}\mathrm{Ti}$ \\
&
$(\mathrm{M}_{\odot})$ &
--- &
$(\mathrm{M}_{\odot})$ &
$(\mathrm{M}_{\odot})$ &
$(\mathrm{M}_{\odot})$ &
$(\mathrm{M}_{\odot})$ &
$(\mathrm{M}_{\odot})$ &
$(\mathrm{B})$ &
$(\mathrm{M}_{\odot})$ &
$(\mathrm{M}_{\odot})$ &
$(\mathrm{M}_{\odot})$ &
$(\mathrm{M}_{\odot})$ &
$(\mathrm{M}_{\odot})$ \\
\hline
Ms18 &
18 &
0.26 &
10.08 &
1.49 &
5.50 &
7.41 &
2.67 &
0.97 &
1.61 &
5.63E-02 &
1.82E-03 &
2.59E-03 &
2.86E-05 \\
Mb18 &
18 &
0.34 &
5.74 &
1.52 &
4.31 &
5.74 &
0.00 &
1.34 &
1.75 &
7.78E-02 &
2.78E-03 &
4.48E-03 &
3.75E-05 \\
\hline

\end{tabular}

\end{table*}

\subsection{Impact of Binary Stripping on Elemental Yields}
\label{Production Factors}

\begin{figure*}
    \centering
    \includegraphics[width=\textwidth]{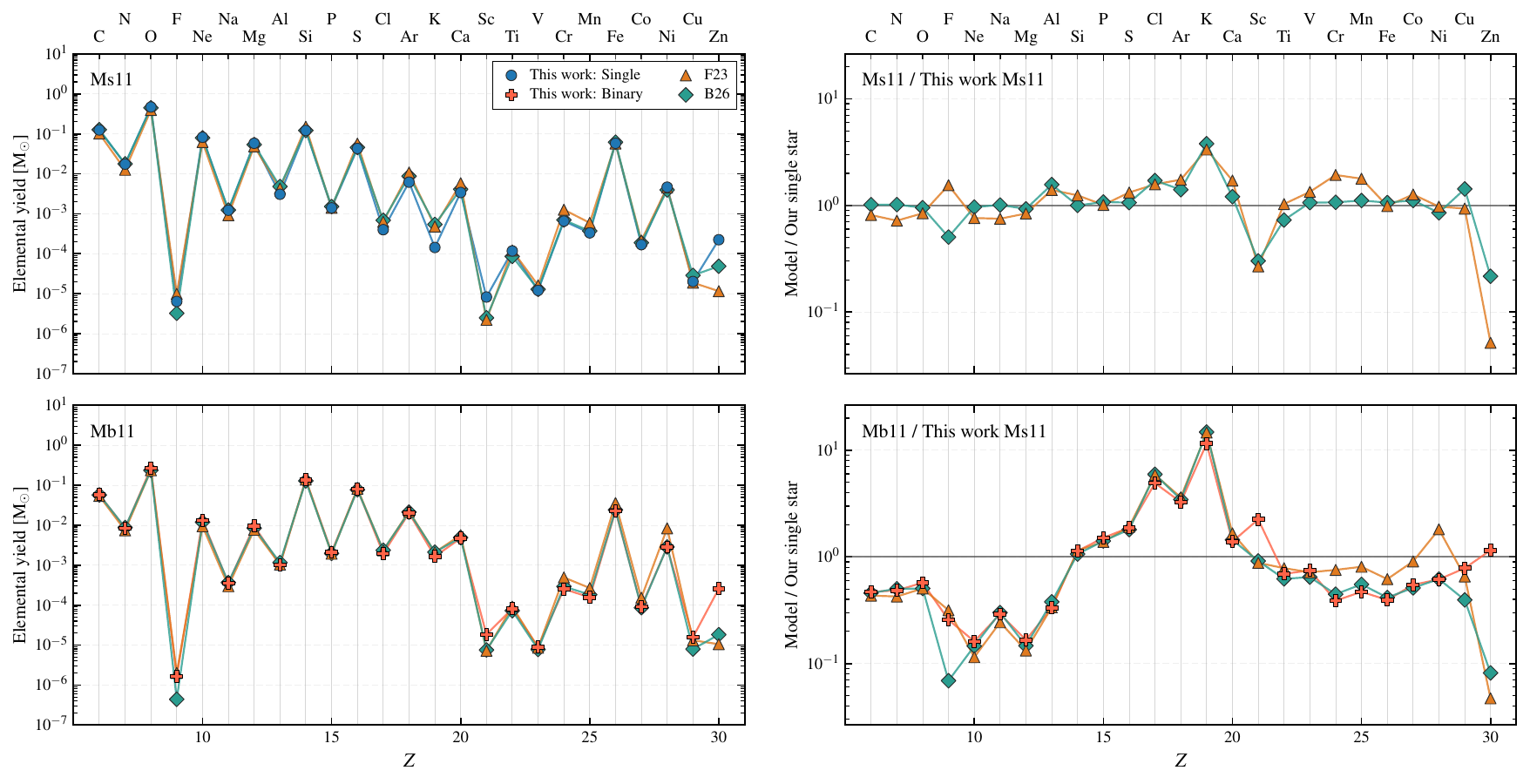}
\caption{
Elemental yields for the Ms11 (top) and Mb11 (bottom) models compared with those of \citet{Farmer2023} and \citet{Boccioli2026}.
The left-hand panels show the absolute elemental yields, whereas the right-hand panels show the corresponding yields relative to the Ms11 yields from this work (our single-star model).
Note that in the upper-right panel only the \citet{Farmer2023} and \citet{Boccioli2026} models are shown and that the solid horizontal line marks equal yields with the Ms11 model from this work.
}
    \label{fig:Ms11_Mb11_absolute_relative_yields}
\end{figure*}

In this section, we investigate the impact of binarity on the nucleosynthetic yields by comparing Ms and Mb progenitors at the same ZAMS mass. We first place our nucleosynthesis calculations in the context of previous studies. Figure~\ref{fig:Ms11_Mb11_absolute_relative_yields} compares the yields obtained in this work with those of \citet{Farmer2023} and \citet{Boccioli2026}. We focus on the Ms11 and Mb11 models, as $11\,\mass$ is the only ZAMS mass for which both the model series produce successful explosions in all three treatments.
The three calculations employ substantially different explosion and nucleosynthesis treatments. In particular, \citet{Farmer2023} adjust the thermal-bomb energy input to achieve a common total explosion energy of $10^{51}\,\mathrm{erg}$, whereas the explosion energies in \citet{Boccioli2026} and our PUSH calculations emerge from the hydrodynamic evolution under the adopted explosion prescriptions.
Differences in the yields may therefore reflect differences in explosion energetics and dynamics, as well as the adopted mass cuts and nucleosynthesis treatments. We post-process the thermodynamic trajectories with \texttt{CFNET} using a network of 2902 isotopes extending to $^{211}$At, including neutrino and antineutrino captures. \citet{Boccioli2026} post-process their explosions with \textsc{SkyNet} using 1500 isotopes up to $A=100$, without neutrino interactions during the post-processing, while \citet{Farmer2023} employ the smaller 162-isotope \textsc{mesa} network up to $^{64}$Zn.

Despite differences in approach, all three recover similar yield patterns, with most individual cases agreeing to within a factor of a few. The largest differences occur for the iron-group elements, particularly Zn. These elements are synthesised predominantly in the innermost ejecta through complete and incomplete Si burning, making their yields particularly sensitive to the explosion dynamics, mass cut, thermodynamic conditions, and local electron fraction.
\citet{Boccioli2026} compared previous PUSH calculations \citep{push3} with their \textsc{GR1D+} explosion and nucleosynthesis calculations and found that differences between the treatments can be traced to contributions from the $\nu p$-process, weak $r$-process elements, and the adopted NSE threshold. 
Nevertheless, when all three treatments agree that a given model explodes, the predicted yields agree remarkably well, with all three showing that binary stripping can substantially alter the yields of particular elements.

\begin{figure*}
\centering

\begin{minipage}{\textwidth}
    \centering
    \includegraphics[width=0.87\textwidth]{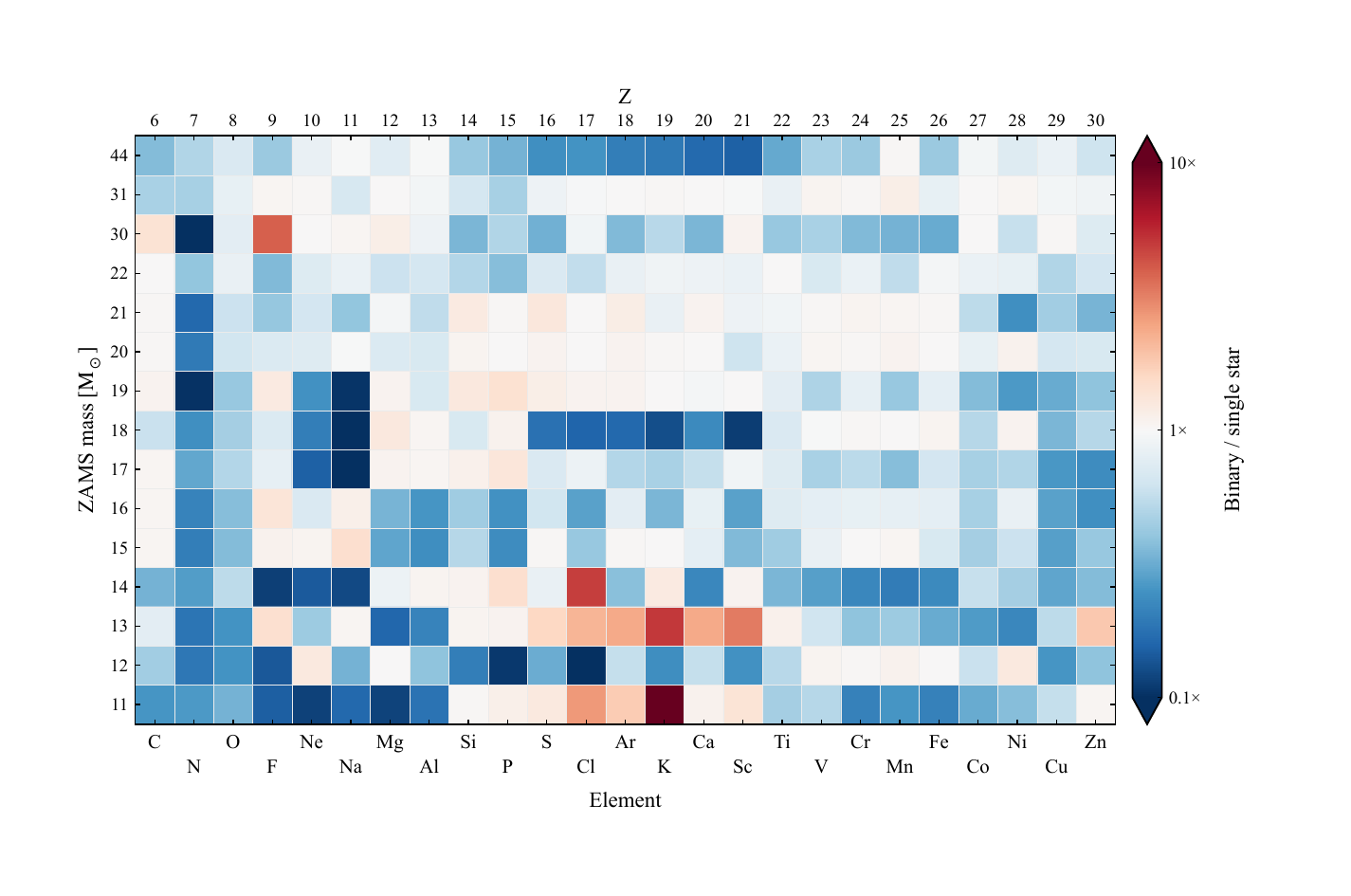}
    \captionof{figure}{
    Ratio of the total elemental yields from binary-stripped progenitors to those from their single-star counterparts as a function of ZAMS mass and element.
    Red indicates larger yields in the binary-stripped model, while blue indicates larger yields in the single-star model. Ratios are shown only for ZAMS masses at which both progenitors explode successfully.
    }
    \label{fig:Mb_over_Ms_elemental_yields_raw}
\end{minipage}

\vspace{0.5cm}

\begin{minipage}{\textwidth}
    \centering
    \includegraphics[width=0.87\textwidth]{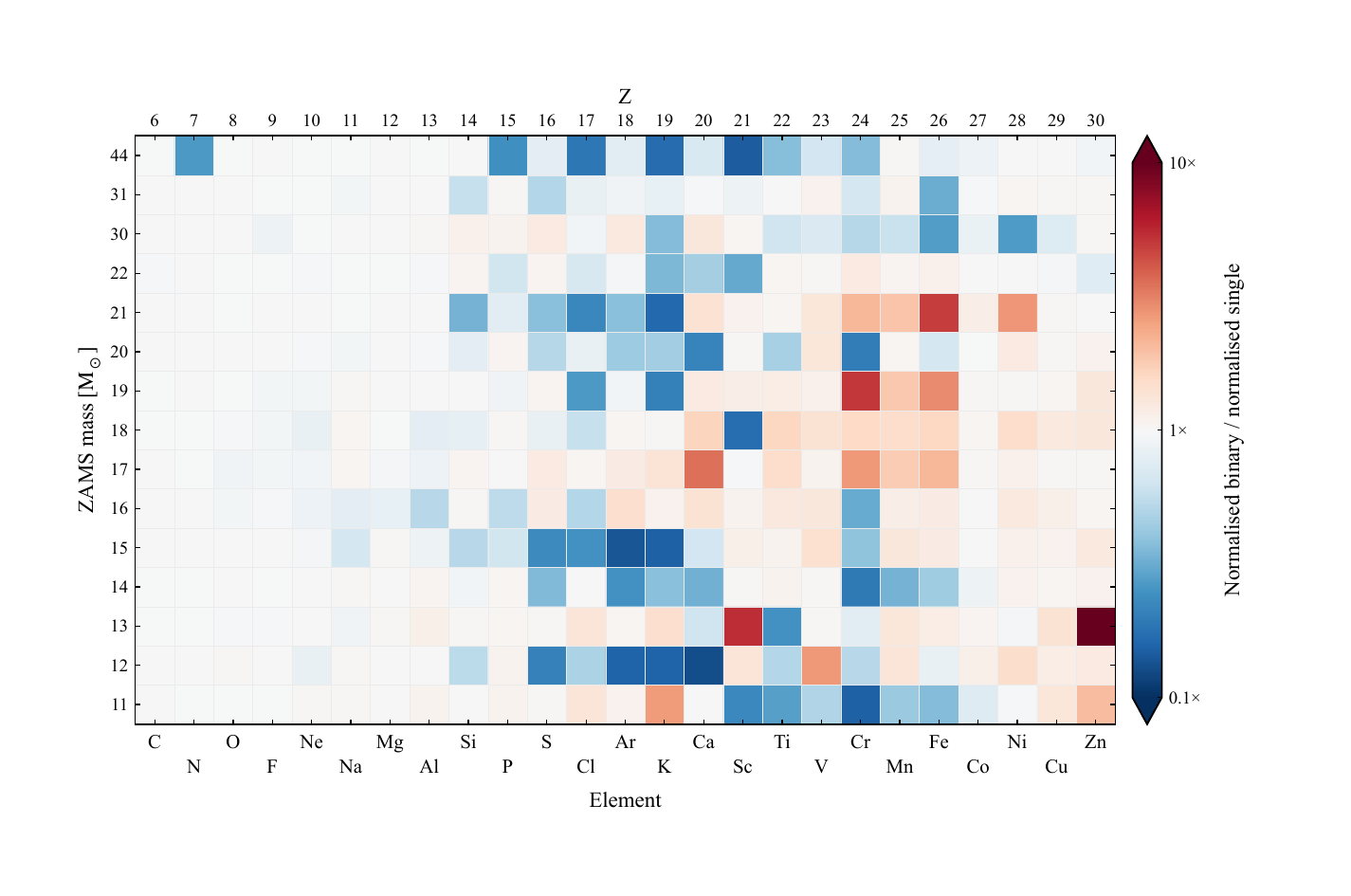}
    \captionof{figure}{
    Similar to Figure~\ref{fig:Mb_over_Ms_elemental_yields_raw}, but with each elemental yield normalised by its corresponding pre-explosion value to compare the relative changes between the model sets.
    }
    \label{fig:Mb_over_Ms_elemental_yields_normalised}
\end{minipage}

\end{figure*}

\begin{figure*}
\centering
    \includegraphics[width=0.87\textwidth]{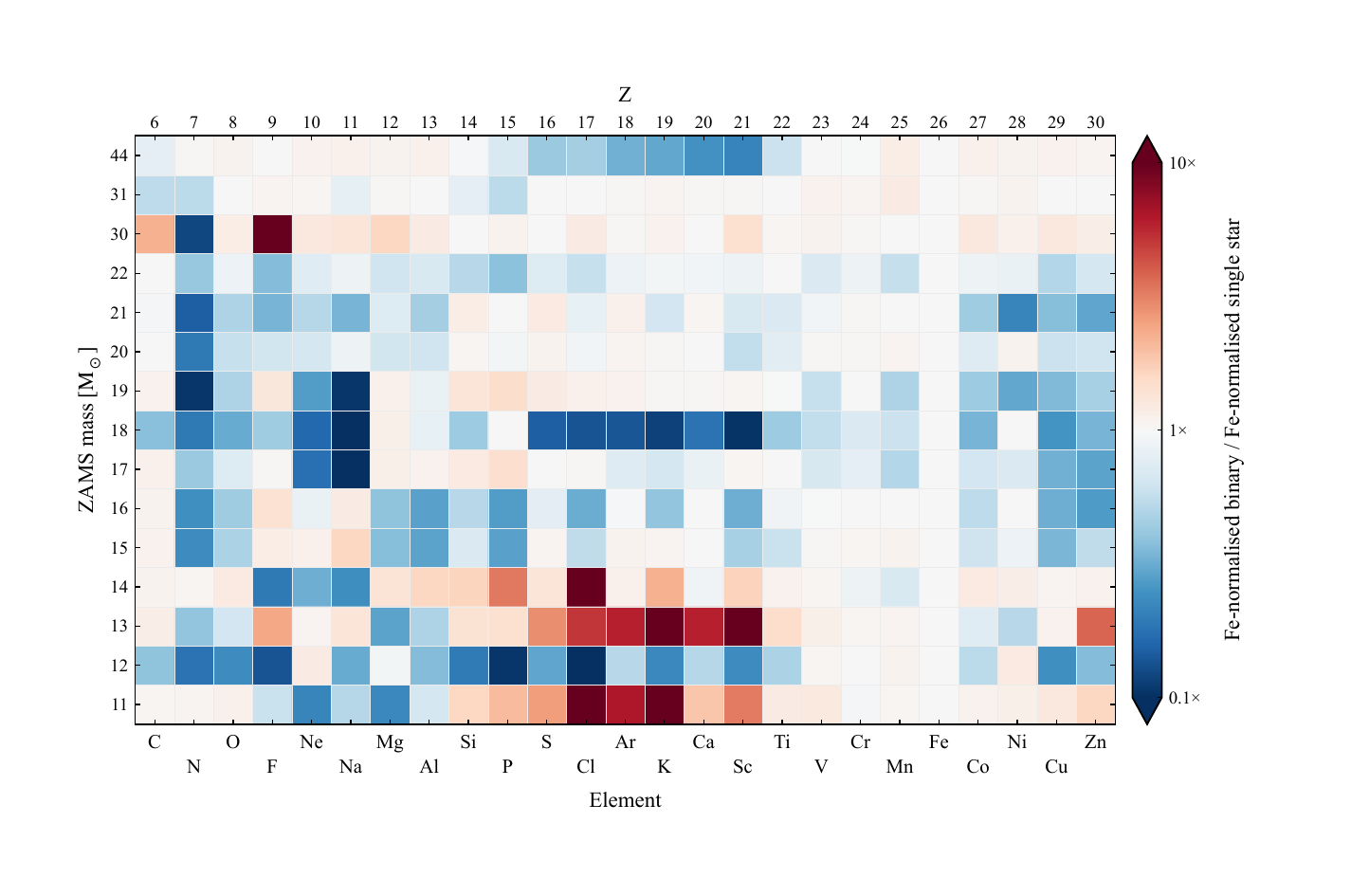}
    \caption{
    Similar to Figure~\ref{fig:Mb_over_Ms_elemental_yields_raw}, but with each elemental yield normalised by the corresponding final Fe yield. Red indicates a larger abundance relative to Fe in the binary-stripped model, while blue indicates a larger abundance relative to Fe in the single-star model.
    }
    \label{fig:Mb_over_Ms_elemental_yields_fe}
\end{figure*}

\begin{figure}
\centering
\includegraphics[width=\columnwidth]{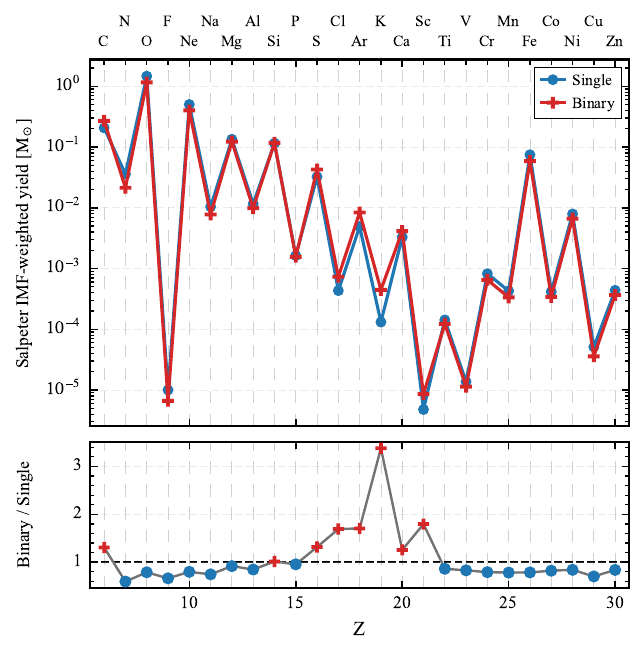}
    \caption{    
    Salpeter IMF-weighted elemental yields for the single-star and binary-stripped progenitor populations (top) and the binary-stripped to single-star yield ratio (bottom). Red and blue points in the lower panel indicate larger yields from the binary-stripped and single-star populations, respectively. The dashed horizontal line marks equal yields.
    }
\label{fig:Salpeter_IMF_elemental_yields}
\end{figure}

Figure~\ref{fig:Mb_over_Ms_elemental_yields_raw} shows the ratio of final elemental yields from binary-stripped progenitors to those from their single-star counterparts across the ZAMS-mass range. These yields reflect both pre-supernova evolution and subsequent reprocessing during collapse and explosion. Single-star progenitors typically have larger yields of lighter elements, which are predominantly produced during hydrostatic evolution and reside in the outer layers. However, binarity also strongly affects the yields of heavier elements, with clear differences for elements from S to Sc in the lowest ZAMS-mass models ($11$--$14\,\mass$). This is particularly interesting as S, Ar, and Ca are major products of explosive O burning and incomplete Si burning, while the production of Sc probes deeper and hotter layers, where it is largely produced through the decay of $^{45}$Ti synthesised in high $Y_e$ regions. This behaviour is consistent with \citet{Farmer2023}, who found particularly strong effects of binarity on yields from Al to Sc, while \citet{Boccioli2026} also found sensitivity of iron-group yields to the adopted explosion treatment.

To account for differences in the pre-explosion mass of each element, we normalise each yield by the corresponding progenitor's pre-explosion yield in Figure~\ref{fig:Mb_over_Ms_elemental_yields_normalised}.
The lighter elements up to Al remain broadly similar, except in the $44\,\mass$ progenitor, where a small increase in N produces a considerably larger relative enhancement.
Between Si and Sc, the single-star models typically show greater relative enhancement, whereas at intermediate ZAMS masses the binary-stripped models can show greater enhancement from Cr to Fe.
The heaviest elements, which are particularly sensitive to the explosion treatment, generally show only slight enhancement in the binary-stripped models, with Zn in the $13\,\mass$ progenitor the clearest exception. 
Zn is produced in complete Si burning and in material with $Y_e \gtrsim 0.5$ close to the mass cut. Mb13 contains approximately twice as much Si-rich material near the mass cut as Ms13, partly accounting for its enhanced Zn production.
However, differences in compactness and the associated PUSH heating term complicate these trends (see Section~\ref{sec:yield_trends} for a discussion of their dependence on compactness).

Figure~\ref{fig:Mb_over_Ms_elemental_yields_fe} shows the ratio of final elemental abundances relative to Fe in each binary-stripped model to those in its corresponding single-star counterpart.
Normalising to Fe partly removes differences in the overall scale of the yields and highlights variations in the relative abundance patterns.
We find no systematic enhancement across the full mass range; instead, the patterns vary considerably with ZAMS mass. At the lowest masses ($11$, $13$, and $14\,\mass$), the binary-stripped models generally show enhanced intermediate-mass and several iron-group element abundances relative to Fe.
At higher masses, abundances relative to Fe are more commonly lower in the binary-stripped models, with several IMEs enhanced in the single-star models at $44\,\mass$, although considerable scatter remains.
This further demonstrates that binarity does not produce a uniform shift in the abundance pattern, with differences depending on both ZAMS mass and the individual element. We also compare several representative species with their corresponding solar abundance ratios in Section~\ref{sec:alpha_elements}, providing further context for the predicted yields.

We conclude this subsection by assessing the population-level impact of binarity on nucleosynthetic yields, weighting each progenitor by a Salpeter IMF \citep[][]{Salpeter1955}, normalised over the progenitor grid. The yields include pre-supernova winds and final PUSH ejecta for successful explosions, but only wind yields for failed explosions. Following \citet{Boccioli2026}, we exclude the four binary-stripped progenitors that did not reach core collapse ($24$, $25$, $28$, and $29\,\mass$). We assume the same initial mass distribution for both populations. 
Figure~\ref{fig:Salpeter_IMF_elemental_yields} shows that the IMF-weighted yields of the two populations remain broadly similar, although binarity produces the largest differences for elements from Cl to Ti, with K enhanced by a factor of $\sim3.4$ in the binary-stripped population. \citet{Farmer2023} linked enhanced pre-supernova production of $^{39}$K in low-mass binary-stripped progenitors to the extent of their Si-rich layers. The greater weight assigned to these progenitors by the Salpeter IMF amplifies their contribution to the integrated K yield. Recently, \citet{Rodriguez2023} inferred a mean CCSN Fe yield of $0.058\pm0.007\,\mass$, subsequently used by \citet{Weinberg2024} to anchor the absolute scale of CCSN nucleosynthetic yields. Our IMF-weighted Fe yields are $0.0747\,\mass$ and $0.0589\,\mass$ for the single-star and binary-stripped populations, respectively. The binary-stripped population therefore shows closer agreement with this benchmark.

\subsection{Trends in Nucleosynthetic Yields with Compactness}
\label{sec:yield_trends}

\begin{figure*}
    \centering    \includegraphics[width=0.9\linewidth]{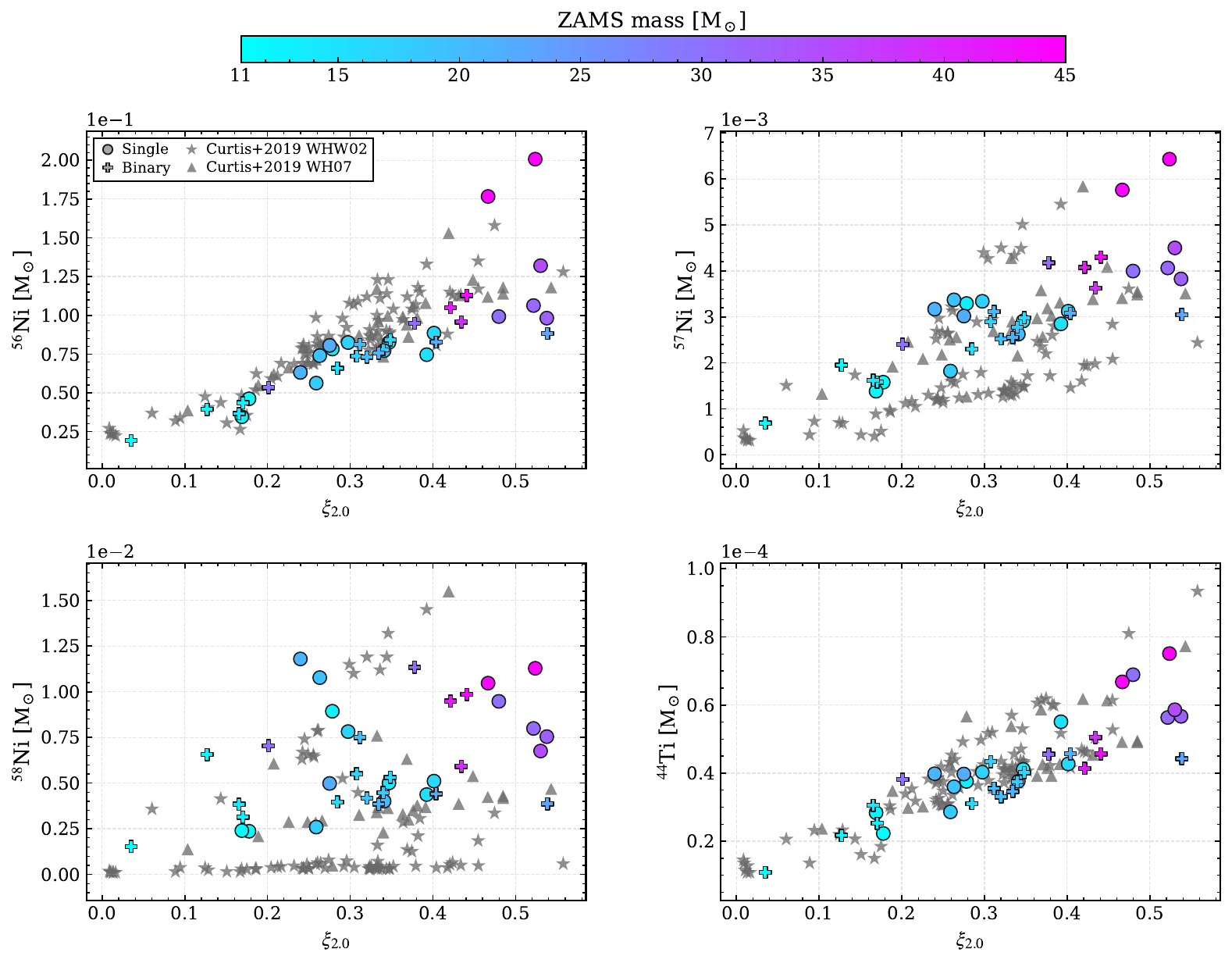}
\caption{
    Isotopic yields of $^{56}$Ni (top left), $^{57}$Ni (top right), $^{58}$Ni (bottom left), and $^{44}$Ti (bottom right) after explosive processing, shown as a function of compactness for the single-star (circles) and binary-stripped (pluses) model series, with symbol colour indicating ZAMS mass. The corresponding yields from \protect\citet{push3} are overplotted for the WHW02 and WH07 model series as grey stars and triangles, respectively.
}
\label{fig:Ms_Mb_rad_masses_vs_xi2_cbar}
\end{figure*}

We now examine how the yields of key isotopes and iron-group elements in the exploding models depend on progenitor compactness, while also considering differences associated with the pre-explosion structure. 
Figure~\ref{fig:Ms_Mb_rad_masses_vs_xi2_cbar} shows the yields of $^{56}$Ni, $^{57}$Ni, $^{58}$Ni, and $^{44}$Ti as a function of $\xi_{2.0}$, with the models colour-coded by ZAMS mass.
This allows the yield trends to be considered alongside the progenitor structures shown in Figures~\ref{fig:mass_components_single} and~\ref{fig:mass_components_binary}. 
The yields of $^{56}$Ni and $^{44}$Ti generally increase with compactness, while $^{57}$Ni follows a similar trend but with greater scatter.
At higher compactness, the single-star and binary-stripped models investigated here show somewhat more scatter in these three isotopes than the previous PUSH calculations \citep{push3}.
The $^{58}$Ni yields exhibit substantial scatter across the entire compactness range. These variations highlight the sensitivity of the yields to differences in progenitor structure that are not captured by compactness alone.

The scatter at fixed compactness is particularly apparent among the single-star models. For example, Ms35 and Ms45 both have $\xi_{2.0}\simeq0.52$, yet Ms45 produces almost twice as much $^{56}$Ni. This difference reflects their pre-explosion structures: Ms45 has a larger and more compressed Si-rich region available for explosive processing, allowing more material to undergo Si burning and contribute to the radioactive Ni yield. The binary-stripped sequence also shows differences in yields at comparable compactness, although these are generally smaller than those among the single-star models.

The weakest dependence on compactness is found for the stable, neutron-rich isotope $^{58}$Ni. Although the most compact models tend, on average, to produce somewhat larger $^{58}$Ni yields, the models do not follow a single well-defined trend. This is broadly consistent with \citet{push3}, who found that neutron-rich isotope production depends on the local electron fraction. In their calculations, $^{58}$Ni yields separate into multiple branches associated with different $Y_e$ conditions, with lower $Y_e$ generally favouring greater production of neutron-rich isotopes such as $^{57}$Ni and $^{58}$Ni.

\begin{figure*}
    \centering    \includegraphics[width=0.9\linewidth]{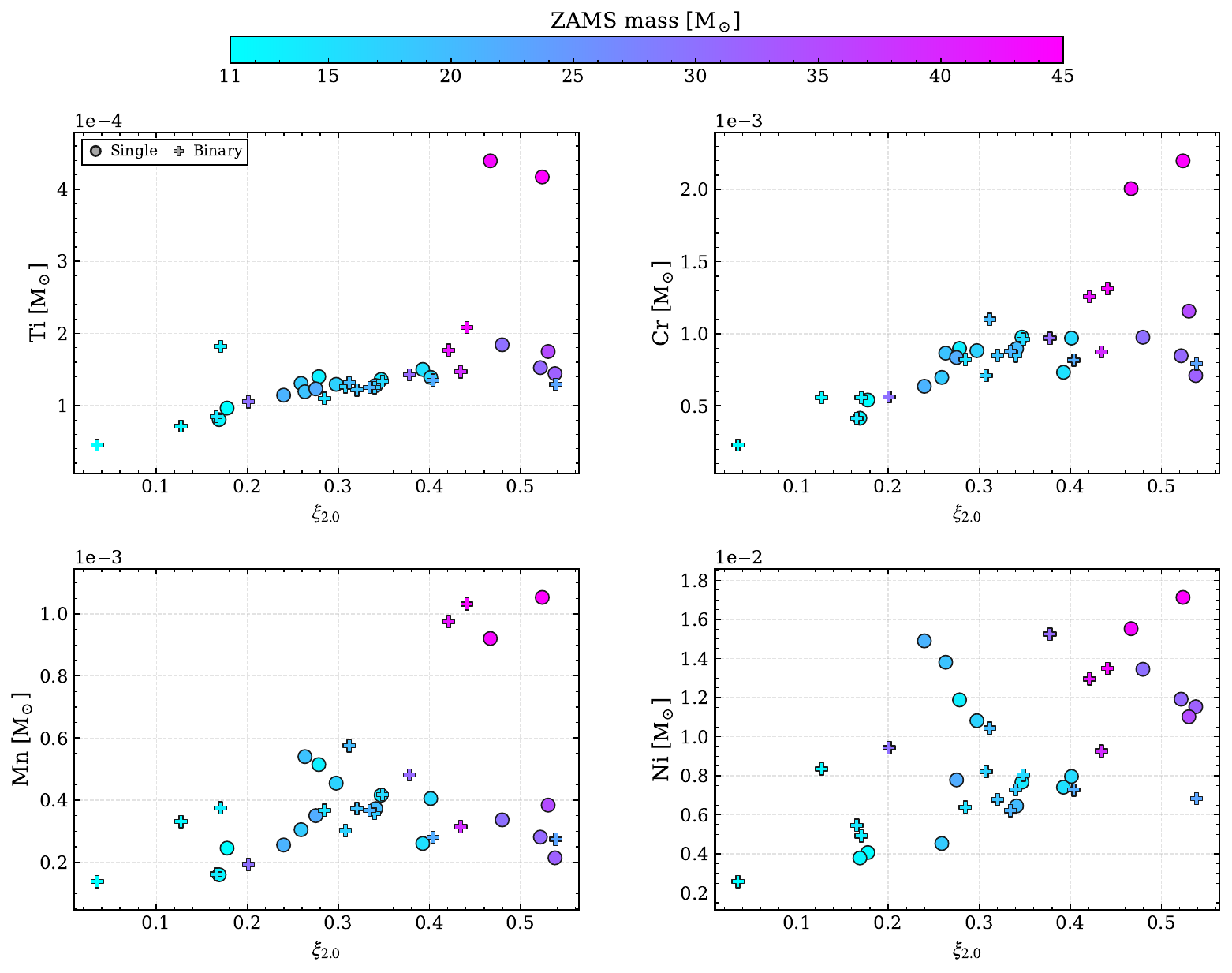}
\caption{
Similar to Figure~\ref{fig:Ms_Mb_rad_masses_vs_xi2_cbar}, but showing the final elemental yields of Ti (top left), Cr (top right), Mn (bottom left), and Ni (bottom right).
}

\label{fig:Ms_Mb_elemental_yields_vs_xi2_cbar}
\end{figure*}

Figure~\ref{fig:Ms_Mb_elemental_yields_vs_xi2_cbar} shows the final elemental yields of Ti, Cr, Mn, and Ni as a function of compactness, with the models colour-coded by ZAMS mass. Ti, Cr, and Ni receive substantial contributions from complete Si burning, so their yields depend in part on how much material undergoes this processing and is subsequently ejected. Mn, however, is produced mainly through the decay of $^{55}$Co synthesised during incomplete Si burning, and its yield is sensitive to the local electron fraction \citep[see table~2 of][]{push3}. Differences in the extent of these burning regions can therefore affect the four elements differently.
The comparison between Ms44 and Mb44 illustrates how differences in pre-explosion structure can contribute to the yield variations. In Mb44, the Si-dominated region above the mass cut extends from $\sim1.8\,\mass$ to $\sim2.3\,\mass$, beyond which O-rich material becomes dominant. In Ms44, the Si-rich region extends farther out, to $\sim2.6\,\mass$, providing a larger reservoir available for explosive processing. The higher Ti and Cr yields, together with the more modest increase in Ni, are consistent with more material undergoing complete Si burning in Ms44. The similar Mn yields suggest more comparable contributions from incomplete Si burning in the two models. 
We note that the models have slightly different compactnesses and are therefore assigned slightly different values of $k_{\rm push}$, meaning that differences in the additional heating may also contribute to the yield differences alongside their pre-explosion structures.

\section{Comparison with Observational Constraints}
\label{sec:Comparison_With_Observations}

To place the predicted nucleosynthetic yields in an observational context, we follow \citet{push3} in comparing our single-star and binary-stripped models with observational estimates of $^{56}$Ni masses and Ni/Fe ratios in individual CCSNe in Section~\ref{sec:observed_ccsne}, and with the iron-group abundance pattern of the metal-poor star HD~84937 in Section~\ref{sec:MPS}. We then examine $\alpha$-element abundance ratios and their observational constraints in Section~\ref{sec:alpha_elements}, before comparing our IMF-weighted yields with empirical CCSN [X/Mg] ratios inferred from APOGEE in Section~\ref{sec:empirical_xmg}.

\subsection{Comparison with the Observed Landscape}
\label{sec:observed_ccsne}

\begin{figure}
    \centering    \includegraphics[width=0.99\columnwidth]{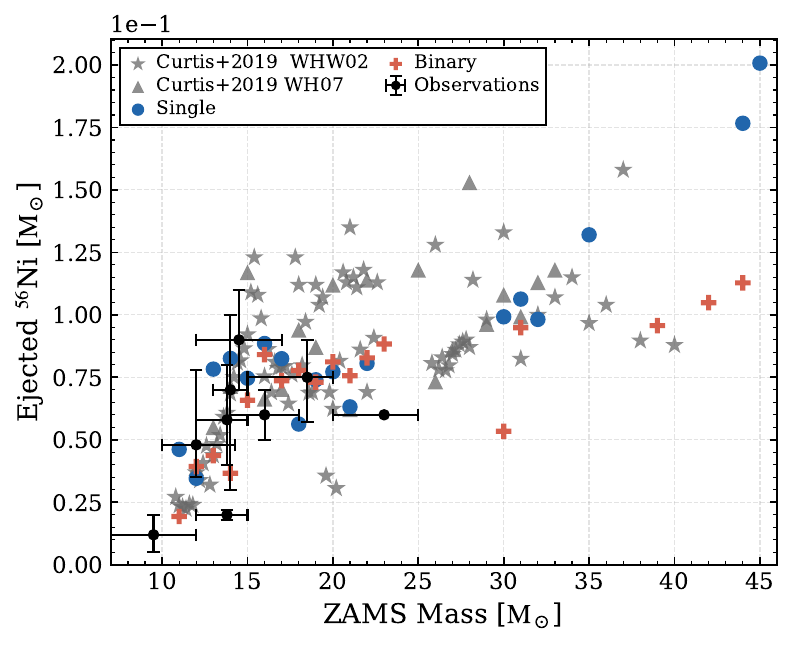}
\caption{
Ejected $^{56}$Ni mass as a function of ZAMS mass for the single-star (blue circles) and binary-stripped (red pluses) models. Observational estimates for CCSNe, excluding hypernovae, are shown as black crosses with error bars \citep[see][]{nomoto13,Bruenn2016}. The WHW02 and WH07 yields from \citet{push3} are shown in grey for comparison.
}
\label{fig:56ni_obs_zams}
\end{figure}

Radioactive $^{56}$Ni is an important energy source in CCSNe, with its decay through $^{56}$Co to $^{56}$Fe heating the ejecta, influencing their ionisation and spectral evolution, as well as the radioactive tail of the light curve. The ejected $^{56}$Ni mass therefore provides an observational constraint on explosive nucleosynthesis. Observationally, this is done by relating the bolometric luminosity during the radioactive tail to the decay energy deposited in the ejecta, although uncertainties in distance, extinction, and time of explosion affect the inferred mass \citep{Bruenn2016}. Figure~\ref{fig:56ni_obs_zams} shows the ejected $^{56}$Ni mass as a function of ZAMS mass for the single-star and binary-stripped explosion models. For comparison, we include the nucleosynthetic yields calculated by \citet{push3} for the explosion models investigated in \citet{push2}. The observational estimates are taken from \citet{nomoto13} and \citet{Bruenn2016}, excluding hypernovae. This sample spans inferred progenitor ZAMS masses of $\sim10$--$23\,\mass$ and $^{56}$Ni masses of $\sim0.01$--$0.09\,\mass$, with no constraints above $\sim25\,\mass$.

Both model series broadly cover the observed yield range, although differences emerge at individual ZAMS masses. Between $11$ and $15\,\mass$, the binary-stripped models generally produce weaker explosions and smaller $^{56}$Ni yields. The Mb11 model, in particular, lies close to the two lowest observationally inferred $^{56}$Ni masses, whereas Ms11 ejects considerably more $^{56}$Ni. 
The smaller $^{56}$Ni yield of Mb11 is consistent with its lower compactness, in line with the general trends in $^{56}$Ni production discussed in Section~\ref{sec:yield_trends}.
More broadly, our models show less scatter in $^{56}$Ni yields than those of \citet{push3} up to $\sim23\,\mass$, but substantially greater variation at higher ZAMS masses, where the previous calculations found relatively constant yields.
This is particularly evident for Ms44 and Ms45, which produce approximately twice as much $^{56}$Ni as similar-mass binary-stripped models. As discussed above, this difference is associated with variations in pre-explosion structure and the amount of Si-rich material available for explosive processing.

\begin{figure}
    \centering    \includegraphics[width=0.99\columnwidth]{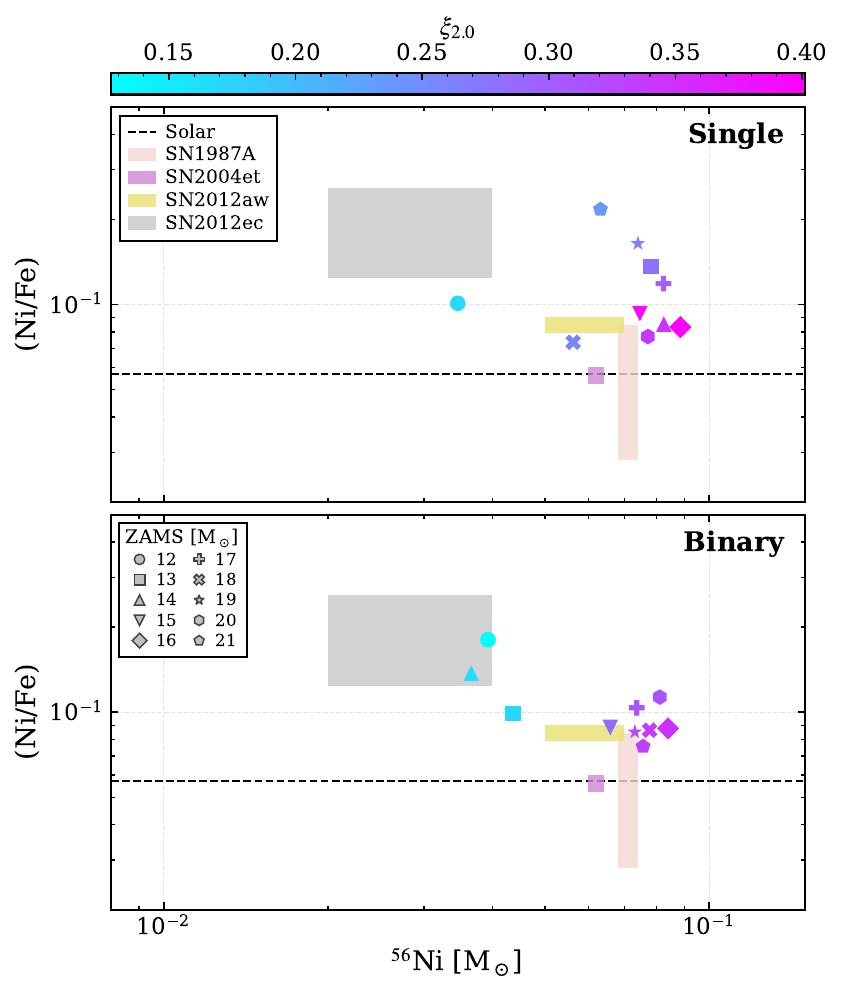}
\caption{
Ni/Fe ratio as a function of ejected $^{56}$Ni mass for the single-star (top) and binary-stripped (bottom) models with ZAMS masses of $12$--$21\,\mathrm{M}_\odot$, compared with observational estimates, shown as shaded regions, for SN~1987A, SN~2004et, SN~2012aw, and SN~2012ec \citep{Jerkstrand2015}.
Marker symbols denote ZAMS mass, while colour indicates the compactness. The dashed horizontal line marks the solar Ni/Fe ratio. 
Note that no Ni/Fe uncertainties were reported for SN~2004et and SN~2012aw; these points are therefore shown with error bars comparable to the marker size for visibility.
}
\label{fig:Ni_Fe_56Ni_zams}
\end{figure}

Figure~\ref{fig:Ni_Fe_56Ni_zams} compares the predicted Ni/Fe ratios and $^{56}$Ni yields of the single-star and binary-stripped models with the observational estimates compiled by \citet{Jerkstrand2015} for SN~1987A, SN~2004et, SN~2012aw, and SN~2012ec.
The comparison includes models with ZAMS masses between $12$ and $21\,\mass$, with marker colour indicating the bounce compactness, $\xi_{2.0}$, and the dashed horizontal line marking the solar Ni/Fe ratio.
The low-compactness Mb12 and Mb14 models provide the closest overall agreement with SN~2012ec, reproducing both its supersolar Ni/Fe ratio and its comparatively large $^{56}$Ni yield.
The Ms12 and Mb13 models produce comparable $^{56}$Ni yields, but their predicted Ni/Fe ratios differ from that inferred for SN~2012ec. 
Several higher-compactness models are consistent with both the $^{56}$Ni yields and Ni/Fe ratios inferred for SN~1987A and SN~2012aw. However, no model produces a Ni/Fe ratio low enough to match SN~2004et. Overall, the models span much of the observed range in both quantities.

\subsection{Comparison with Metal-Poor Stars}
\label{sec:MPS}

\begin{figure}
    \centering\includegraphics[width=0.99\columnwidth]{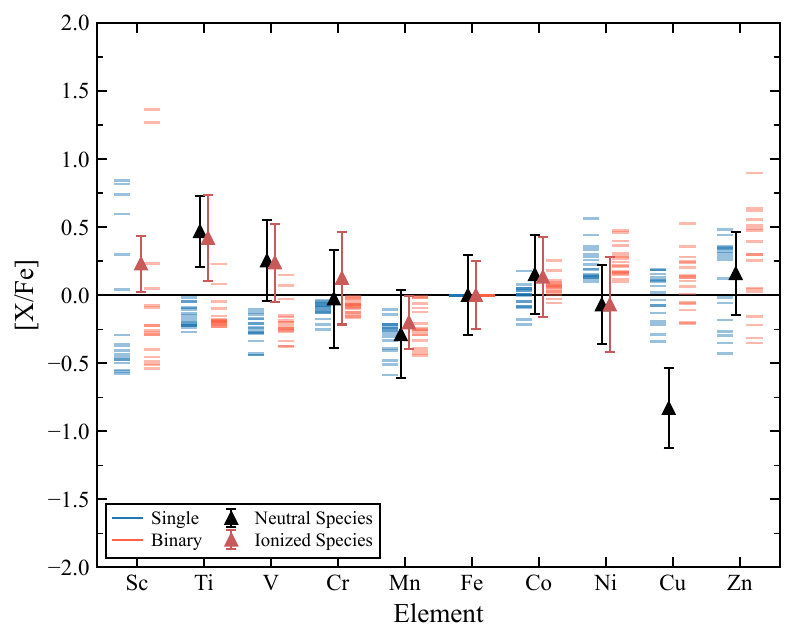}
\caption{
Elemental abundance ratios, $[\mathrm{X}/\mathrm{Fe}]$, for the single-star (blue) and binary-stripped (red) models, compared with abundances inferred from neutral and singly ionised species in HD~84937 (black and red triangles, respectively), with $3\sigma$ uncertainties \citep{Sneden2016}.
Each horizontal segment represents an individual model, offset to the left (single-star) or right (binary-stripped) of each element for clarity. 
Overlapping segments appear darker, indicating where multiple models predict similar abundance ratios. The opacity is scaled separately for each series so that complete overlap reproduces the corresponding legend colour.
}
\label{fig:Ms_Mb_xfe}
\end{figure}

The iron-group abundances in the atmospheres of metal-poor stars preserve a record of enrichment by earlier supernovae and therefore provide a test of predicted nucleosynthetic yields. Numerous photometric and spectroscopic surveys have identified metal-poor stars and characterised their atmospheric abundances \citep[e.g.][]{Caffau2013,Norris2013,Keller2014,Li2015a,Li2015b,Howes2015,Placco2015,Jacobson2015,Hansen2015,Roederer2016,Abohalima2018,Starkenburg2017,Andrievsky2018,DaCosta2019,Marino2019,Venn2020,Yong2021,Aoki2022,Li2022,Zepeda2022,Arentsen2023,Griffith2023,Martin2024,Griffith2026}.
We adopt the abundance measurements of \citet{Sneden2016} for HD~84937, a very metal-poor main-sequence turnoff star with $[\mathrm{Fe}/\mathrm{H}] = -2.32$. Their analysis used improved laboratory transition data to determine abundances from neutral and singly ionised iron-group species. Following the comparison introduced by \citet{push3}, Figure~\ref{fig:Ms_Mb_xfe} places our predicted elemental $[\mathrm{X}/\mathrm{Fe}]$ ratios alongside these measurements to assess whether binary stripping improves agreement with the observed abundance pattern.
The triangles show the abundances inferred from neutral and singly ionised species, with $3\sigma$ uncertainties, while the semi-transparent horizontal segments represent individual progenitor models. We show these separately rather than applying an IMF weighting to highlight how agreement with the observed abundances varies within each model series.

Overall, the predicted iron-group abundance patterns broadly agree with that observed in HD~84937, although the level of agreement varies between elements and individual models. 
The largest spread is seen for Sc and Zn, both of which have traditionally been challenging to reproduce in sufficient quantities in CCSN calculations. Nevertheless, subsets of both the single and binary-stripped models reproduce the observed Sc and Zn abundances.
Their production is particularly sensitive to conditions in the innermost ejecta. The largest $[\mathrm{Sc}/\mathrm{Fe}]$ ratios are obtained for the Mb11 and Mb13 models, consistent with the enhanced Sc yields seen in Figure~\ref{fig:Mb_over_Ms_elemental_yields_raw}. We note that neutrino interactions can enhance Sc production in CCSN ejecta by altering the electron fraction of the innermost layers \citep[see][for a discussion]{frohlich2006b}.
The clearest discrepancy with HD~84937 occurs for $[\mathrm{Cu}/\mathrm{Fe}]$, with our models systematically predicting ratios above the observed value. Cu production is metallicity dependent, and our progenitors have solar metallicity, whereas HD~84937 is very metal-poor, so some disagreement is not unexpected. The low-metallicity PUSH calculations of \citet{push4} obtained Cu abundances consistent with the observations.  Finally, both model series produce a range of $[\mathrm{Mn}/\mathrm{Fe}]$ ratios, many of which lie within the observational constraints.

\subsection{Comparison of Alpha-Element Abundances}
\label{sec:alpha_elements}

We now examine whether single and binary-stripped progenitors produce systematically different alpha-to-iron abundance ratios, focusing on four representative alpha nuclei: $^{16}$O, $^{24}$Mg, $^{28}$Si, and $^{40}$Ca.
In metal-poor stars the observed surface abundances largely reflect enrichment of the interstellar medium by earlier core-collapse supernovae, particularly when $[\mathrm{Fe}/\mathrm{H}]\lesssim-1$. Metal-poor stars typically exhibit alpha enhancements of $[\alpha/\mathrm{Fe}]\simeq+0.3$--$+0.5$ dex, although there is considerable scatter between individual elements and stellar populations \citep{cayrel04,nomoto13}. At higher metallicities, the delayed contribution of Type~Ia supernovae becomes increasingly important and lowers the characteristic alpha-to-iron ratio, as they produce a considerable amount of Fe-group elements. For an isotope $^{A}\mathrm{X}$, we calculate its abundance relative to Fe and normalise this ratio to the corresponding solar value. We adopt the solar elemental abundances of \citet{asplund2009} and the isotopic fractions of \citet{lodders2009}.

\begin{figure}
    \centering\includegraphics[width=0.95\columnwidth]{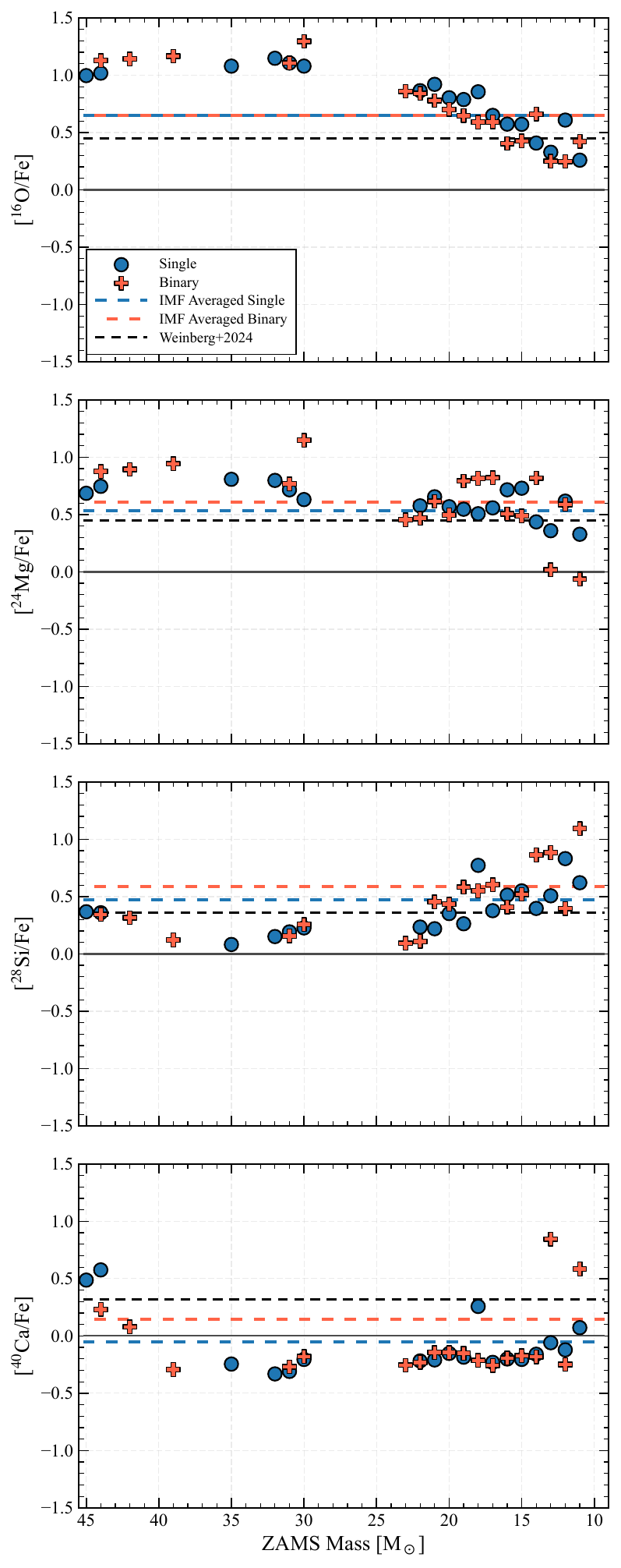}
\caption{
Abundance ratios of $^{16}$O, $^{24}$Mg, $^{28}$Si, and $^{40}$Ca (top to bottom) to Fe, relative to their solar ratios, as functions of ZAMS mass for the single-star (blue circles) and binary-stripped (red pluses) models. Coloured horizontal lines show the corresponding Salpeter IMF-averaged isotopic abundance ratios.
Black dashed lines mark the adopted low-metallicity elemental abundance plateaus for O, Mg, and Si from \citet{Weinberg2024}, and for Ca derived here using the same method.
}
\label{fig:solar_abundance}
\end{figure}

Figure~\ref{fig:solar_abundance} shows the resulting abundance ratios. The clearest trend occurs for $[^{16}\mathrm{O}/\mathrm{Fe}]$, which generally increases with ZAMS mass, rising from $\sim0.2$--$0.4$ dex at $11\,\mass$ to $\sim1.0$--$1.3$ dex for the higher-mass models. Most exploding models between $11$ and $22\,\mass$ have ratios of $\sim0.2$--$0.8$ dex. This trend reflects the hydrostatic origin of most ejected $^{16}$O: as the ZAMS mass and CO-core mass increase, the oxygen-rich layers generally grow. In contrast, the final Fe yield is dominated by the decay of explosively synthesised $^{56}$Ni and depends strongly on the explosion properties and mass cut. As shown in Figure~\ref{fig:56ni_obs_zams}, the $^{56}$Ni yield initially increases with ZAMS mass before broadly levelling off around $15\,\mass$, although substantial variation remains at the highest masses. The continued growth of the oxygen-rich layers without a corresponding increase in Fe therefore explains the general rise in $[^{16}\mathrm{O}/\mathrm{Fe}]$. The single-star and binary-stripped models exhibit broadly similar behaviour, typically differing by less than $0.2$ dex at the same ZAMS mass. This similarity is consistent with most oxygen residing in the deeper CO-rich layers rather than the outer envelope removed by binary stripping.

The $[^{24}\mathrm{Mg}/\mathrm{Fe}]$ ratio also broadly increases with ZAMS mass, although the trend is less regular than for $[^{16}\mathrm{O}/\mathrm{Fe}]$. Most models between $11$ and $23\,\mass$ have ratios of $\sim0.3$--$0.8$ dex, while the higher-mass models generally reach $\sim0.6$--$1.1$ dex. This increase is consistent with greater ejection of Mg-rich material from carbon and neon burning relative to Fe, while variations in the burning-shell structure and subsequent explosive processing can account for the less regular mass dependence. The binary-stripped models show greater variation, including ratios near or below solar at $11$ and $13\,\mass$. Unlike O, Mg therefore shows increased differences between the two progenitor populations at some masses, with no consistent ordering across the full mass range.

The $[^{28}\mathrm{Si}/\mathrm{Fe}]$ ratio generally decreases with ZAMS mass, although neither progenitor series follows a smooth trend. Ratios reach $\sim1.0$ dex in the lowest-mass models, fall to $\sim0.1$--$0.3$ dex at $22$--$35\,\mass$, and rise slightly to $\sim0.3$--$0.4$ dex in the highest-mass explosions. This behaviour reflects the competing contributions to the final $^{28}$Si yield. Some is inherited from the progenitor, while additional $^{28}$Si is produced through explosive oxygen burning. Material exposed to higher temperatures can instead undergo further processing into Fe-group nuclei. Changes in shock heating and the mass cut therefore affect both the surviving $^{28}$Si and the Fe yield, producing a less straightforward dependence on ZAMS mass than for $[^{16}\mathrm{O}/\mathrm{Fe}]$.

The $[^{40}\mathrm{Ca}/\mathrm{Fe}]$ ratio shows the greatest scatter with ZAMS mass. Below $\sim15\,\mass$, both progenitor series contain positive and negative ratios, while most intermediate-mass explosions cluster around $-0.2$ dex, with several of the highest-mass models producing positive values. The production of $^{40}$Ca through incomplete silicon burning is sensitive to the amount of ejecta reaching the required temperatures \citep[see table~3 of][]{push3}. Stronger shock heating can increase the Ca yield while also processing more material through complete silicon burning into Fe-group nuclei. If the Fe yield increases more rapidly, $[^{40}\mathrm{Ca}/\mathrm{Fe}]$ decreases even when the absolute Ca yield rises. The resulting ratios therefore depend on both the thermal history of the ejecta and the final mass cut.

We now consider the IMF-averaged abundance ratios of the two progenitor populations and compare them with the observed low-metallicity [$\alpha$/Fe] plateau, which we assume represents pure CCSN enrichment prior to the onset of Type Ia supernovae.
We use O, Mg, and Si plateaus of 0.45, 0.45, and 0.36 dex, respectively, as adopted in \citet{Weinberg2024} based on SDSS DR17 APOGEE data \citep{majewski2017} and other observational studies, and extend their methods to determine a corresponding plateau value for Ca of 0.32 dex. These values can be compared to the IMF-averaged abundance ratios for our single-star and binary-stripped yield sets, weighted by a Salpeter IMF.

We plot both the IMF-averaged abundance ratios and the adopted low-metallicity plateaus as horizontal lines in Figure~\ref{fig:solar_abundance}. Both populations exceed the O plateau by $\sim0.20$ dex. For Mg, the single-star and binary-stripped populations lie above the plateau by $\sim0.08$ and $\sim0.16$ dex, respectively, while the corresponding Si excesses are $\sim0.11$ and $\sim0.23$ dex. In contrast, Ca remains below the adopted plateau by $\sim0.37$ dex for the single-star population and $\sim0.17$ dex for the binary-stripped population. As discussed by \citet{Weinberg2024}, the adopted O and Mg plateaus have substantial systematic uncertainty, with plausible values spanning $0.30$--$0.60$ dex. The Mg predictions lie within or close to this range, while both populations remain above it for O. Binary stripping brings the predicted Ca ratio closer to the adopted plateau, although a deficit remains.

\subsection{Comparison with Empirical [X/Mg] Abundances}
\label{sec:empirical_xmg}

To further compare our single-star and binary-stripped IMF-averaged yields with observations, we follow the methodology of \citet{griffith2021} and measure the offset between the theoretical CCSN abundance ratios and empirical measurements of the CCSN component of observed stellar abundances. The empirical values are taken from \citet{griffith2024}, who fit a two-process model to SDSS DR17 APOGEE abundances \citep{majewski2017}. The model describes each abundance of each star as the sum of a CCSN and SN~Ia process, allowing the CCSN component of the observed abundance space to be separated from the more delayed SN~Ia enrichment. Following \citet[][see Appendix~B]{griffith2021}, the two-process model parameters can be translated into an empirical CCSN abundance ratio and compared to the theoretical yields. We define $\Delta[\rm X/Mg]_{\rm cc}$ as the difference between the theoretical and empirical yields, relative to Mg. We use Mg as the reference element as it is produced almost entirely by CCSNe and is well measured observationally. 

\begin{figure}
    \centering
    \includegraphics[width=1\linewidth]{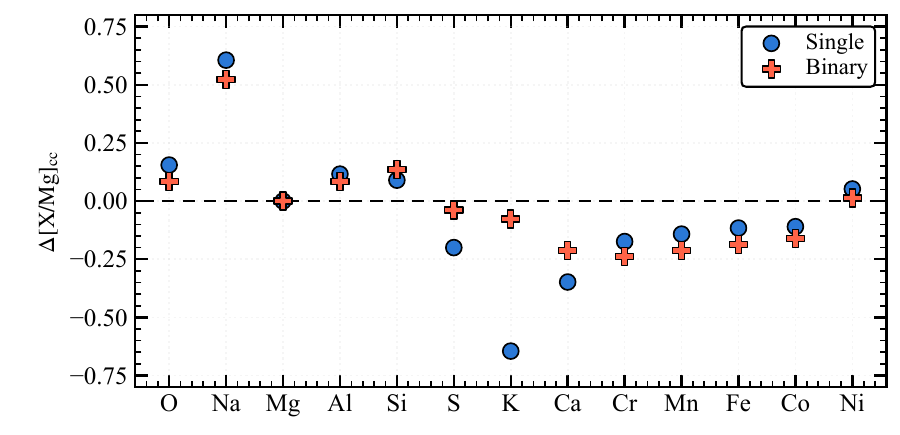}
    \caption{Offset between the theoretical CCSN abundance ratios from this work and empirical measurements of the CCSN contribution to [X/Mg] for the Ms models (blue circles) and Mb models (red pluses). Empirical constraints from \citet{griffith2024}. The dashed line at 0.0 represents agreement between the theoretical and empirical values.}
    \label{fig:delta_xmg}
\end{figure}

The $\Delta[\rm X/Mg]_{\rm cc}$ for the single-star and binary-stripped yield sets are shown in Figure~\ref{fig:delta_xmg}. If $\Delta[\rm X/Mg]_{\rm cc}=0$, then the model reproduces the empirical APOGEE CCSN yields. Positive/negative values indicate that the models over/underproduce a given element relative to Mg. We find that most elements have a $\Delta[\rm X/Mg]_{\rm cc}$ between $-0.25$ and 0.15 dex, where O, Na, Al, and Si are overproduced relative to Mg while S, K, Ca, Cr, Mn, Fe, and Co are underproduced in both the single and binary models. 
For all elements except S, K, and Ca, the $\Delta[\rm X/Mg]_{\rm cc}$ values for the single-star and binary-stripped models are within 0.1 dex of one another.

The apparent overproduction of elements O, Al, and Si by $\sim$0.15 dex may be a manifestation of the underproduction of Mg in the models, as seen in many CCSN yield sets \citep[e.g.,][]{rybizki2017}. The overproduction of Na, however, is much more drastic, with a $\Delta[\rm X/Mg]_{\rm cc}$ of 0.5 dex for the binary-stripped models and 0.6 dex for single-star models. \citet{griffith2021} also find a drastic overproduction of Na when comparing yields from \citet{Sukhbold2016} with observational constraints. Conversely, K is significantly underproduced relative to Mg---but only in the single-star models. S and Ca also show more significant underproduction in the single-star models, but to a lesser degree.

\section{Discussion and Conclusions}
\label{sec:Discussion_and_Conclusions}

In this work, we have used the PUSH framework to investigate the explodability, remnant properties, and nucleosynthetic yields of the single-star and binary-stripped progenitors of \citet{Farmer2023}. We considered 35 single-star and 31 binary-stripped models spanning ZAMS masses of $11$--$45\,\mass$, using the standard PUSH calibration \citep{push2} and the \texttt{CFNET} reaction network \citep{frohlich2006b} for nucleosynthesis post-processing. We compared our results with previous PUSH calculations \citep{push2,push3} and those of \citet{Farmer2023} and \citet{Boccioli2026}. Our main conclusions are as follows:

\begin{enumerate}

\item We find that binary stripping alters the explodability landscape at higher ZAMS masses.
Overall, 18 of the 35 single-star progenitors and 18 of the 31 binary-stripped progenitors explode successfully, corresponding to success rates of 51.4 and 58.1 per cent, respectively.
The two series produce identical outcomes up to and including $22\,\mass$, with all models successfully exploding, while their outcomes diverge at higher masses. 
Binary stripping can therefore alter the pre-collapse structure sufficiently to change the explosion outcome at a given ZAMS mass, enabling explosions in some models and preventing them in others. However, within our adopted parametrisation, the slightly higher overall success rate of the binary-stripped models does not indicate a difference in explosion outcomes at the lowest ZAMS masses, where both series explode successfully.

\item The predicted explodability landscape depends strongly on the adopted explosion treatment. Our PUSH calculations and those of \citet{Farmer2023} and \citet{Boccioli2026} agree on successful explosions for only 5 of the 66 models, with $11\,\mass$ the only ZAMS mass at which all predict successful explosions for both the single-star and binary-stripped progenitors. Where the same models explode, their predicted remnant masses are broadly comparable. We find that our predicted outcomes are also sensitive to the adopted calibration, as changing the PUSH factor can be sufficient to reverse whether a model explodes. We therefore stress that our explodability landscape depends on the assumption that compactness is the appropriate quantity with which to parameterise the additional PUSH heating.

\item Successfully exploding single-star and binary-stripped models occupy broadly similar regions of explosion parameter space, with energies of $\sim0.4$--$1.3\,\mathrm{B}$, baryonic remnant masses of $\sim1.4$--$2.0\,\mass$, and explosion times of $\sim0.30$--$0.65\,\mathrm{s}$. The differences introduced by binary stripping are comparable to the scatter among the progenitor sets used in previous PUSH calculations \citep{push2}, which were evolved using nuclear reaction networks of different sizes. We find that explosion properties show clearer trends with compactness than with ZAMS mass: more compact progenitors generally produce more massive remnants and explode later, while explosion energies peak at intermediate compactness and decrease towards both lower and higher values.

\item After nucleosynthesis post-processing of the successful explosions, we find that the largest systematic differences between the single-star and binary-stripped yields occur for the lightest elements, with single stars ejecting more material from their retained envelopes. Differences remain when the yields are normalised by the total pre-explosion elemental yields or the Fe yield. At ZAMS masses of $\sim11$--$14\,\mass$, binary stripping substantially alters the yields of elements from S to Sc, which probe material processed through explosive O and Si burning. 
While some Fe-group elements exhibit scatter in their yields, this can reflect differences in the structure close to the mass cut. At higher ZAMS masses, the yields of the two progenitor series are generally more similar.

\item Using a Salpeter IMF weighting, we find that the single-star and binary-stripped populations produce broadly similar elemental yield patterns, with most elements differing by only $\sim20$ per cent. The binary-stripped population shows moderate enhancements in Cl, Ar, and Sc, while N, F, Cu, and several Fe-group elements are suppressed. The clearest difference occurs for K, which is enhanced by a factor of $\sim3.4$, corresponding to an increase of $\sim0.63$ dex in [K/Fe]. This enhancement reflects increased $^{39}$K production in the lowest-mass progenitors, whose contributions receive greater weight under the Salpeter IMF, consistent with \citet{Farmer2023}.

\item Our predicted $^{56}$Ni yields span much of the observed CCSN range below $\sim23\,\mass$, and the iron-group abundance patterns broadly agree with those of HD~84937. The clearest discrepancy occurs for Cu, with predicted abundances above the observed value. However, Cu production is metallicity dependent, and our progenitors have solar metallicity whereas HD~84937 is very metal-poor, so some disagreement is not unexpected.
The $[\alpha/\mathrm{Fe}]$ ratios show the expected behaviour with ZAMS mass, with $[^{16}\mathrm{O}/\mathrm{Fe}]$ increasing most clearly and the Si and Ca ratios showing greater sensitivity to the explosive processing.
Our IMF-averaged $[\alpha/\mathrm{Fe}]$ ratios show broadly similar agreement with the observational constraints \citep{Weinberg2024} for both populations, with the clearest difference occurring for Ca. For $[^{40}\mathrm{Ca}/\mathrm{Fe}]$, the single-star population lies substantially below the plateau by $0.37$ dex, whereas the binary-stripped population shows a smaller deficit of $0.17$ dex.

\item We compared the IMF-averaged $[\mathrm{X}/\mathrm{Mg}]$ ratios of our single-star and binary-stripped progenitors with empirical CCSN abundance ratios inferred by \citet{griffith2024} using a two-process model of SDSS DR17 APOGEE stellar abundances. We found that most elements show offsets from the empirical values between $-0.25$ and $0.15$ dex. The largest discrepancy occurs for Na, with excesses of $0.5$ dex for the binary-stripped population and $0.6$ dex for the single-star population, while the smaller excesses in O, Al, and Si may reflect an underproduction of Mg. The two populations differ by less than $0.1$ dex for most elements, with the clearest differences occurring for S, K, and Ca. 
Although both populations show similar abundance patterns overall, the closer agreement of the binary-stripped models with the empirical S, K, and Ca ratios highlights the importance of accounting for binarity when predicting the CCSN contribution to galactic chemical enrichment.

\end{enumerate}

\section*{Acknowledgements}

APJ is supported by NSF AST-2510795 and acknowledges the Alfred P. Sloan Research Fellowship and the University of Chicago's Research Computing Center for their support.
The work at NC State was supported by the United States Department of Energy, Office of Science, Office of Nuclear Physics (Award No. DE-FG02-02ER41216).
E.J.G. acknowledges support for this work provided by NASA through the NASA Hubble Fellowship Program grant No. HST-HF2-51576.001-A awarded by the Space Telescope Science Institute, which is operated by the Association of Universities for Research in Astronomy, Inc., for NASA, under the contract NAS 5-26555. 
JMP thanks those who attended the 2026 Rise\_Time workshop and Aysha Aamer for useful discussions on core-collapse supernovae.
The authors acknowledge the use of generative AI tools for editorial assistance, including improving clarity, concision, and organisation. All suggested edits were reviewed by the authors, who developed and verified all scientific content, interpretations, and conclusions.
We acknowledge the use of NumPy \citep{harris2020array}, SciPy \citep{2020SciPy-NMeth} and Matplotlib \citep{Hunter:2007}.

\section*{Data Availability}

The explosion properties and nucleosynthesis data are available from the corresponding author upon reasonable request. They will also be made publicly available on Zenodo and via the Heidelberg Supernova Model Archive (HESMA; \citealt{Kromer2017}; https://hesma.hits.org).

\bibliographystyle{mnras}
\bibliography{bibliography} 
\appendix

\section{Compactness as a Function of ZAMS Mass}
\label{apen:compactness}

\begin{figure}
\centering
\includegraphics[width=0.89\columnwidth]{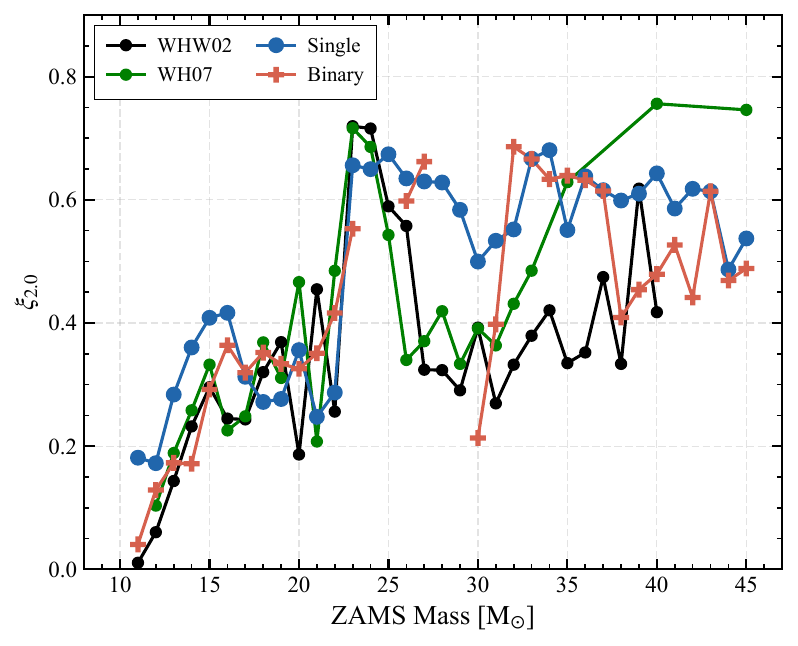} 
\caption{Similar to Figure~\protect\ref{fig:Method_compactness}, but for the compactness $\xi_{2.0}$.}
\label{apen:compactness_2.0_fig}
\end{figure}

Figure~\ref{apen:compactness_2.0_fig} shows the compactness evaluated at $2.0\,\mass$ at bounce, $\xi_{2.0}$, which is used to determine the additional PUSH heating through $k_{\rm push}$, as a function of ZAMS mass \citep[see][]{push2}.

\onecolumn

\section{Isotopic Yields of the Ms18 and Mb18 Models}
\label{MsandMbyields}

Table~\ref{tab:ms18_mb18_isotopic_yields} presents a sample of isotopic yields of the Ms18 and Mb18 models. These models are used throughout
Section~\ref{sec:Nucleosynthesis} to illustrate the differences in
nucleosynthetic yields between the single-star and binary-stripped progenitors.
The table includes isotopic yields of the first 10 isotopes. For the remaining
isotopic yields for these two models and the complete yield tables for all other
successfully exploding models, see the Data Availability section.
\input{Ms18_Mb18_pre_explosive_final_yields}

\bsp	
\label{lastpage}
\end{document}

%% file: Ms18_Mb18_pre_explosive_final_yields.tex
\begingroup
\setlength\LTleft{\fill}
\setlength\LTright{\fill}
\begin{longtable}{|l|ccc|ccc|}
\caption{Pre-explosion, net change, and final isotopic yields, in units of \mass, for the representative Ms18.00 and Mb18.00 models.
The net change is defined as the final isotopic mass minus the corresponding pre-supernova isotopic mass. Negative and positive values therefore indicate a decrease or increase, respectively, in the total mass of a given isotope over the course of collapse and explosion. Note that these net changes can also reflect the reprocessing of material during collapse and explosion, such that substantial destruction and subsequent re-production of an isotope may result in only a small net change in its final abundance.}
\label{tab:ms18_mb18_isotopic_yields}\\
\hline
 & \multicolumn{3}{c|}{Ms18} & \multicolumn{3}{c|}{Mb18} \\ \cline{2-7}
Isotope & Pre-explosion & Net change & Final & Pre-explosion & Net change & Final \\ \hline
\endfirsthead

\multicolumn{7}{c}{\tablename\ \thetable{} -- continued from previous page}\\
\hline
 & \multicolumn{3}{c|}{Ms18} & \multicolumn{3}{c|}{Mb18} \\ \cline{2-7}
Isotope & Pre-explosion & Explosive & Final & Pre-explosion & Explosive & Final \\ \hline
\endhead

\hline
\multicolumn{7}{r}{Continued on next page}\\
\endfoot

\hline
\endlastfoot

$^{1}\mathrm{H}$ & 1.76E+00 & -4.71E-04 & 1.76E+00 & 7.28E-04 & -5.52E-04 & 1.76E-04 \\
$^{2}\mathrm{H}$ & 9.64E-09 & -9.64E-09 & 4.58E-17 & 1.57E-08 & -1.57E-08 & 7.09E-20 \\
$^{3}\mathrm{He}$ & 5.16E-05 & -2.98E-11 & 5.16E-05 & 5.22E-11 & -5.08E-11 & 1.37E-12 \\
$^{4}\mathrm{He}$ & 2.67E+00 & -2.11E-03 & 2.67E+00 & 1.28E+00 & -3.66E-03 & 1.28E+00 \\
$^{6}\mathrm{Li}$ &  & 1.36E-19 & 1.36E-19 &  & 5.52E-20 & 5.52E-20 \\
$^{7}\mathrm{Li}$ & 2.41E-12 & 1.19E-11 & 1.43E-11 & 3.72E-12 & 3.32E-11 & 3.69E-11 \\
$^{9}\mathrm{Be}$ & 4.03E-11 & -1.87E-12 & 3.84E-11 & 3.20E-12 & -3.20E-12 & 1.66E-19 \\
$^{10}\mathrm{B}$ & 1.99E-10 & -1.42E-13 & 1.99E-10 & 3.44E-13 & -3.43E-13 & 1.65E-16 \\
$^{11}\mathrm{B}$ & 8.92E-10 & 6.01E-11 & 9.52E-10 & 1.59E-12 & 2.13E-11 & 2.29E-11 \\
$^{12}\mathrm{C}$ & 3.06E-01 & 7.05E-05 & 3.07E-01 & 2.46E-01 & -6.36E-04 & 2.45E-01 \\
\end{longtable}
\endgroup

%% file: main.bbl
\begin{thebibliography}{}
\makeatletter
\relax
\def\mn@urlcharsother{\let\do\@makeother \do\$\do\&\do\#\do\^\do\_\do\%\do\~}
\def\mn@doi{\begingroup\mn@urlcharsother \@ifnextchar [ {\mn@doi@}
  {\mn@doi@[]}}
\def\mn@doi@[#1]#2{\def\@tempa{#1}\ifx\@tempa\@empty \href
  {http://dx.doi.org/#2} {doi:#2}\else \href {http://dx.doi.org/#2} {#1}\fi
  \endgroup}
\def\mn@eprint#1#2{\mn@eprint@#1:#2::\@nil}
\def\mn@eprint@arXiv#1{\href {http://arxiv.org/abs/#1} {{\tt arXiv:#1}}}
\def\mn@eprint@dblp#1{\href {http://dblp.uni-trier.de/rec/bibtex/#1.xml}
  {dblp:#1}}
\def\mn@eprint@#1:#2:#3:#4\@nil{\def\@tempa {#1}\def\@tempb {#2}\def\@tempc
  {#3}\ifx \@tempc \@empty \let \@tempc \@tempb \let \@tempb \@tempa \fi \ifx
  \@tempb \@empty \def\@tempb {arXiv}\fi \@ifundefined
  {mn@eprint@\@tempb}{\@tempb:\@tempc}{\expandafter \expandafter \csname
  mn@eprint@\@tempb\endcsname \expandafter{\@tempc}}}

\bibitem[\protect\citeauthoryear{{Abohalima} \& {Frebel}}{{Abohalima} \&
  {Frebel}}{2018}]{Abohalima2018}
{Abohalima} A.,  {Frebel} A.,  2018, \mn@doi [\apjs]
  {10.3847/1538-4365/aadfe9}, \href
  {https://ui.adsabs.harvard.edu/abs/2018ApJS..238...36A} {238, 36}

\bibitem[\protect\citeauthoryear{{Abt}}{{Abt}}{1983}]{Abt1983}
{Abt} H.~A.,  1983, \mn@doi [\araa] {10.1146/annurev.aa.21.090183.002015},
  \href {https://ui.adsabs.harvard.edu/abs/1983ARA&A..21..343A} {21, 343}

\bibitem[\protect\citeauthoryear{{Aloy} \& {Obergaulinger}}{{Aloy} \&
  {Obergaulinger}}{2021}]{Aloy2021}
{Aloy} M.~{\'A}.,  {Obergaulinger} M.,  2021, \mn@doi [\mnras]
  {10.1093/mnras/staa3273}, \href
  {https://ui.adsabs.harvard.edu/abs/2021MNRAS.500.4365A} {500, 4365}

\bibitem[\protect\citeauthoryear{{Andrews}, {Fryer}, {Even}, {Jones}  \&
  {Pignatari}}{{Andrews} et~al.}{2020}]{Andrews2020}
{Andrews} S.,  {Fryer} C.,  {Even} W.,  {Jones} S.,   {Pignatari} M.,  2020,
  \mn@doi [\apj] {10.3847/1538-4357/ab64f8}, \href
  {https://ui.adsabs.harvard.edu/abs/2020ApJ...890...35A} {890, 35}

\bibitem[\protect\citeauthoryear{{Andrievsky}, {Bonifacio}, {Caffau},
  {Korotin}, {Spite}, {Spite}, {Sbordone}  \& {Zhukova}}{{Andrievsky}
  et~al.}{2018}]{Andrievsky2018}
{Andrievsky} S.,  {Bonifacio} P.,  {Caffau} E.,  {Korotin} S.,  {Spite} M.,
  {Spite} F.,  {Sbordone} L.,   {Zhukova} A.~V.,  2018, \mn@doi [\mnras]
  {10.1093/mnras/stx2526}, \href
  {https://ui.adsabs.harvard.edu/abs/2018MNRAS.473.3377A} {473, 3377}

\bibitem[\protect\citeauthoryear{{Aoki} et~al.,}{{Aoki}
  et~al.}{2022}]{Aoki2022}
{Aoki} W.,  et~al., 2022, \mn@doi [\apj] {10.3847/1538-4357/ac6515}, \href
  {https://ui.adsabs.harvard.edu/abs/2022ApJ...931..146A} {931, 146}

\bibitem[\protect\citeauthoryear{{Arentsen}, {Aguado}, {Sestito}, {Gonz{\'a}lez
  Hern{\'a}ndez}, {Martin}, {Starkenburg}, {Jablonka}  \& {Yuan}}{{Arentsen}
  et~al.}{2023}]{Arentsen2023}
{Arentsen} A.,  {Aguado} D.~S.,  {Sestito} F.,  {Gonz{\'a}lez Hern{\'a}ndez}
  J.~I.,  {Martin} N.~F.,  {Starkenburg} E.,  {Jablonka} P.,   {Yuan} Z.,
  2023, \mn@doi [\mnras] {10.1093/mnras/stad043}, \href
  {https://ui.adsabs.harvard.edu/abs/2023MNRAS.519.5554A} {519, 5554}

\bibitem[\protect\citeauthoryear{{Asplund}, {Grevesse}, {Sauval}  \&
  {Scott}}{{Asplund} et~al.}{2009}]{asplund2009}
{Asplund} M.,  {Grevesse} N.,  {Sauval} A.~J.,   {Scott} P.,  2009, \mn@doi
  [\araa] {10.1146/annurev.astro.46.060407.145222}, \href
  {https://ui.adsabs.harvard.edu/abs/2009ARA&A..47..481A} {47, 481}

\bibitem[\protect\citeauthoryear{{Aufderheide}, {Baron}  \&
  {Thielemann}}{{Aufderheide} et~al.}{1991}]{Aufderheide1991}
{Aufderheide} M.~B.,  {Baron} E.,   {Thielemann} F.-K.,  1991, \mn@doi [\apj]
  {10.1086/169849}, \href
  {https://ui.adsabs.harvard.edu/abs/1991ApJ...370..630A} {370, 630}

\bibitem[\protect\citeauthoryear{{Blondin}, {Mezzacappa}  \&
  {DeMarino}}{{Blondin} et~al.}{2003}]{Blondin2003}
{Blondin} J.~M.,  {Mezzacappa} A.,   {DeMarino} C.,  2003, \mn@doi [\apj]
  {10.1086/345812}, \href
  {https://ui.adsabs.harvard.edu/abs/2003ApJ...584..971B} {584, 971}

\bibitem[\protect\citeauthoryear{{Boccioli} \& {Roberti}}{{Boccioli} \&
  {Roberti}}{2026}]{Boccioli2026}
{Boccioli} L.,  {Roberti} L.,  2026, \mn@doi [\aap]
  {10.1051/0004-6361/202557714}, \href
  {https://ui.adsabs.harvard.edu/abs/2026A&A...709A.201B} {709, A201}

\bibitem[\protect\citeauthoryear{{Boccioli}, {Mathews}  \&
  {O'Connor}}{{Boccioli} et~al.}{2021}]{Boccioli2021}
{Boccioli} L.,  {Mathews} G.~J.,   {O'Connor} E.~P.,  2021, \mn@doi [\apj]
  {10.3847/1538-4357/abe767}, \href
  {https://ui.adsabs.harvard.edu/abs/2021ApJ...912...29B} {912, 29}

\bibitem[\protect\citeauthoryear{{Braun} \& {Langer}}{{Braun} \&
  {Langer}}{1995a}]{Braun1995b}
{Braun} H.,  {Langer} N.,  1995a, in {van der Hucht} K.~A.,  {Williams} P.~M.,
  eds,  IAU Symposium Vol. 163, Wolf-Rayet Stars: Binaries; Colliding Winds;
  Evolution. p.~305

\bibitem[\protect\citeauthoryear{{Braun} \& {Langer}}{{Braun} \&
  {Langer}}{1995b}]{Braun1995a}
{Braun} H.,  {Langer} N.,  1995b, \aap, \href
  {https://ui.adsabs.harvard.edu/abs/1995A&A...297..483B} {297, 483}

\bibitem[\protect\citeauthoryear{{Brown}, {Weingartner}  \& {Wijers}}{{Brown}
  et~al.}{1996}]{Brown1996}
{Brown} G.~E.,  {Weingartner} J.~C.,   {Wijers} R. A.~M.~J.,  1996, \mn@doi
  [\apj] {10.1086/177241}, \href
  {https://ui.adsabs.harvard.edu/abs/1996ApJ...463..297B} {463, 297}

\bibitem[\protect\citeauthoryear{{Bruenn} et~al.,}{{Bruenn}
  et~al.}{2016}]{Bruenn2016}
{Bruenn} S.~W.,  et~al., 2016, \mn@doi [\apj] {10.3847/0004-637X/818/2/123},
  \href {https://ui.adsabs.harvard.edu/abs/2016ApJ...818..123B} {818, 123}

\bibitem[\protect\citeauthoryear{{Burrows}}{{Burrows}}{2013}]{Burrows2013}
{Burrows} A.,  2013, \mn@doi [Reviews of Modern Physics]
  {10.1103/RevModPhys.85.245}, \href
  {https://ui.adsabs.harvard.edu/abs/2013RvMP...85..245B} {85, 245}

\bibitem[\protect\citeauthoryear{{Burrows}, {Vartanyan}, {Dolence}, {Skinner}
  \& {Radice}}{{Burrows} et~al.}{2018}]{Burrows2018}
{Burrows} A.,  {Vartanyan} D.,  {Dolence} J.~C.,  {Skinner} M.~A.,   {Radice}
  D.,  2018, \mn@doi [\ssr] {10.1007/s11214-017-0450-9}, \href
  {https://ui.adsabs.harvard.edu/abs/2018SSRv..214...33B} {214, 33}

\bibitem[\protect\citeauthoryear{{Burrows}, {Radice}, {Vartanyan}, {Nagakura},
  {Skinner}  \& {Dolence}}{{Burrows} et~al.}{2020}]{Burrows2020}
{Burrows} A.,  {Radice} D.,  {Vartanyan} D.,  {Nagakura} H.,  {Skinner} M.~A.,
   {Dolence} J.~C.,  2020, \mn@doi [\mnras] {10.1093/mnras/stz3223}, \href
  {https://ui.adsabs.harvard.edu/abs/2020MNRAS.491.2715B} {491, 2715}

\bibitem[\protect\citeauthoryear{{Burrows}, {Wang}  \& {Vartanyan}}{{Burrows}
  et~al.}{2024}]{Burrows2024}
{Burrows} A.,  {Wang} T.,   {Vartanyan} D.,  2024, \mn@doi [\apjl]
  {10.3847/2041-8213/ad319e}, \href
  {https://ui.adsabs.harvard.edu/abs/2024ApJ...964L..16B} {964, L16}

\bibitem[\protect\citeauthoryear{{Burrows}, {Wang}  \& {Vartanyan}}{{Burrows}
  et~al.}{2026}]{Burrows2026}
{Burrows} A.,  {Wang} T.,   {Vartanyan} D.,  2026, \mn@doi [arXiv e-prints]
  {10.48550/arXiv.2607.06664}, \href
  {https://ui.adsabs.harvard.edu/abs/2026arXiv260706664B} {p. arXiv:2607.06664}

\bibitem[\protect\citeauthoryear{{Caffau} et~al.,}{{Caffau}
  et~al.}{2013}]{Caffau2013}
{Caffau} E.,  et~al., 2013, \mn@doi [\aap] {10.1051/0004-6361/201322488}, \href
  {https://ui.adsabs.harvard.edu/abs/2013A&A...560A..71C} {560, A71}

\bibitem[\protect\citeauthoryear{{Calvert}, {Redle}, {Gautam}, {Stapleford},
  {Fr{\"o}hlich}, {Kneller}  \& {Liebendorfer}}{{Calvert}
  et~al.}{2025}]{Calvert2025}
{Calvert} D.,  {Redle} M.,  {Gautam} B.,  {Stapleford} C.~J.,  {Fr{\"o}hlich}
  C.,  {Kneller} J.~P.,   {Liebendorfer} M.,  2025, \mn@doi [\apj]
  {10.3847/1538-4357/ae1475}, \href
  {https://ui.adsabs.harvard.edu/abs/2025ApJ...995..109C} {995, 109}

\bibitem[\protect\citeauthoryear{{Calvert}, {Redle}, {Gautam}, {Stapleford},
  {Frohlich}, {Kneller}  \& {Liebendoerfer}}{{Calvert}
  et~al.}{2026}]{Calvert2026}
{Calvert} D.,  {Redle} M.,  {Gautam} B.,  {Stapleford} C.~J.,  {Frohlich} C.,
  {Kneller} J.~P.,   {Liebendoerfer} M.,  2026, \mn@doi [arXiv e-prints]
  {10.48550/arXiv.2608.24578}, \href
  {https://ui.adsabs.harvard.edu/abs/2026arXiv260824578C} {p. arXiv:2608.24578}

\bibitem[\protect\citeauthoryear{{Cayrel} et~al.,}{{Cayrel}
  et~al.}{2004}]{cayrel04}
{Cayrel} R.,  et~al., 2004, \mn@doi [\aap] {10.1051/0004-6361:20034074}, \href
  {http://adsabs.harvard.edu/abs/2004A%26A...416.1117C} {416, 1117}

\bibitem[\protect\citeauthoryear{{Couch}, {Warren}  \& {O'Connor}}{{Couch}
  et~al.}{2020}]{Couch2020}
{Couch} S.~M.,  {Warren} M.~L.,   {O'Connor} E.~P.,  2020, \mn@doi [\apj]
  {10.3847/1538-4357/ab609e}, \href
  {https://ui.adsabs.harvard.edu/abs/2020ApJ...890..127C} {890, 127}

\bibitem[\protect\citeauthoryear{{Curtis}, {Ebinger}, {Fr{\"o}hlich}, {Hempel},
  {Perego}, {Liebend{\"o}rfer}  \& {Thielemann}}{{Curtis} et~al.}{2019}]{push3}
{Curtis} S.,  {Ebinger} K.,  {Fr{\"o}hlich} C.,  {Hempel} M.,  {Perego} A.,
  {Liebend{\"o}rfer} M.,   {Thielemann} F.-K.,  2019, \mn@doi [\apj]
  {10.3847/1538-4357/aae7d2}, \href
  {http://adsabs.harvard.edu/abs/2019ApJ...870....2C} {870, 2}

\bibitem[\protect\citeauthoryear{{Curtis}, {Wolfe}, {Fr{\"o}hlich}, {Miller},
  {Wollaeger}  \& {Ebinger}}{{Curtis} et~al.}{2021}]{Curtis2021}
{Curtis} S.,  {Wolfe} N.,  {Fr{\"o}hlich} C.,  {Miller} J.~M.,  {Wollaeger} R.,
    {Ebinger} K.,  2021, \mn@doi [\apj] {10.3847/1538-4357/ac0dc5}, \href
  {https://ui.adsabs.harvard.edu/abs/2021ApJ...921..143C} {921, 143}

\bibitem[\protect\citeauthoryear{{Cyburt} et~al.,}{{Cyburt}
  et~al.}{2010}]{reaclib}
{Cyburt} R.~H.,  et~al., 2010, \mn@doi [\apjs] {10.1088/0067-0049/189/1/240},
  \href {http://adsabs.harvard.edu/abs/2010ApJS..189..240C} {189, 240}

\bibitem[\protect\citeauthoryear{{Da Costa} et~al.,}{{Da Costa}
  et~al.}{2019}]{DaCosta2019}
{Da Costa} G.~S.,  et~al., 2019, \mn@doi [\mnras] {10.1093/mnras/stz2550},
  \href {https://ui.adsabs.harvard.edu/abs/2019MNRAS.489.5900D} {489, 5900}

\bibitem[\protect\citeauthoryear{{Dessart}}{{Dessart}}{2025}]{Dessart2025}
{Dessart} L.,  2025, \mn@doi [\aap] {10.1051/0004-6361/202555664}, \href
  {https://ui.adsabs.harvard.edu/abs/2025A&A...703A.287D} {703, A287}

\bibitem[\protect\citeauthoryear{{Ebinger}, {Curtis}, {Fr{\"o}hlich}, {Hempel},
  {Perego}, {Liebend{\"o}rfer}  \& {Thielemann}}{{Ebinger}
  et~al.}{2019}]{push2}
{Ebinger} K.,  {Curtis} S.,  {Fr{\"o}hlich} C.,  {Hempel} M.,  {Perego} A.,
  {Liebend{\"o}rfer} M.,   {Thielemann} F.-K.,  2019, \mn@doi [\apj]
  {10.3847/1538-4357/aae7c9}, \href
  {http://adsabs.harvard.edu/abs/2019ApJ...870....1E} {870, 1}

\bibitem[\protect\citeauthoryear{{Ebinger}, {Curtis}, {Ghosh}, {Fr{\"o}hlich},
  {Hempel}, {Perego}, {Liebend{\"o}rfer}  \& {Thielemann}}{{Ebinger}
  et~al.}{2020}]{push4}
{Ebinger} K.,  {Curtis} S.,  {Ghosh} S.,  {Fr{\"o}hlich} C.,  {Hempel} M.,
  {Perego} A.,  {Liebend{\"o}rfer} M.,   {Thielemann} F.-K.,  2020, \mn@doi
  [\apj] {10.3847/1538-4357/ab5dcb}, \href
  {https://ui.adsabs.harvard.edu/abs/2020ApJ...888...91E} {888, 91}

\bibitem[\protect\citeauthoryear{{Eldridge}, {Izzard}  \& {Tout}}{{Eldridge}
  et~al.}{2008}]{Eldridge2008}
{Eldridge} J.~J.,  {Izzard} R.~G.,   {Tout} C.~A.,  2008, \mn@doi [\mnras]
  {10.1111/j.1365-2966.2007.12738.x}, \href
  {https://ui.adsabs.harvard.edu/abs/2008MNRAS.384.1109E} {384, 1109}

\bibitem[\protect\citeauthoryear{{Ertl}, {Janka}, {Woosley}, {Sukhbold}  \&
  {Ugliano}}{{Ertl} et~al.}{2016}]{Ertl2016}
{Ertl} T.,  {Janka} H.-T.,  {Woosley} S.~E.,  {Sukhbold} T.,   {Ugliano} M.,
  2016, \mn@doi [\apj] {10.3847/0004-637X/818/2/124}, \href
  {https://ui.adsabs.harvard.edu/abs/2016ApJ...818..124E} {818, 124}

\bibitem[\protect\citeauthoryear{{Farmer}, {Fields}, {Petermann}, {Dessart},
  {Cantiello}, {Paxton}  \& {Timmes}}{{Farmer} et~al.}{2016}]{Farmer2016}
{Farmer} R.,  {Fields} C.~E.,  {Petermann} I.,  {Dessart} L.,  {Cantiello} M.,
  {Paxton} B.,   {Timmes} F.~X.,  2016, \mn@doi [\apjs]
  {10.3847/1538-4365/227/2/22}, \href
  {https://ui.adsabs.harvard.edu/abs/2016ApJS..227...22F} {227, 22}

\bibitem[\protect\citeauthoryear{{Farmer}, {Laplace}, {de Mink}  \&
  {Justham}}{{Farmer} et~al.}{2021}]{Farmer2021}
{Farmer} R.,  {Laplace} E.,  {de Mink} S.~E.,   {Justham} S.,  2021, \mn@doi
  [\apj] {10.3847/1538-4357/ac2f44}, \href
  {https://ui.adsabs.harvard.edu/abs/2021ApJ...923..214F} {923, 214}

\bibitem[\protect\citeauthoryear{{Farmer}, {Laplace}, {Ma}, {de Mink}  \&
  {Justham}}{{Farmer} et~al.}{2023}]{Farmer2023}
{Farmer} R.,  {Laplace} E.,  {Ma} J.-z.,  {de Mink} S.~E.,   {Justham} S.,
  2023, \mn@doi [\apj] {10.3847/1538-4357/acc315}, \href
  {https://ui.adsabs.harvard.edu/abs/2023ApJ...948..111F} {948, 111}

\bibitem[\protect\citeauthoryear{{Fischer}, {Mart{\'\i}nez-Pinedo}, {Hempel}
  \& {Liebend{\"o}rfer}}{{Fischer} et~al.}{2012}]{Fischer2012}
{Fischer} T.,  {Mart{\'\i}nez-Pinedo} G.,  {Hempel} M.,   {Liebend{\"o}rfer}
  M.,  2012, \mn@doi [\prd] {10.1103/PhysRevD.85.083003}, \href
  {https://ui.adsabs.harvard.edu/abs/2012PhRvD..85h3003F} {85, 083003}

\bibitem[\protect\citeauthoryear{{Foglizzo}, {Galletti}, {Scheck}  \&
  {Janka}}{{Foglizzo} et~al.}{2007}]{Foglizzo2007}
{Foglizzo} T.,  {Galletti} P.,  {Scheck} L.,   {Janka} H.-T.,  2007, \mn@doi
  [\apj] {10.1086/509612}, \href
  {https://ui.adsabs.harvard.edu/abs/2007ApJ...654.1006F} {654, 1006}

\bibitem[\protect\citeauthoryear{{Fr{\"o}hlich} et~al.,}{{Fr{\"o}hlich}
  et~al.}{2006}]{frohlich2006b}
{Fr{\"o}hlich} C.,  et~al., 2006, \mn@doi [\apj] {10.1086/498224}, \href
  {https://ui.adsabs.harvard.edu/abs/2006ApJ...637..415F} {637, 415}

\bibitem[\protect\citeauthoryear{{Grevesse} \& {Sauval}}{{Grevesse} \&
  {Sauval}}{1998}]{Grevesse1998}
{Grevesse} N.,  {Sauval} A.~J.,  1998, \mn@doi [\ssr]
  {10.1023/A:1005161325181}, \href
  {https://ui.adsabs.harvard.edu/abs/1998SSRv...85..161G} {85, 161}

\bibitem[\protect\citeauthoryear{{Griffith}, {Sukhbold}, {Weinberg}, {Johnson},
  {Johnson}  \& {Vincenzo}}{{Griffith} et~al.}{2021}]{griffith2021}
{Griffith} E.~J.,  {Sukhbold} T.,  {Weinberg} D.~H.,  {Johnson} J.~A.,
  {Johnson} J.~W.,   {Vincenzo} F.,  2021, \mn@doi [\apj]
  {10.3847/1538-4357/ac1bac}, \href
  {https://ui.adsabs.harvard.edu/abs/2021ApJ...921...73G} {921, 73}

\bibitem[\protect\citeauthoryear{{Griffith}, {Johnson}, {Weinberg}, {Ilyin},
  {Johnson}, {Rodriguez-Martinez}  \& {Strassmeier}}{{Griffith}
  et~al.}{2023}]{Griffith2023}
{Griffith} E.~J.,  {Johnson} J.~A.,  {Weinberg} D.~H.,  {Ilyin} I.,  {Johnson}
  J.~W.,  {Rodriguez-Martinez} R.,   {Strassmeier} K.~G.,  2023, \mn@doi [\apj]
  {10.3847/1538-4357/aca659}, \href
  {https://ui.adsabs.harvard.edu/abs/2023ApJ...944...47G} {944, 47}

\bibitem[\protect\citeauthoryear{{Griffith}, {Hogg}, {Dalcanton},
  {Hasselquist}, {Ratcliffe}, {Ness}  \& {Weinberg}}{{Griffith}
  et~al.}{2024}]{griffith2024}
{Griffith} E.~J.,  {Hogg} D.~W.,  {Dalcanton} J.~J.,  {Hasselquist} S.,
  {Ratcliffe} B.,  {Ness} M.,   {Weinberg} D.~H.,  2024, \mn@doi [\aj]
  {10.3847/1538-3881/ad19c7}, \href
  {https://ui.adsabs.harvard.edu/abs/2024AJ....167...98G} {167, 98}

\bibitem[\protect\citeauthoryear{{Griffith}, {Blum}, {Weinberg}, {Johnson},
  {Sit}, {Ilyin}  \& {Strassmeier}}{{Griffith} et~al.}{2026}]{Griffith2026}
{Griffith} E.~J.,  {Blum} M.,  {Weinberg} D.~H.,  {Johnson} J.~A.,  {Sit} T.,
  {Ilyin} I.,   {Strassmeier} K.~G.,  2026, \mn@doi [\apj]
  {10.3847/1538-4357/ae4e24}, \href
  {https://ui.adsabs.harvard.edu/abs/2026ApJ..1001..193G} {1001, 193}

\bibitem[\protect\citeauthoryear{{Halevi} \& {M{\"o}sta}}{{Halevi} \&
  {M{\"o}sta}}{2018}]{Halevi2018}
{Halevi} G.,  {M{\"o}sta} P.,  2018, \mn@doi [\mnras] {10.1093/mnras/sty797},
  \href {https://ui.adsabs.harvard.edu/abs/2018MNRAS.477.2366H} {477, 2366}

\bibitem[\protect\citeauthoryear{{Hansen} et~al.,}{{Hansen}
  et~al.}{2015}]{Hansen2015}
{Hansen} T.,  et~al., 2015, \mn@doi [\apj] {10.1088/0004-637X/807/2/173}, \href
  {https://ui.adsabs.harvard.edu/abs/2015ApJ...807..173H} {807, 173}

\bibitem[\protect\citeauthoryear{Harris et~al.,}{Harris
  et~al.}{2020}]{harris2020array}
Harris C.~R.,  et~al., 2020, \mn@doi [Nature] {10.1038/s41586-020-2649-2}, 585,
  357

\bibitem[\protect\citeauthoryear{{Heger} \& {Woosley}}{{Heger} \&
  {Woosley}}{2002}]{Heger2002b}
{Heger} A.,  {Woosley} S.~E.,  2002, \mn@doi [\apj] {10.1086/338487}, \href
  {https://ui.adsabs.harvard.edu/abs/2002ApJ...567..532H} {567, 532}

\bibitem[\protect\citeauthoryear{{Heger}, {Woosley}, {Mart{\'\i}nez-Pinedo}  \&
  {Langanke}}{{Heger} et~al.}{2001}]{Heger2001}
{Heger} A.,  {Woosley} S.~E.,  {Mart{\'\i}nez-Pinedo} G.,   {Langanke} K.,
  2001, \mn@doi [\apj] {10.1086/324092}, \href
  {https://ui.adsabs.harvard.edu/abs/2001ApJ...560..307H} {560, 307}

\bibitem[\protect\citeauthoryear{{Heger}, {Woosley}, {Rauscher}, {Hoffman}  \&
  {Boyes}}{{Heger} et~al.}{2002}]{Heger2002a}
{Heger} A.,  {Woosley} S.~E.,  {Rauscher} T.,  {Hoffman} R.~D.,   {Boyes}
  M.~M.,  2002, \mn@doi [\nar] {10.1016/S1387-6473(02)00184-7}, \href
  {https://ui.adsabs.harvard.edu/abs/2002NewAR..46..463H} {46, 463}

\bibitem[\protect\citeauthoryear{Heger, Kolbe, Haxton, Langanke,
  Martínez-Pinedo  \& Woosley}{Heger et~al.}{2005}]{Heger2005}
Heger A.,  Kolbe E.,  Haxton W.,  Langanke K.,  Martínez-Pinedo G.,   Woosley
  S.,  2005, \mn@doi [Physics Letters B]
  {https://doi.org/10.1016/j.physletb.2004.12.017}, 606, 258

\bibitem[\protect\citeauthoryear{{Hempel} \& {Schaffner-Bielich}}{{Hempel} \&
  {Schaffner-Bielich}}{2010}]{Hempel.SchaffnerBielich:2010}
{Hempel} M.,  {Schaffner-Bielich} J.,  2010, \mn@doi [Nuclear Physics A]
  {10.1016/j.nuclphysa.2010.02.010}, \href
  {http://adsabs.harvard.edu/abs/2010NuPhA.837..210H} {837, 210}

\bibitem[\protect\citeauthoryear{{Herant}, {Benz}, {Hix}, {Fryer}  \&
  {Colgate}}{{Herant} et~al.}{1994}]{Herant1994}
{Herant} M.,  {Benz} W.,  {Hix} W.~R.,  {Fryer} C.~L.,   {Colgate} S.~A.,
  1994, \mn@doi [\apj] {10.1086/174817}, \href
  {https://ui.adsabs.harvard.edu/abs/1994ApJ...435..339H} {435, 339}

\bibitem[\protect\citeauthoryear{{Hirschi}}{{Hirschi}}{2015}]{Hirschi2015}
{Hirschi} R.,  2015, in {Vink} J.~S.,  ed.,  Astrophysics and Space Science
  Library Vol. 412, Very Massive Stars in the Local Universe. p.~157
  (\mn@eprint {arXiv} {1409.7053}), \mn@doi{10.1007/978-3-319-09596-7_6}

\bibitem[\protect\citeauthoryear{{Hix} et~al.,}{{Hix} et~al.}{2016}]{Hix2016}
{Hix} W.~R.,  et~al., 2016, \mn@doi [Acta Physica Polonica B]
  {10.5506/APhysPolB.47.645}, \href
  {https://ui.adsabs.harvard.edu/abs/2016AcPPB..47..645H} {47, 645}

\bibitem[\protect\citeauthoryear{{Howes} et~al.,}{{Howes}
  et~al.}{2015}]{Howes2015}
{Howes} L.~M.,  et~al., 2015, \mn@doi [\nat] {10.1038/nature15747}, \href
  {https://ui.adsabs.harvard.edu/abs/2015Natur.527..484H} {527, 484}

\bibitem[\protect\citeauthoryear{Hunter}{Hunter}{2007}]{Hunter:2007}
Hunter J.~D.,  2007, \mn@doi [Computing in Science \& Engineering]
  {10.1109/MCSE.2007.55}, 9, 90

\bibitem[\protect\citeauthoryear{{Jacobson} et~al.,}{{Jacobson}
  et~al.}{2015}]{Jacobson2015}
{Jacobson} H.~R.,  et~al., 2015, \mn@doi [\apj] {10.1088/0004-637X/807/2/171},
  \href {https://ui.adsabs.harvard.edu/abs/2015ApJ...807..171J} {807, 171}

\bibitem[\protect\citeauthoryear{{Jalin}, {Arcones}, {O'Connor}, {Ercolino}  \&
  {Langer}}{{Jalin} et~al.}{2026}]{Jalin2026}
{Jalin} G.,  {Arcones} A.,  {O'Connor} E.~P.,  {Ercolino} A.,   {Langer} N.,
  2026, arXiv e-prints, \href
  {https://ui.adsabs.harvard.edu/abs/2026arXiv261008617J} {p. arXiv:2610.08617}

\bibitem[\protect\citeauthoryear{{Janka}}{{Janka}}{2012}]{janka12}
{Janka} H.-T.,  2012, \mn@doi [Annual Review of Nuclear and Particle Science]
  {10.1146/annurev-nucl-102711-094901}, \href
  {http://adsabs.harvard.edu/abs/2012ARNPS..62..407J} {62, 407}

\bibitem[\protect\citeauthoryear{{Janka}}{{Janka}}{2025}]{Janka2025}
{Janka} H.-T.,  2025, \mn@doi [Annual Review of Nuclear and Particle Science]
  {10.1146/annurev-nucl-121423-100945}, \href
  {https://ui.adsabs.harvard.edu/abs/2025ARNPS..75..425J} {75, 425}

\bibitem[\protect\citeauthoryear{{Janka} \& {Mueller}}{{Janka} \&
  {Mueller}}{1996}]{Janka1996}
{Janka} H.-T.,  {Mueller} E.,  1996, \aap, \href
  {https://ui.adsabs.harvard.edu/abs/1996A&A...306..167J} {306, 167}

\bibitem[\protect\citeauthoryear{{Janka}, {Melson}  \& {Summa}}{{Janka}
  et~al.}{2016}]{Janka2016}
{Janka} H.-T.,  {Melson} T.,   {Summa} A.,  2016, \mn@doi [Annual Review of
  Nuclear and Particle Science] {10.1146/annurev-nucl-102115-044747}, \href
  {https://ui.adsabs.harvard.edu/abs/2016ARNPS..66..341J} {66, 341}

\bibitem[\protect\citeauthoryear{{Jerkstrand} et~al.,}{{Jerkstrand}
  et~al.}{2015}]{Jerkstrand2015}
{Jerkstrand} A.,  et~al., 2015, \mn@doi [\mnras] {10.1093/mnras/stv087}, \href
  {https://ui.adsabs.harvard.edu/abs/2015MNRAS.448.2482J} {448, 2482}

\bibitem[\protect\citeauthoryear{{Jermyn} et~al.,}{{Jermyn}
  et~al.}{2023}]{Jermyn2023}
{Jermyn} A.~S.,  et~al., 2023, \mn@doi [\apjs] {10.3847/1538-4365/acae8d},
  \href {https://ui.adsabs.harvard.edu/abs/2023ApJS..265...15J} {265, 15}

\bibitem[\protect\citeauthoryear{{Keller} et~al.,}{{Keller}
  et~al.}{2014}]{Keller2014}
{Keller} S.~C.,  et~al., 2014, \mn@doi [\nat] {10.1038/nature12990}, \href
  {https://ui.adsabs.harvard.edu/abs/2014Natur.506..463K} {506, 463}

\bibitem[\protect\citeauthoryear{{Kromer}, {Ohlmann}  \& {R{\"o}pke}}{{Kromer}
  et~al.}{2017}]{Kromer2017}
{Kromer} M.,  {Ohlmann} S.,   {R{\"o}pke} F.~K.,  2017, \mn@doi [\memsai]
  {10.48550/arXiv.1706.09879}, \href
  {https://ui.adsabs.harvard.edu/abs/2017MmSAI..88..312K} {88, 312}

\bibitem[\protect\citeauthoryear{{Kuroda}, {Arcones}, {Takiwaki}  \&
  {Kotake}}{{Kuroda} et~al.}{2020}]{Kuroda2020}
{Kuroda} T.,  {Arcones} A.,  {Takiwaki} T.,   {Kotake} K.,  2020, \mn@doi
  [\apj] {10.3847/1538-4357/ab9308}, \href
  {https://ui.adsabs.harvard.edu/abs/2020ApJ...896..102K} {896, 102}

\bibitem[\protect\citeauthoryear{{Langanke} \&
  {Mart{\'{\i}}nez-Pinedo}}{{Langanke} \& {Mart{\'{\i}}nez-Pinedo}}{2001}]{lmp}
{Langanke} K.,  {Mart{\'{\i}}nez-Pinedo} G.,  2001, \mn@doi [Atomic Data and
  Nuclear Data Tables] {10.1006/adnd.2001.0865}, \href
  {http://adsabs.harvard.edu/abs/2001ADNDT..79....1L} {79, 1}

\bibitem[\protect\citeauthoryear{{Langer}}{{Langer}}{2012}]{Langer2012}
{Langer} N.,  2012, \mn@doi [\araa] {10.1146/annurev-astro-081811-125534},
  \href {https://ui.adsabs.harvard.edu/abs/2012ARA&A..50..107L} {50, 107}

\bibitem[\protect\citeauthoryear{{Laplace}, {Justham}, {Renzo}, {G{\"o}tberg},
  {Farmer}, {Vartanyan}  \& {de Mink}}{{Laplace} et~al.}{2021}]{Laplace2021}
{Laplace} E.,  {Justham} S.,  {Renzo} M.,  {G{\"o}tberg} Y.,  {Farmer} R.,
  {Vartanyan} D.,   {de Mink} S.~E.,  2021, \mn@doi [\aap]
  {10.1051/0004-6361/202140506}, \href
  {https://ui.adsabs.harvard.edu/abs/2021A&A...656A..58L} {656, A58}

\bibitem[\protect\citeauthoryear{{Lentz} et~al.,}{{Lentz}
  et~al.}{2015}]{Lentz2015}
{Lentz} E.~J.,  et~al., 2015, \mn@doi [\apjl] {10.1088/2041-8205/807/2/L31},
  \href {https://ui.adsabs.harvard.edu/abs/2015ApJ...807L..31L} {807, L31}

\bibitem[\protect\citeauthoryear{{Li}, {Aoki}, {Honda}, {Zhao}, {Christlieb}
  \& {Suda}}{{Li} et~al.}{2015a}]{Li2015a}
{Li} H.-N.,  {Aoki} W.,  {Honda} S.,  {Zhao} G.,  {Christlieb} N.,   {Suda} T.,
   2015a, \mn@doi [Research in Astronomy and Astrophysics]
  {10.1088/1674-4527/15/8/011}, \href
  {https://ui.adsabs.harvard.edu/abs/2015RAA....15.1264L} {15, 1264}

\bibitem[\protect\citeauthoryear{{Li}, {Aoki}, {Zhao}, {Honda}, {Christlieb}
  \& {Suda}}{{Li} et~al.}{2015b}]{Li2015b}
{Li} H.,  {Aoki} W.,  {Zhao} G.,  {Honda} S.,  {Christlieb} N.,   {Suda} T.,
  2015b, \mn@doi [\pasj] {10.1093/pasj/psv053}, \href
  {https://ui.adsabs.harvard.edu/abs/2015PASJ...67...84L} {67, 84}

\bibitem[\protect\citeauthoryear{{Li} et~al.,}{{Li} et~al.}{2022}]{Li2022}
{Li} H.,  et~al., 2022, \mn@doi [\apj] {10.3847/1538-4357/ac6514}, \href
  {https://ui.adsabs.harvard.edu/abs/2022ApJ...931..147L} {931, 147}

\bibitem[\protect\citeauthoryear{{Liebend{\"o}rfer}, {Mezzacappa}  \&
  {Thielemann}}{{Liebend{\"o}rfer} et~al.}{2001}]{Liebendoerfer.Agile}
{Liebend{\"o}rfer} M.,  {Mezzacappa} A.,   {Thielemann} F.-K.,  2001, \mn@doi
  [\prd] {10.1103/PhysRevD.63.104003}, \href
  {http://adsabs.harvard.edu/abs/2001PhRvD..63j4003L} {63, 104003}

\bibitem[\protect\citeauthoryear{{Liebend{\"o}rfer}, {Messer}, {Mezzacappa},
  {Bruenn}, {Cardall}  \& {Thielemann}}{{Liebend{\"o}rfer}
  et~al.}{2004}]{Liebendoerfer2004}
{Liebend{\"o}rfer} M.,  {Messer} O.~E.~B.,  {Mezzacappa} A.,  {Bruenn} S.~W.,
  {Cardall} C.~Y.,   {Thielemann} F.-K.,  2004, \mn@doi [\apjs]
  {10.1086/380191}, \href
  {https://ui.adsabs.harvard.edu/abs/2004ApJS..150..263L} {150, 263}

\bibitem[\protect\citeauthoryear{{Limongi} \& {Chieffi}}{{Limongi} \&
  {Chieffi}}{2006}]{limongi06}
{Limongi} M.,  {Chieffi} A.,  2006, \mn@doi [\apj] {10.1086/505164}, \href
  {http://adsabs.harvard.edu/abs/2006ApJ...647..483L} {647, 483}

\bibitem[\protect\citeauthoryear{{Limongi} \& {Chieffi}}{{Limongi} \&
  {Chieffi}}{2012}]{limongi12}
{Limongi} M.,  {Chieffi} A.,  2012, \mn@doi [\apjs]
  {10.1088/0067-0049/199/2/38}, \href
  {http://adsabs.harvard.edu/abs/2012ApJS..199...38L} {199, 38}

\bibitem[\protect\citeauthoryear{{Limongi} \& {Chieffi}}{{Limongi} \&
  {Chieffi}}{2018}]{Limongi2018}
{Limongi} M.,  {Chieffi} A.,  2018, \mn@doi [\apjs] {10.3847/1538-4365/aacb24},
  \href {https://ui.adsabs.harvard.edu/abs/2018ApJS..237...13L} {237, 13}

\bibitem[\protect\citeauthoryear{{Lodders}, {Palme}  \& {Gail}}{{Lodders}
  et~al.}{2009}]{lodders2009}
{Lodders} K.,  {Palme} H.,   {Gail} H.-P.,  2009, \mn@doi [Landolt
  B{\"o}rnstein] {10.1007/978-3-540-88055-4_34}, \href
  {https://ui.adsabs.harvard.edu/abs/2009LanB...4B..712L} {4B, 712}

\bibitem[\protect\citeauthoryear{{Majewski} et~al.,}{{Majewski}
  et~al.}{2017}]{majewski2017}
{Majewski} S.~R.,  et~al., 2017, \mn@doi [\aj] {10.3847/1538-3881/aa784d},
  \href {https://ui.adsabs.harvard.edu/abs/2017AJ....154...94M} {154, 94}

\bibitem[\protect\citeauthoryear{{Marino} et~al.,}{{Marino}
  et~al.}{2019}]{Marino2019}
{Marino} A.~F.,  et~al., 2019, \mn@doi [\mnras] {10.1093/mnras/stz645}, \href
  {https://ui.adsabs.harvard.edu/abs/2019MNRAS.485.5153M} {485, 5153}

\bibitem[\protect\citeauthoryear{{Martin} et~al.,}{{Martin}
  et~al.}{2024}]{Martin2024}
{Martin} N.~F.,  et~al., 2024, \mn@doi [\aap] {10.1051/0004-6361/202347633},
  \href {https://ui.adsabs.harvard.edu/abs/2024A&A...692A.115M} {692, A115}

\bibitem[\protect\citeauthoryear{{Mason}, {Hartkopf}, {Gies}, {Henry}  \&
  {Helsel}}{{Mason} et~al.}{2009}]{Mason2009}
{Mason} B.~D.,  {Hartkopf} W.~I.,  {Gies} D.~R.,  {Henry} T.~J.,   {Helsel}
  J.~W.,  2009, \mn@doi [\aj] {10.1088/0004-6256/137/2/3358}, \href
  {https://ui.adsabs.harvard.edu/abs/2009AJ....137.3358M} {137, 3358}

\bibitem[\protect\citeauthoryear{{Melson}, {Janka}  \& {Marek}}{{Melson}
  et~al.}{2015a}]{Melson2015a}
{Melson} T.,  {Janka} H.-T.,   {Marek} A.,  2015a, \mn@doi [\apjl]
  {10.1088/2041-8205/801/2/L24}, \href
  {https://ui.adsabs.harvard.edu/abs/2015ApJ...801L..24M} {801, L24}

\bibitem[\protect\citeauthoryear{{Melson}, {Janka}, {Bollig}, {Hanke}, {Marek}
  \& {M{\"u}ller}}{{Melson} et~al.}{2015b}]{Melson2015b}
{Melson} T.,  {Janka} H.-T.,  {Bollig} R.,  {Hanke} F.,  {Marek} A.,
  {M{\"u}ller} B.,  2015b, \mn@doi [\apjl] {10.1088/2041-8205/808/2/L42}, \href
  {https://ui.adsabs.harvard.edu/abs/2015ApJ...808L..42M} {808, L42}

\bibitem[\protect\citeauthoryear{{Moe} \& {Di Stefano}}{{Moe} \& {Di
  Stefano}}{2017}]{Moe2017}
{Moe} M.,  {Di Stefano} R.,  2017, \mn@doi [\apjs] {10.3847/1538-4365/aa6fb6},
  \href {https://ui.adsabs.harvard.edu/abs/2017ApJS..230...15M} {230, 15}

\bibitem[\protect\citeauthoryear{{M{\"o}ller}, {Nix}, {Myers}  \&
  {Swiatecki}}{{M{\"o}ller} et~al.}{1995}]{Moller1995}
{M{\"o}ller} P.,  {Nix} J.~R.,  {Myers} W.~D.,   {Swiatecki} W.~J.,  1995,
  \mn@doi [Atomic Data and Nuclear Data Tables] {10.1006/adnd.1995.1002}, \href
  {https://ui.adsabs.harvard.edu/abs/1995ADNDT..59..185M} {59, 185}

\bibitem[\protect\citeauthoryear{{Morozova}, {Piro}, {Renzo}, {Ott}, {Clausen},
  {Couch}, {Ellis}  \& {Roberts}}{{Morozova} et~al.}{2015}]{Morozova2015}
{Morozova} V.,  {Piro} A.~L.,  {Renzo} M.,  {Ott} C.~D.,  {Clausen} D.,
  {Couch} S.~M.,  {Ellis} J.,   {Roberts} L.~F.,  2015, \mn@doi [\apj]
  {10.1088/0004-637X/814/1/63}, \href
  {https://ui.adsabs.harvard.edu/abs/2015ApJ...814...63M} {814, 63}

\bibitem[\protect\citeauthoryear{{M{\"o}sta} et~al.,}{{M{\"o}sta}
  et~al.}{2014}]{Mosta2014}
{M{\"o}sta} P.,  et~al., 2014, \mn@doi [\apjl] {10.1088/2041-8205/785/2/L29},
  \href {https://ui.adsabs.harvard.edu/abs/2014ApJ...785L..29M} {785, L29}

\bibitem[\protect\citeauthoryear{{M{\"o}sta}, {Ott}, {Radice}, {Roberts},
  {Schnetter}  \& {Haas}}{{M{\"o}sta} et~al.}{2015}]{Mosta2015}
{M{\"o}sta} P.,  {Ott} C.~D.,  {Radice} D.,  {Roberts} L.~F.,  {Schnetter} E.,
   {Haas} R.,  2015, \mn@doi [\nat] {10.1038/nature15755}, \href
  {https://ui.adsabs.harvard.edu/abs/2015Natur.528..376M} {528, 376}

\bibitem[\protect\citeauthoryear{{M{\"u}ller}}{{M{\"u}ller}}{2015}]{Muller2015a}
{M{\"u}ller} B.,  2015, \mn@doi [\mnras] {10.1093/mnras/stv1611}, \href
  {https://ui.adsabs.harvard.edu/abs/2015MNRAS.453..287M} {453, 287}

\bibitem[\protect\citeauthoryear{{Nakamura}, {Takiwaki}, {Kuroda}  \&
  {Kotake}}{{Nakamura} et~al.}{2015}]{Nakamura2015}
{Nakamura} K.,  {Takiwaki} T.,  {Kuroda} T.,   {Kotake} K.,  2015, \mn@doi
  [\pasj] {10.1093/pasj/psv073}, \href
  {https://ui.adsabs.harvard.edu/abs/2015PASJ...67..107N} {67, 107}

\bibitem[\protect\citeauthoryear{{Neopane}, {Sandoval}, {Hix}, {Harris},
  {Messer}  \& {Lentz}}{{Neopane} et~al.}{2026}]{Neopane2026}
{Neopane} S.,  {Sandoval} M.~A.,  {Hix} W.~R.,  {Harris} J.~A.,  {Messer}
  O.~E.~B.,   {Lentz} E.~J.,  2026, \mn@doi [\apj] {10.3847/1538-4357/ae8a44},
  \href {https://ui.adsabs.harvard.edu/abs/2026ApJ..1008...36N} {1008, 36}

\bibitem[\protect\citeauthoryear{{Nishimura}, {Takiwaki}  \&
  {Thielemann}}{{Nishimura} et~al.}{2015}]{Nishimura2015}
{Nishimura} N.,  {Takiwaki} T.,   {Thielemann} F.-K.,  2015, \mn@doi [\apj]
  {10.1088/0004-637X/810/2/109}, \href
  {https://ui.adsabs.harvard.edu/abs/2015ApJ...810..109N} {810, 109}

\bibitem[\protect\citeauthoryear{{Nishimura}, {Sawai}, {Takiwaki}, {Yamada}  \&
  {Thielemann}}{{Nishimura} et~al.}{2017}]{Nishimura2017}
{Nishimura} N.,  {Sawai} H.,  {Takiwaki} T.,  {Yamada} S.,   {Thielemann}
  F.-K.,  2017, \mn@doi [\apjl] {10.3847/2041-8213/aa5dee}, \href
  {https://ui.adsabs.harvard.edu/abs/2017ApJ...836L..21N} {836, L21}

\bibitem[\protect\citeauthoryear{{Nomoto}, {Kobayashi}  \& {Tominaga}}{{Nomoto}
  et~al.}{2013}]{nomoto13}
{Nomoto} K.,  {Kobayashi} C.,   {Tominaga} N.,  2013, \mn@doi [\araa]
  {10.1146/annurev-astro-082812-140956}, \href
  {http://adsabs.harvard.edu/abs/2013ARA%26A..51..457N} {51, 457}

\bibitem[\protect\citeauthoryear{{Norris} et~al.,}{{Norris}
  et~al.}{2013}]{Norris2013}
{Norris} J.~E.,  et~al., 2013, \mn@doi [\apj] {10.1088/0004-637X/762/1/25},
  \href {https://ui.adsabs.harvard.edu/abs/2013ApJ...762...25N} {762, 25}

\bibitem[\protect\citeauthoryear{{O'Connor} \& {Ott}}{{O'Connor} \&
  {Ott}}{2011}]{oconnor11}
{O'Connor} E.,  {Ott} C.~D.,  2011, \mn@doi [\apj]
  {10.1088/0004-637X/730/2/70}, \href
  {http://adsabs.harvard.edu/abs/2011ApJ...730...70O} {730, 70}

\bibitem[\protect\citeauthoryear{{Obergaulinger} \& {Aloy}}{{Obergaulinger} \&
  {Aloy}}{2020}]{Obergaulinger2020}
{Obergaulinger} M.,  {Aloy} M.~{\'A}.,  2020, \mn@doi [\mnras]
  {10.1093/mnras/staa096}, \href
  {https://ui.adsabs.harvard.edu/abs/2020MNRAS.492.4613O} {492, 4613}

\bibitem[\protect\citeauthoryear{{Paxton}, {Bildsten}, {Dotter}, {Herwig},
  {Lesaffre}  \& {Timmes}}{{Paxton} et~al.}{2011}]{Paxton2011}
{Paxton} B.,  {Bildsten} L.,  {Dotter} A.,  {Herwig} F.,  {Lesaffre} P.,
  {Timmes} F.,  2011, \mn@doi [\apjs] {10.1088/0067-0049/192/1/3}, \href
  {https://ui.adsabs.harvard.edu/abs/2011ApJS..192....3P} {192, 3}

\bibitem[\protect\citeauthoryear{{Paxton} et~al.,}{{Paxton}
  et~al.}{2013}]{Paxton2013}
{Paxton} B.,  et~al., 2013, \mn@doi [\apjs] {10.1088/0067-0049/208/1/4}, \href
  {https://ui.adsabs.harvard.edu/abs/2013ApJS..208....4P} {208, 4}

\bibitem[\protect\citeauthoryear{{Paxton} et~al.,}{{Paxton}
  et~al.}{2015}]{Paxton2015}
{Paxton} B.,  et~al., 2015, \mn@doi [\apjs] {10.1088/0067-0049/220/1/15}, \href
  {https://ui.adsabs.harvard.edu/abs/2015ApJS..220...15P} {220, 15}

\bibitem[\protect\citeauthoryear{{Paxton} et~al.,}{{Paxton}
  et~al.}{2018}]{Paxton2018}
{Paxton} B.,  et~al., 2018, \mn@doi [\apjs] {10.3847/1538-4365/aaa5a8}, \href
  {https://ui.adsabs.harvard.edu/abs/2018ApJS..234...34P} {234, 34}

\bibitem[\protect\citeauthoryear{{Paxton} et~al.,}{{Paxton}
  et~al.}{2019}]{Paxton2019}
{Paxton} B.,  et~al., 2019, \mn@doi [\apjs] {10.3847/1538-4365/ab2241}, \href
  {https://ui.adsabs.harvard.edu/abs/2019ApJS..243...10P} {243, 10}

\bibitem[\protect\citeauthoryear{{Perego}, {Hempel}, {Fr{\"o}hlich}, {Ebinger},
  {Eichler}, {Casanova}, {Liebend{\"o}rfer}  \& {Thielemann}}{{Perego}
  et~al.}{2015}]{push1}
{Perego} A.,  {Hempel} M.,  {Fr{\"o}hlich} C.,  {Ebinger} K.,  {Eichler} M.,
  {Casanova} J.,  {Liebend{\"o}rfer} M.,   {Thielemann} F.-K.,  2015, \mn@doi
  [Astrophys.\ J.] {10.1088/0004-637X/806/2/275}, \href
  {http://adsabs.harvard.edu/abs/2015ApJ...806..275P} {806, 275}

\bibitem[\protect\citeauthoryear{{Pignatari} et~al.,}{{Pignatari}
  et~al.}{2016}]{Pignatari2016}
{Pignatari} M.,  et~al., 2016, \mn@doi [\apjs] {10.3847/0067-0049/225/2/24},
  \href {https://ui.adsabs.harvard.edu/abs/2016ApJS..225...24P} {225, 24}

\bibitem[\protect\citeauthoryear{{Placco}, {Frebel}, {Lee}, {Jacobson},
  {Beers}, {Pena}, {Chan}  \& {Heger}}{{Placco} et~al.}{2015}]{Placco2015}
{Placco} V.~M.,  {Frebel} A.,  {Lee} Y.~S.,  {Jacobson} H.~R.,  {Beers} T.~C.,
  {Pena} J.~M.,  {Chan} C.,   {Heger} A.,  2015, \mn@doi [\apj]
  {10.1088/0004-637X/809/2/136}, \href
  {https://ui.adsabs.harvard.edu/abs/2015ApJ...809..136P} {809, 136}

\bibitem[\protect\citeauthoryear{{Podsiadlowski}, {Joss}  \&
  {Hsu}}{{Podsiadlowski} et~al.}{1992}]{Podsiadlowski1992}
{Podsiadlowski} P.,  {Joss} P.~C.,   {Hsu} J.~J.~L.,  1992, \mn@doi [\apj]
  {10.1086/171341}, \href
  {https://ui.adsabs.harvard.edu/abs/1992ApJ...391..246P} {391, 246}

\bibitem[\protect\citeauthoryear{{Pols}, {Tout}, {Eggleton}  \& {Han}}{{Pols}
  et~al.}{1995}]{Pols1995}
{Pols} O.~R.,  {Tout} C.~A.,  {Eggleton} P.~P.,   {Han} Z.,  1995, \mn@doi
  [\mnras] {10.1093/mnras/274.3.964}, \href
  {https://ui.adsabs.harvard.edu/abs/1995MNRAS.274..964P} {274, 964}

\bibitem[\protect\citeauthoryear{{Rauscher} \& {Thielemann}}{{Rauscher} \&
  {Thielemann}}{2000}]{Rauscher.FKT:2000}
{Rauscher} T.,  {Thielemann} F.-K.,  2000, \mn@doi [Atomic Data and Nuclear
  Data Tables] {10.1006/adnd.2000.0834}, \href
  {http://adsabs.harvard.edu/abs/2000ADNDT..75....1R} {75, 1}

\bibitem[\protect\citeauthoryear{{Rauscher}, {Heger}, {Hoffman}  \&
  {Woosley}}{{Rauscher} et~al.}{2002}]{rauscher02}
{Rauscher} T.,  {Heger} A.,  {Hoffman} R.~D.,   {Woosley} S.~E.,  2002, \mn@doi
  [\apj] {10.1086/341728}, \href
  {http://adsabs.harvard.edu/abs/2002ApJ...576..323R} {576, 323}

\bibitem[\protect\citeauthoryear{{Rodr{\'\i}guez}, {Maoz}  \&
  {Nakar}}{{Rodr{\'\i}guez} et~al.}{2023}]{Rodriguez2023}
{Rodr{\'\i}guez} {\'O}.,  {Maoz} D.,   {Nakar} E.,  2023, \mn@doi [\apj]
  {10.3847/1538-4357/ace2bd}, \href
  {https://ui.adsabs.harvard.edu/abs/2023ApJ...955...71R} {955, 71}

\bibitem[\protect\citeauthoryear{{Roederer}, {Placco}  \& {Beers}}{{Roederer}
  et~al.}{2016}]{Roederer2016}
{Roederer} I.~U.,  {Placco} V.~M.,   {Beers} T.~C.,  2016, \mn@doi [\apjl]
  {10.3847/2041-8205/824/2/L19}, \href
  {https://ui.adsabs.harvard.edu/abs/2016ApJ...824L..19R} {824, L19}

\bibitem[\protect\citeauthoryear{{Rybizki}, {Just}  \& {Rix}}{{Rybizki}
  et~al.}{2017}]{rybizki2017}
{Rybizki} J.,  {Just} A.,   {Rix} H.-W.,  2017, \mn@doi [\aap]
  {10.1051/0004-6361/201730522}, \href
  {https://ui.adsabs.harvard.edu/abs/2017A&A...605A..59R} {605, A59}

\bibitem[\protect\citeauthoryear{{Salpeter}}{{Salpeter}}{1955}]{Salpeter1955}
{Salpeter} E.~E.,  1955, \mn@doi [\apj] {10.1086/145971}, \href
  {https://ui.adsabs.harvard.edu/abs/1955ApJ...121..161S} {121, 161}

\bibitem[\protect\citeauthoryear{{Sana} et~al.,}{{Sana}
  et~al.}{2012}]{Sana2012}
{Sana} H.,  et~al., 2012, \mn@doi [Science] {10.1126/science.1223344}, \href
  {https://ui.adsabs.harvard.edu/abs/2012Sci...337..444S} {337, 444}

\bibitem[\protect\citeauthoryear{{Sandoval}, {Hix}, {Messer}, {Lentz}  \&
  {Harris}}{{Sandoval} et~al.}{2021}]{Sandoval2021}
{Sandoval} M.~A.,  {Hix} W.~R.,  {Messer} O.~E.~B.,  {Lentz} E.~J.,   {Harris}
  J.~A.,  2021, \mn@doi [\apj] {10.3847/1538-4357/ac1d49}, \href
  {https://ui.adsabs.harvard.edu/abs/2021ApJ...921..113S} {921, 113}

\bibitem[\protect\citeauthoryear{{Sawada} \& {Maeda}}{{Sawada} \&
  {Maeda}}{2019}]{Sawada2019}
{Sawada} R.,  {Maeda} K.,  2019, \mn@doi [\apj] {10.3847/1538-4357/ab4da3},
  \href {https://ui.adsabs.harvard.edu/abs/2019ApJ...886...47S} {886, 47}

\bibitem[\protect\citeauthoryear{{Sieverding}, {Kresse}  \&
  {Janka}}{{Sieverding} et~al.}{2023}]{Sieverding2023}
{Sieverding} A.,  {Kresse} D.,   {Janka} H.-T.,  2023, \mn@doi [\apjl]
  {10.3847/2041-8213/ad045b}, \href
  {https://ui.adsabs.harvard.edu/abs/2023ApJ...957L..25S} {957, L25}

\bibitem[\protect\citeauthoryear{{Sneden}, {Cowan}, {Kobayashi}, {Pignatari},
  {Lawler}, {Den Hartog}  \& {Wood}}{{Sneden} et~al.}{2016}]{Sneden2016}
{Sneden} C.,  {Cowan} J.~J.,  {Kobayashi} C.,  {Pignatari} M.,  {Lawler} J.~E.,
   {Den Hartog} E.~A.,   {Wood} M.~P.,  2016, \mn@doi [\apj]
  {10.3847/0004-637X/817/1/53}, \href
  {https://ui.adsabs.harvard.edu/abs/2016ApJ...817...53S} {817, 53}

\bibitem[\protect\citeauthoryear{{Starkenburg} et~al.,}{{Starkenburg}
  et~al.}{2017}]{Starkenburg2017}
{Starkenburg} E.,  et~al., 2017, \mn@doi [\mnras] {10.1093/mnras/stx1068},
  \href {https://ui.adsabs.harvard.edu/abs/2017MNRAS.471.2587S} {471, 2587}

\bibitem[\protect\citeauthoryear{{Steiner}, {Hempel}  \& {Fischer}}{{Steiner}
  et~al.}{2013}]{Steiner2013}
{Steiner} A.~W.,  {Hempel} M.,   {Fischer} T.,  2013, \mn@doi [\apj]
  {10.1088/0004-637X/774/1/17}, \href
  {https://ui.adsabs.harvard.edu/abs/2013ApJ...774...17S} {774, 17}

\bibitem[\protect\citeauthoryear{{Sukhbold}, {Ertl}, {Woosley}, {Brown}  \&
  {Janka}}{{Sukhbold} et~al.}{2016}]{Sukhbold2016}
{Sukhbold} T.,  {Ertl} T.,  {Woosley} S.~E.,  {Brown} J.~M.,   {Janka} H.-T.,
  2016, \mn@doi [\apj] {10.3847/0004-637X/821/1/38}, \href
  {https://ui.adsabs.harvard.edu/abs/2016ApJ...821...38S} {821, 38}

\bibitem[\protect\citeauthoryear{{Takiwaki}, {Kotake}  \& {Suwa}}{{Takiwaki}
  et~al.}{2014}]{Takiwaki2014}
{Takiwaki} T.,  {Kotake} K.,   {Suwa} Y.,  2014, \mn@doi [\apj]
  {10.1088/0004-637X/786/2/83}, \href
  {https://ui.adsabs.harvard.edu/abs/2014ApJ...786...83T} {786, 83}

\bibitem[\protect\citeauthoryear{{Thielemann}, {Nomoto}  \&
  {Hashimoto}}{{Thielemann} et~al.}{1996}]{thielemann96}
{Thielemann} F.-K.,  {Nomoto} K.,   {Hashimoto} M.-A.,  1996, \mn@doi [\apj]
  {10.1086/176980}, \href {http://adsabs.harvard.edu/abs/1996ApJ...460..408T}
  {460, 408}

\bibitem[\protect\citeauthoryear{{Timmes} \& {Swesty}}{{Timmes} \&
  {Swesty}}{2000}]{Timmes.Swesty:2000}
{Timmes} F.~X.,  {Swesty} F.~D.,  2000, \mn@doi [\apjs] {10.1086/313304}, \href
  {http://adsabs.harvard.edu/abs/2000ApJS..126..501T} {126, 501}

\bibitem[\protect\citeauthoryear{{Tout}, {Pols}, {Eggleton}  \& {Han}}{{Tout}
  et~al.}{1996}]{Tout1996}
{Tout} C.~A.,  {Pols} O.~R.,  {Eggleton} P.~P.,   {Han} Z.,  1996, \mn@doi
  [\mnras] {10.1093/mnras/281.1.257}, \href
  {https://ui.adsabs.harvard.edu/abs/1996MNRAS.281..257T} {281, 257}

\bibitem[\protect\citeauthoryear{{Ugliano}, {Janka}, {Marek}  \&
  {Arcones}}{{Ugliano} et~al.}{2012}]{Ugliano2012}
{Ugliano} M.,  {Janka} H.-T.,  {Marek} A.,   {Arcones} A.,  2012, \mn@doi
  [\apj] {10.1088/0004-637X/757/1/69}, \href
  {https://ui.adsabs.harvard.edu/abs/2012ApJ...757...69U} {757, 69}

\bibitem[\protect\citeauthoryear{{Umeda} \& {Nomoto}}{{Umeda} \&
  {Nomoto}}{2008}]{umeda08}
{Umeda} H.,  {Nomoto} K.,  2008, \mn@doi [\apj] {10.1086/524767}, \href
  {http://adsabs.harvard.edu/abs/2008ApJ...673.1014U} {673, 1014}

\bibitem[\protect\citeauthoryear{{Vartanyan} \& {Burrows}}{{Vartanyan} \&
  {Burrows}}{2023}]{Vartanyan2023}
{Vartanyan} D.,  {Burrows} A.,  2023, \mn@doi [\mnras]
  {10.1093/mnras/stad2887}, \href
  {https://ui.adsabs.harvard.edu/abs/2023MNRAS.526.5900V} {526, 5900}

\bibitem[\protect\citeauthoryear{{Vartanyan}, {Burrows}, {Radice}, {Skinner}
  \& {Dolence}}{{Vartanyan} et~al.}{2018}]{Vartanyan2018}
{Vartanyan} D.,  {Burrows} A.,  {Radice} D.,  {Skinner} M.~A.,   {Dolence} J.,
  2018, \mn@doi [\mnras] {10.1093/mnras/sty809}, \href
  {https://ui.adsabs.harvard.edu/abs/2018MNRAS.477.3091V} {477, 3091}

\bibitem[\protect\citeauthoryear{{Vartanyan}, {Laplace}, {Renzo},
  {G{\"o}tberg}, {Burrows}  \& {de Mink}}{{Vartanyan}
  et~al.}{2021}]{Vartanyan2021}
{Vartanyan} D.,  {Laplace} E.,  {Renzo} M.,  {G{\"o}tberg} Y.,  {Burrows} A.,
  {de Mink} S.~E.,  2021, \mn@doi [\apjl] {10.3847/2041-8213/ac0b42}, \href
  {https://ui.adsabs.harvard.edu/abs/2021ApJ...916L...5V} {916, L5}

\bibitem[\protect\citeauthoryear{{Venn} et~al.,}{{Venn}
  et~al.}{2020}]{Venn2020}
{Venn} K.~A.,  et~al., 2020, \mn@doi [\mnras] {10.1093/mnras/stz3546}, \href
  {https://ui.adsabs.harvard.edu/abs/2020MNRAS.492.3241V} {492, 3241}

\bibitem[\protect\citeauthoryear{Virtanen et~al.,}{Virtanen
  et~al.}{2020}]{2020SciPy-NMeth}
Virtanen P.,  et~al., 2020, \mn@doi [Nature Methods]
  {10.1038/s41592-019-0686-2}, \href {https://rdcu.be/b08Wh} {17, 261}

\bibitem[\protect\citeauthoryear{{Weinberg}, {Griffith}, {Johnson}  \&
  {Thompson}}{{Weinberg} et~al.}{2024}]{Weinberg2024}
{Weinberg} D.~H.,  {Griffith} E.~J.,  {Johnson} J.~W.,   {Thompson} T.~A.,
  2024, \mn@doi [\apj] {10.3847/1538-4357/ad6313}, \href
  {https://ui.adsabs.harvard.edu/abs/2024ApJ...973..122W} {973, 122}

\bibitem[\protect\citeauthoryear{{Wellstein} \& {Langer}}{{Wellstein} \&
  {Langer}}{1999}]{Wellstein1999}
{Wellstein} S.,  {Langer} N.,  1999, \mn@doi [\aap]
  {10.48550/arXiv.astro-ph/9904256}, \href
  {https://ui.adsabs.harvard.edu/abs/1999A&A...350..148W} {350, 148}

\bibitem[\protect\citeauthoryear{{Winteler}, {K{\"a}ppeli}, {Perego},
  {Arcones}, {Vasset}, {Nishimura}, {Liebend{\"o}rfer}  \&
  {Thielemann}}{{Winteler} et~al.}{2012}]{Winteler2012}
{Winteler} C.,  {K{\"a}ppeli} R.,  {Perego} A.,  {Arcones} A.,  {Vasset} N.,
  {Nishimura} N.,  {Liebend{\"o}rfer} M.,   {Thielemann} F.-K.,  2012, \mn@doi
  [\apjl] {10.1088/2041-8205/750/1/L22}, \href
  {https://ui.adsabs.harvard.edu/abs/2012ApJ...750L..22W} {750, L22}

\bibitem[\protect\citeauthoryear{{Wollaeger}, {van Rossum}, {Graziani},
  {Couch}, {Jordan}, {Lamb}  \& {Moses}}{{Wollaeger}
  et~al.}{2013}]{Wollaeger2013}
{Wollaeger} R.~T.,  {van Rossum} D.~R.,  {Graziani} C.,  {Couch} S.~M.,
  {Jordan} George~C. I.,  {Lamb} D.~Q.,   {Moses} G.~A.,  2013, \mn@doi [\apjs]
  {10.1088/0067-0049/209/2/36}, \href
  {https://ui.adsabs.harvard.edu/abs/2013ApJS..209...36W} {209, 36}

\bibitem[\protect\citeauthoryear{Woosley \& Heger}{Woosley \&
  Heger}{2007}]{Woosley2007a}
Woosley S.,  Heger A.,  2007, \mn@doi [Physics Reports]
  {https://doi.org/10.1016/j.physrep.2007.02.009}, 442, 269

\bibitem[\protect\citeauthoryear{{Woosley} \& {Weaver}}{{Woosley} \&
  {Weaver}}{1995}]{Woosley1995}
{Woosley} S.~E.,  {Weaver} T.~A.,  1995, \mn@doi [\apjs] {10.1086/192237},
  \href {https://ui.adsabs.harvard.edu/abs/1995ApJS..101..181W} {101, 181}

\bibitem[\protect\citeauthoryear{{Woosley}, {Hartmann}, {Hoffman}  \&
  {Haxton}}{{Woosley} et~al.}{1990}]{Woosley1990}
{Woosley} S.~E.,  {Hartmann} D.~H.,  {Hoffman} R.~D.,   {Haxton} W.~C.,  1990,
  \mn@doi [\apj] {10.1086/168839}, \href
  {https://ui.adsabs.harvard.edu/abs/1990ApJ...356..272W} {356, 272}

\bibitem[\protect\citeauthoryear{{Woosley}, {Langer}  \& {Weaver}}{{Woosley}
  et~al.}{1993}]{Woosley1993}
{Woosley} S.~E.,  {Langer} N.,   {Weaver} T.~A.,  1993, \mn@doi [\apj]
  {10.1086/172886}, \href
  {https://ui.adsabs.harvard.edu/abs/1993ApJ...411..823W} {411, 823}

\bibitem[\protect\citeauthoryear{{Woosley}, {Heger}  \& {Weaver}}{{Woosley}
  et~al.}{2002}]{Woosley2002}
{Woosley} S.~E.,  {Heger} A.,   {Weaver} T.~A.,  2002, \mn@doi [Reviews of
  Modern Physics] {10.1103/RevModPhys.74.1015}, \href
  {http://adsabs.harvard.edu/abs/2002RvMP...74.1015W} {74, 1015}

\bibitem[\protect\citeauthoryear{{Yong} et~al.,}{{Yong}
  et~al.}{2021}]{Yong2021}
{Yong} D.,  et~al., 2021, \mn@doi [\mnras] {10.1093/mnras/stab2001}, \href
  {https://ui.adsabs.harvard.edu/abs/2021MNRAS.507.4102Y} {507, 4102}

\bibitem[\protect\citeauthoryear{{Zepeda}, {Rasmussen}, {Beers}, {Placco},
  {Huang}  \& {Depagne}}{{Zepeda} et~al.}{2022}]{Zepeda2022}
{Zepeda} J.,  {Rasmussen} K.~C.,  {Beers} T.~C.,  {Placco} V.~M.,  {Huang} Y.,
   {Depagne} {\'E}.,  2022, \mn@doi [\apj] {10.3847/1538-4357/ac3a79}, \href
  {https://ui.adsabs.harvard.edu/abs/2022ApJ...927...13Z} {927, 13}

\bibitem[\protect\citeauthoryear{{de Mink}, {Langer}, {Izzard}, {Sana}  \& {de
  Koter}}{{de Mink} et~al.}{2013}]{demink2013}
{de Mink} S.~E.,  {Langer} N.,  {Izzard} R.~G.,  {Sana} H.,   {de Koter} A.,
  2013, \mn@doi [\apj] {10.1088/0004-637X/764/2/166}, \href
  {https://ui.adsabs.harvard.edu/abs/2013ApJ...764..166D} {764, 166}

\makeatother
\end{thebibliography}
